
%
%
%
%
%
%
\input harvmac.tex
\input epsf.tex
\def\Titlemz#1#2{\nopagenumbers\abstractfont\hsize=\hstitle\rightline{#1}%
\vskip 0.4in\centerline{\titlefont #2}\abstractfont\vskip 0.3in\pageno=0}
%


\writetoc 

\Titlemz{\vbox{\baselineskip 12pt \hbox{LAVAL-PHY-21/94}}}
{\vbox {\centerline{ Structure of the conservation laws
in integrable}
\smallskip
\centerline{spin chains with short range interactions}
}}
\centerline{M. P. Grabowski and P. Mathieu$^*$}
\smallskip
\centerline{ \it D\'epartement de
Physique, Universit\'e Laval, Qu\'ebec, Canada G1K 7P4}
\bigskip
\bigskip
\centerline{\bf Abstract}
\smallskip

We present a detailed analysis of the structure of the
conservation laws in
quantum integrable chains of the XYZ-type and in the Hubbard model.
 The essential tool for the former class of models
is the boost operator,
which provides a recursive way of calculation of  the
integrals of motion.
 With its  help,  we establish
the general form of the XYZ  conserved charges
in terms of simple polynomials in spin variables and derive
recursion relations for the relative coefficients of these polynomials.
Although these relations are difficult to solve in general,  a subset
of the coefficients can be determined.
Moreover, for two submodels of the XYZ chain - namely  the XXX and
XY cases, all the charges can be calculated in closed form.
Using this approach,
we rederive the known expressions for the XY charges in a novel way.
For the XXX case, a simple description of conserved charges
is found in terms of a Catalan tree.
This construction is generalized for the $su(M)$ invariant
integrable chain. We also investigate the
circumstances permitting the existence of a recursive (ladder)
operator in general quantum integrable systems.
We indicate that a quantum ladder operator can be traced back to
the presence of a hamiltonian mastersymmetry of degree one in the
classical continuous version of the model. In this way, quantum
chains endowed with a recursive structure can be identified
from the properties of their classical relatives.
We also show that in the quantum continuous limits of the
XYZ model, the ladder property of the boost operator disappears.
For the Hubbard model we demonstrate the non-existence of
a ladder operator. Nevertheless, the general structure
of the conserved charges is indicated,
and the expression for the terms linear in the model's free parameter for all
charges is derived in closed form.

\bigskip
\bigskip
{\noindent $^*$ Work supported by NSERC (Canada).  }
\noindent
\bigskip
\Date{10/94}

\bigskip
\vfill\eject

\newcount\eqnum \eqnum=1
\def\eq{
\eqno(\secsym\the\meqno)
\global\advance\meqno by1
 }
\def\eqlabel#1{
{\xdef#1{\secsym\the\meqno}}
\eq
}

\newwrite\refs
\def\startreferences{
 \immediate\openout\refs=references
 \immediate\write\refs{\baselineskip=14pt \parindent=16pt \parskip=2pt}
}
\startreferences

\refno=0
\def\aref#1{\global\advance\refno by1
 \immediate\write\refs{\noexpand\item{\the\refno.}#1\hfil\par}}
\def\ref#1{\aref{#1}\the\refno}
\def\refname#1{\xdef#1{\the\refno}}

\def\s{\sigma}
\def\t{\tau}
\def\la{\lambda}
\def\vp{\varphi}
\def\ep{\epsilon}
\def\ua{\uparrow}
\def\da{\downarrow}
\def\lra{\leftrightarrow}
\def\ra{\rightarrow}
\def\dg{\dagger}
\let\n=\noindent
\def\frac#1#2{{\textstyle{#1\over #2}}}
\def\sgn{{\rm sgn}}

\def\const{{\rm const}}
\def\ln{{\rm ln}}
\def\diag{{\rm diag}}
\def\max{{\rm max}}
\font\smallcapfont=cmr9
\def\sc#1{{\smallcapfont\uppercase{#1}}}
\def\hf{\frac12}

\def\Tn#1{({\buildrel \leftarrow\over{T^{#1}}})}
\def\Rc{{\check R}}
\def\Tr{{\rm Tr}}

\def\text#1{\quad\hbox{#1}\quad}


\def\ubrackfill#1{$\mathsurround=0pt
	\kern2.5pt\vrule depth#1\leaders\hrule\hfill\vrule depth#1\kern2.5pt$}
\def\contract#1{\mathop{\vbox{\ialign{##\crcr\noalign{\kern3pt}
	\ubrackfill{4pt}\crcr\noalign{\kern3pt\nointerlineskip}
	$\hfil\displaystyle{#1}\hfil$\crcr}}}\limits
}

\def\ubrack#1{$\mathsurround=0pt
	\vrule depth#1\leaders\hrule\hfill\vrule depth#1$}
\def\dbrack#1{$\mathsurround=0pt
	\vrule height#1\leaders\hrule\hfill\vrule height#1$}
\def\ucontract#1#2{\mathop{\vbox{\ialign{##\crcr\noalign{\kern 4pt}
	\ubrack{#2}\crcr\noalign{\kern 4pt\nointerlineskip}
	$\hskip #1\relax$\crcr}}}\limits
}
\def\dcontract#1#2{\mathop{\vbox{\ialign{##\crcr
	$\hskip #1\relax$\crcr\noalign{\kern0pt}
	\dbrack{#2}\crcr\noalign{\kern0pt\nointerlineskip}
	}}}\limits
}

\def\ucont#1#2#3{^{\kern-#3\ucontract{#1}{#2}\kern #3\kern-#1}}
\def\dcont#1#2#3{_{\kern-#3\dcontract{#1}{#2}\kern #3\kern-#1}}

\def \sumL{{\sum_{j\in\Lambda}}}
\def \S{{\bf {\sigma} }}
\def \a{{\la_x}}
\def \b{{\la_y}}
\def \c {{\la_z}}


\font\tenmib=cmmib10
\font\sevenmib=cmmib10 at 7pt
\font\fivemib=cmmib10 at 5pt
\newfam\mibfam 

\textfont\mibfam=\tenmib
\scriptfont\mibfam=\sevenmib
\scriptscriptfont\mibfam=\fivemib
\mathchardef\alphaB="080B
\mathchardef\betaB="080C
\mathchardef\gammaB="080D
\mathchardef\deltaB="080E
\mathchardef\epsilonB="080F
\mathchardef\zetaB="0810
\mathchardef\etaB="0811
\mathchardef\thetaB="0812
\mathchardef\iotaB="0813
\mathchardef\kappaB="0814
\mathchardef\lambdaB="0815
\mathchardef\muB="0816
\mathchardef\nuB="0817
\mathchardef\xiB="0818
\mathchardef\piB="0819
\mathchardef\rhoB="081A
\mathchardef\sigmaB="081B
\mathchardef\tauB="081C
\mathchardef\upsilonB="081D
\mathchardef\phiB="081E
\mathchardef\chiB="081F
\mathchardef\psiB="0820
\mathchardef\omegaB="0821
\mathchardef\varepsilonB="0822
\mathchardef\varthetaB="0823
\mathchardef\varpiB="0824
\mathchardef\varrhoB="0825
\mathchardef\varsigmaB="0826
\mathchardef\varphiB="0827

\def\subsubsec#1{\medskip\goodbreak\noindent
        {\sl #1}\par\nobreak}

{\ifx\answ\bigans
\noindent {1.} {Introduction} \leaderfill{3} \par
\noindent \quad{1.1.} {The problem of calculating conservation laws and ladder
operators in integrable systems} \leaderfill{3} \par
\noindent \quad{1.2.} {An overview of the present results} \leaderfill{9} \par
\noindent \quad{1.3.} {Motivation} \leaderfill{10} \par
\noindent {2.} {The Yang-Baxter equation, commuting transfer matrices and the
boost operator} \leaderfill{11} \par
\noindent {3.} {The conserved charges of the XYZ model: notation and
generalities} \leaderfill{15} \par
\noindent \quad{3.1.} {Notation} \leaderfill{15} \par
\noindent \quad{3.2.} {The structure of the XYZ conservation laws}
\leaderfill{18} \par
\noindent {4.} {The explicit form of the conservation laws in the XXX case}
\leaderfill{20} \par
\noindent \quad{4.1.} {The general formula} \leaderfill{20} \par
\noindent \quad{4.2.} {Details of the proof} \leaderfill{22} \par
\noindent {5.} {The XYZ model revisited} \leaderfill{33} \par
\noindent \quad{5.1.} {Proof of the general pattern for the charges}
\leaderfill{33} \par
\noindent \quad{5.2.} {Recursion relations for the $\mathaccent "707E F$'s}
\leaderfill{34} \par
\noindent \quad{5.3.} {Explicit form of the first few conserved charges}
\leaderfill{36} \par
\noindent {6.} {The conserved charges in the XY model} \leaderfill{38} \par
\noindent \quad{6.1.} {Generalities} \leaderfill{38} \par
\noindent \quad{6.2.} {The XX case} \leaderfill{39} \par
\noindent \quad{6.3.} {The general XYh case} \leaderfill{41} \par
\noindent \quad{6.4.} {Higher order ladder operators} \leaderfill{44} \par
\noindent {7.} {The $su(M)$ invariant spin chain} \leaderfill{45} \par
\noindent \quad{7.1.} {Formulation of the model} \leaderfill{45} \par
\noindent \quad{7.2.} {Expressions for the conserved charges} \leaderfill{46}
\par
\noindent \quad{7.3.} {Relation to higher spin $su(2)$ models} \leaderfill{47}
\par
\noindent {8.} {Equivalent representations of charges in spin chains}
\leaderfill{48} \par
\noindent \quad{8.1.} {$su(2)$ spin chains in the Weyl representation}
\leaderfill{48} \par
\noindent \quad{8.2.} {Representations of the XXX charges in terms of braids}
\leaderfill{49} \par
\noindent {9.} {Structure of conserved charges in the Hubbard model}
\leaderfill{50} \par
\noindent \quad{9.1.} {Introduction} \leaderfill{50} \par
\noindent \quad{9.2.} {The non-existence of a ladder operator} \leaderfill{51}
\par
\noindent \quad{9.3.} {Higher order charges} \leaderfill{52} \par
\noindent \quad{9.4.} {A diagrammatic description of the conserved charges}
\leaderfill{56} \par
\noindent \quad{9.5.} {The explicit form of the term linear in $U$}
\leaderfill{57} \par
\noindent {10.} {Concluding remarks} \leaderfill{58} \par
\noindent Appendix {A.} {Mastersymmetries and hamiltonian structures in
classical soliton theory} \leaderfill{60} \par
\noindent \quad{\hbox {A.}1.} {Symmetries and mastersymmetries} \leaderfill{60}
\par
\noindent \quad{\hbox {A.}2.} {Hamiltonian structures in integrable systems}
\leaderfill{61} \par
\noindent \quad{\hbox {A.}3.} {Hamiltonian mastersymmetries for the KdV and NLS
equations} \leaderfill{62} \par
\noindent \quad{\hbox {A.}4.} {Hamiltonian mastersymmetry of the
Landau-Lifshitz equation} \leaderfill{65} \par
\noindent \quad{\hbox {A.}5.} {Higher order mastersymmetries of the classical
XX model} \leaderfill{67} \par
\noindent \quad{\hbox {A.}6.} {Remark on the classical Heisenberg chain}
\leaderfill{69} \par
\noindent Appendix {B.} {No-go theorem for the existence of a ladder operator
for continuous integrable systems related to the XYZ model} \leaderfill{70}
\par
\noindent \quad{\hbox {B.}1.} {The nonlinear Schr{\accent "7F o}dinger
equation} \leaderfill{70} \par
\noindent \quad{\hbox {B.}2.} {The quantum KdV equation} \leaderfill{72} \par
\else
\noindent {1.} {Introduction} \leaderfill{3} \par
\noindent \quad{1.1.} {The problem of calculating conservation laws and ladder
operators in integrable systems} \leaderfill{3} \par
\noindent \quad{1.2.} {An overview of the present results} \leaderfill{10} \par
\noindent \quad{1.3.} {Motivation} \leaderfill{12} \par
\noindent {2.} {The Yang-Baxter equation, commuting transfer matrices and the
boost operator} \leaderfill{13} \par
\noindent {3.} {The conserved charges of the XYZ model: notation and
generalities} \leaderfill{17} \par
\noindent \quad{3.1.} {Notation} \leaderfill{17} \par
\noindent \quad{3.2.} {The structure of the XYZ conservation laws}
\leaderfill{21} \par
\noindent {4.} {The explicit form of the conservation laws in the XXX case}
\leaderfill{22} \par
\noindent \quad{4.1.} {The general formula} \leaderfill{22} \par
\noindent \quad{4.2.} {Details of the proof} \leaderfill{25} \par
\noindent {5.} {The XYZ model revisited} \leaderfill{37} \par
\noindent \quad{5.1.} {Proof of the general pattern for the charges}
\leaderfill{37} \par
\noindent \quad{5.2.} {Recursion relations for the $\mathaccent "707E F$'s}
\leaderfill{39} \par
\noindent \quad{5.3.} {Explicit form of the first few conserved charges}
\leaderfill{40} \par
\noindent {6.} {The conserved charges in the XY model} \leaderfill{43} \par
\noindent \quad{6.1.} {Generalities} \leaderfill{43} \par
\noindent \quad{6.2.} {The XX case} \leaderfill{44} \par
\noindent \quad{6.3.} {The general XYh case} \leaderfill{46} \par
\noindent \quad{6.4.} {Higher order ladder operators} \leaderfill{50} \par
\noindent {7.} {The $su(M)$ invariant spin chain} \leaderfill{51} \par
\noindent \quad{7.1.} {Formulation of the model} \leaderfill{51} \par
\noindent \quad{7.2.} {Expressions for the conserved charges} \leaderfill{52}
\par
\noindent \quad{7.3.} {Relation to higher spin $su(2)$ models} \leaderfill{53}
\par
\noindent {8.} {Equivalent representations of charges in spin chains}
\leaderfill{54} \par
\noindent \quad{8.1.} {$su(2)$ spin chains in the Weyl representation}
\leaderfill{54} \par
\noindent \quad{8.2.} {Representations of the XXX charges in terms of braids}
\leaderfill{55} \par
\noindent {9.} {Structure of conserved charges in the Hubbard model}
\leaderfill{56} \par
\noindent \quad{9.1.} {Introduction} \leaderfill{56} \par
\noindent \quad{9.2.} {The non-existence of a ladder operator} \leaderfill{58}
\par
\noindent \quad{9.3.} {Higher order charges} \leaderfill{59} \par
\noindent \quad{9.4.} {A diagrammatic description of the conserved charges}
\leaderfill{62} \par
\noindent \quad{9.5.} {The explicit form of the term linear in $U$}
\leaderfill{63} \par
\noindent {10.} {Concluding remarks} \leaderfill{65} \par
\noindent Appendix {A.} {Mastersymmetries and hamiltonian structures in
classical soliton theory} \leaderfill{67} \par
\noindent \quad{\hbox {A.}1.} {Symmetries and mastersymmetries} \leaderfill{67}
\par
\noindent \quad{\hbox {A.}2.} {Hamiltonian structures in integrable systems}
\leaderfill{68} \par
\noindent \quad{\hbox {A.}3.} {Hamiltonian mastersymmetries for the KdV and NLS
equations} \leaderfill{70} \par
\noindent \quad{\hbox {A.}4.} {Hamiltonian mastersymmetry of the
Landau-Lifshitz equation} \leaderfill{73} \par
\noindent \quad{\hbox {A.}5.} {Higher order mastersymmetries of the classical
XX model} \leaderfill{75} \par
\noindent \quad{\hbox {A.}6.} {Remark on the classical Heisenberg chain}
\leaderfill{77} \par
\noindent Appendix {B.} {No-go theorem for the existence of a ladder operator
for continuous integrable systems related to the XYZ model} \leaderfill{78}
\par
\noindent \quad{\hbox {B.}1.} {The nonlinear Schr{\accent "7F o}dinger
equation} \leaderfill{78} \par
\noindent \quad{\hbox {B.}2.} {The quantum KdV equation} \leaderfill{81} \par
  \fi}

\vfill\eject

\newsec{Introduction}
This work is concerned with a detailed analysis of the structure of the
conservation laws for quantum integrable chains with short range interactions.
Attention is focused on the XYZ model:
$$H = \sumL
[\a\s^x_j\s^x_{j+1}+\b\s^y_j\s^y_{j+1}+\c\s^z_j\s^z_{j+1}],\eqlabel\xyzham$$
(where
$\Lambda$ is the spin lattice,
$\s^{x,y,z}_i$ are the
Pauli sigma matrices,
acting non-trivially only on the $i$-th site,
 and
$\a,\b,\c$ are constants), the Hubbard model:
$$H = \sumL [\s^x_j\s^x_{j+1}+\s^y_j\s^y_{j+1}+\t^x_j\t^x_{j+1}+
\t^y_j\t^y_{j+1}+U\s^z_j\t^z_{j}],\eqlabel\hubham$$
(with $\s$ and $\t$ standing for two independent sets of
sigma matrices, and
$U$ a constant), and their close relatives.  A brief survey of known
results precedes the  presentation of our findings.


\subsec{The problem of calculating conservation laws and ladder operators in
integrable systems}

Beyond the existence proofs, very few explicit results are known on the
structure of the conserved charges of integrable quantum chains, even for
rather
basic systems such as the XYZ or the Hubbard model.  Most
of the
results concerning these two particular classes of models
were obtained
by means of the Bethe ansatz method, whose immediate
concern is the construction of the hamiltonian eigenstates
(see [\ref{H. Bethe, {\it Z. Phys.} {\bf 71} (1931), 205 (an
English translation can be found in {\rm ``The Many-Body Problem, An
Encyclopedia of
Exactly Solved Models in One Dimension"} (D. C. Mattis, Ed.),
World Scientific, 1993); C.N.
Yang and C. P. Yang, {\it Phys. Rev.} {\bf 150} (1966),  321, 327; {\bf 151}
 258; R.  Baxter, {\it Ann. Phys. (N.Y.)}  {\bf 76} (1973), 1, 25, 48;
M. Gaudin, {\rm ``La fonction d'onde de Bethe,"} Masson, 1983.}]
for the XYZ-type models,
and [\ref {E. Lieb and F. Y. Wu,
{\it Phys. Rev. Lett.} {\bf 20} (1968), 1445;  B. Sutherland, {\it in}
 {\rm ``Exactly Solvable
Problems in Condensed Matter and Field Theory,"} (B. S. Shastry et al, Ed.),
 Lecture Notes in
Physics {\bf 242}, Springer Verlag, 1985.}] for the Hubbard model).
For the XYZ model, a second input came from the discovery of its equivalence
with an exactly solvable two-dimensional classical statistical model, the
eight-vertex model [\ref{R. Baxter, {\rm ``Exactly Solvable
Models in Statistical Mechanics,"} Academic Press, 1982.}].  The row-to-row
transfer matrix
of the latter model,  denoted by $T(\la)$, where $\la$ is a spectral parameter
arising
in the parameterization of the Boltzmann weights, turns out to commute with the
XYZ
hamiltonian [\ref{B. Sutherland, {\it J. Math. Phys.} {\bf 11} (1970), 3183; B.
McCoy and T. T. Wu, {\it Nuovo Cimento}, {\bf 56B} (1968), 311
.}\refname\Suth]:
$$[T(\la),H] = 0.\eq$$
 The expansion of the logarithm of the transfer matrix in
terms of the spectral parameter
yields a family of
conserved charges. This family is infinite in the infinite chain limit.
 Furthermore, the
row-to-row transfer matrices evaluated for different values of the spectral
parameter
were shown to commute [\ref{
R. J. Baxter, {\it Ann. Phys.} {\bf 70} (1972), 323.}\refname\Baxttr]:
$$[T(\la),T(\mu)]=0.\eqlabel\trancom$$
This directly implies the mutual commutativity of all
these conserved charges and thereby
proves the complete integrability of the XYZ
chain.

The last statement embodies implicitly the standard working definition of
integrability
for a quantum system with $N$ degrees of freedom, namely
the existence of $N$ independent, mutually commuting
integrals of motion.
This aspect (the existence of a family of commuting charges)
of a system that can be solved by the  Bethe
ansatz in its original (coordinate) version is not immediate, although
non-questionable.  In the context of the Hubbard model, the absence of a direct
link
between the coordinate Bethe ansatz and integrability, motivated Shastry
[\ref{B. S. Shastry, {\it Phys. Rev. Lett.} {\bf 56} (1986), 1529;
{\it Phys. Rev. Lett.} {\bf 56} (1986), 2453; {\it J. Stat. Phys.} {\bf
50} (1988), 57.}\refname\Shas]
to parallel the integrability proof of the XYZ
model by constructing the two-dimensional classical statistical version of the
Hubbard
model, whose transfer matrix is related to the Hubbard hamiltonian.  The
demonstration
of (\trancom) for this transfer matrix
 establishes then ``directly" the
integrability of the Hubbard model. 
It should be added that in the
quantum inverse scattering reformulation of the Bethe ansatz (the algebraic
Bethe ansatz)
developed in the last 15 years [\ref{E. K. Sklyanin,  L. A.
Takhtajan and L. D. Faddeev, {\it Theor. Math. Phys.} {\bf 40} (1980), 688;  L.
A.
Takhtajan and L. D. Faddeev, {\it Russ. Mat, Surv.} {\bf 34} (1979), 11;
L.D. Faddeev, {\it in}
{\rm Les Houches Session XXXIX 1982, ``Recent Advances in Field Theory and
Statistical
Mechanics,"}
(J.B. Zuber and R. Stora, Eds.), North-Holland, 1984;  L. A. Takhtajan, {\it
in} {\rm
``Exactly Solvable Problems in Condensed Matter and
Field Theory"} (B. S. Shastry et al, Ed.),
Lecture Notes in Physics {\bf 242}, Springer
Verlag, 1985.}, \ref{V. E. Korepin, N. M.
Bogoliubov and A. G. Izergin, {\rm ``Quantum Inverse Scattering Method and
Correlation
Functions,"} Cambridge University Press, 1993.}\refname\KBI], the transfer
matrix is a
central object, which makes algebraic Bethe ansatz
solvability and quantum integrability
in the above sense, essentially equivalent. However, the Hubbard model has not
yet be
reformulated from that point of view.

Recent works on non-abelian symmetries in quantum chains call for a clarifying
comment concerning the relation between the transfer matrix and the conserved
charges.  The expansion of the logarithm of the transfer matrix in terms of
the spectral parameter
at $\la=0$ (or some other suitable finite value)
leads to {\it local} conserved integrals of
motion, which all mutually commute [\ref{M. L\"uscher, {\it Nucl. Phys.} {\bf
B117}
(1976), 475.}\refname\Lush].  The first member of this infinite sequence is
proportional
to the defining hamiltonian of the quantum chain.
Locality
means that interaction involving a certain set of sites disappears when the
distances between them become sufficiently large.
On the other hand, the expansion
of the monodromy matrix at $\la=\infty$ produces
conserved integrals which are {\it non-local} and {\it non-commuting}
 (see e.g. the
review [\ref{D. Bernard, {\it Int. J. Mod. Phys.} {\bf B7} (1993),
3517.}\refname\Bern]
which focuses on the isotropic case, i.e. the XXX model).
More precisely, these charges all commute with the
hamiltonian and the higher
local charges, but not among themselves.  They in fact
generate a highly non-trivial  non-abelian algebra.  For the XXX  model for
instance,
this algebra provides a realization of  a Yangian [\ref{V. G. Drinfeld, {\it
Sov. Math. Dokl.}
{\bf 32} (1985), 1596; {\it in} {\rm ``Proceeedings of the International
Congress of
Mathematicians,"} Berkeley, Am. Math. Soc., 1987.}].  Notice moreover that the
commutativity of these non-local charges with the local ones is true only in
the limit
of an infinite chain, in contrast with the local ones, which are conserved also
for
finite chains with periodic boundary conditions.
 In this work, we are mainly interested
in the local charges and we will have little to say about the non-local ones.

The most direct way of proving the integrability of a quantum chain,
independently of (any) Bethe-ansatz formalism or a relation with a
two-dimensional statistical system, is certainly to display a quantum Lax
equation that reproduces the hamiltonian equation of motion of the chain.  This
construction can be done rather systematically.  For the XYZ chain and the
Hubbard model, the results can be found respectively in [\ref{Yu. A. Bashilov
and S. V.
Pokrovsky, {\it Comm. Math. Phys.} {\bf 76} (1980), 129.}, \ref{K. Sogo and M.
Wadati, {\it Prog.
Theor. Phys.} {\bf 68} (1982), 85.}] and
[\ref{ M. Wadati, E. Olmedilla and Y. Akutsu
{\it J. Phys. Soc. Jap.} {\bf 56} (1987), 1340.}].  However, this framework
does not
lead to a new way of calculating the conserved integrals.  Indeed, the Lax
operator
depends upon a spectral parameter and the conserved laws are recovered from a
series
expansion, in terms of this  spectral parameter, of the product of all the Lax
operators, one for each site.  This product is nothing but the transfer matrix.

Therefore, at our present state of knowledge, a systematic computation of the
conserved integrals in a generic quantum integrable chain must necessarily
proceed through the series expansion of the
generating function, i.e. the transfer matrix.  But for actual
computations, this method is completely impractical, apart from
the evaluation of the
hamiltonian itself and maybe a first few non-trivial conserved laws.
The reason for this is that the transfer matrix
is a formidable object, with
its  size
growing exponentially with the number of spins.
Even relatively small chains ($|\Lambda| \sim 10$) provide a difficult
challenge for computer algebra programs, requiring vast amounts of
running time and memory.

A different approach is thus clearly desirable.
Fortunately, in the XYZ model there exist an alternative, recursive way
of calculating the conserved charges. It is based on the existence
of a ladder operator, i.e. an operator $B$ with the property:
$$[B, Q_n] \sim  Q_{n+1},\eqlabel\heredi$$
where $Q_n$ denotes a charge with at most $n$ adjacent spins interacting,
with $Q_2$ being proportional to the hamiltonian.\foot{Note that
we allow  for the possibility
of a linear combination
of lower order charges $Q_{m\le n}$ on the (rhs)
of \heredi.}
For the XYZ model, such an operator is provided by the first moment of the
hamiltonian density (see e.g.
[\ref{ B. Fuchssteiner, {\it in} ``Lecture Notes in Physics" {\bf 216} (L.
Garrido, Ed.),
p. 305, Springer Verlag, 1985.}, \ref{ E. K.
Sklyanin, {\it in}
{\rm  ``Quantum Groups and Quantum Integrable Systems,"} World Scientific,
Singapore,  1992.}]);
i.e., if one writes the hamiltonian in the form
$$H \sim \sumL h_{j,j+1},\eq$$
it reads
$$ B = \sumL j~h_{j,j+1}.\eqlabel\bovsh$$

The  operator $B$ generates Lorentz transformations,
by acting on the transfer matrix by
differentiation with respect to the spectral parameter [\ref{ M. G.
Tetelman, {\it Sov. Phys. JETP} {\bf 55} (1981), 306.}\refname\Tet, \ref{H. B.
Thacker,
{\it Physica} {\bf 18D} (1986), 348.}]:
$$[B, T(\la)]= {\partial T(\la)\over \partial \la}.\eqlabel\boost$$
Due to (\boost),  $B$ is usually called a boost operator.
For models with  short-range interaction, which are  related to a statistical
system via the
transfer matrix formalism, both (\boost) and the ladder property (\heredi) are
equivalent.
But since we
emphasize the latter property,
we will most often use the more appropriate ladder qualitative.

The formula (\heredi) is a useful tool.
Not only
does it  provide  a convenient recursive way to
calculate explicit expressions for the conserved charges,
avoiding
the computational difficulties with the expansion of the transfer matrix,
but,  as we will see below, it
is also 
helpful in proving general properties of the family $\{Q_n\}$.
However, even with the use of (\heredi), the
complexity of calculations of  higher order charges
is still considerable.
Moreover, the problem is not only in the
computational complexity,
but also that the results, at first sight, seem hopelessly
complicated and no pattern can be easily discerned.
In one of the earlier works
concerned with calculations of the explicit form\foot{The
three-spin charge $Q_3$ for the XYZ model was calculated in
[\Lush].
The explicit form of the four-spin charge appears in
[\ref{H. Q. Zhou,
L. J. Jiang and J. G. Tang, {\it J. Phys. A:
Math. Gen.} {\bf 23} (1990), 213.}\refname\Zhou].
For the isotropic (XXX) model
it can be also obtained, as  a special case, from the
 four-point charge of the t-J model, written down in
[\ref{F. H. L. Essler, V. E.  Korepin, Phys. Rev. B {\bf 46} (1992), 9147.}].}
of the conservation laws
[\ref{B. Fuchssteiner and U. Falck, {\it in} {\rm ``Symmetries and
Nonlinear Phenomena"} (D. Levi and P. Winternitz, Eds.), p. 22,
World Scientific,
1988.}\refname\FF],
the charge  $Q_6$ is described in the following way:
[it] ``is really a monster representing interactions of up to 6 neighbors and
being a sum of roughly 100 terms (each being an infinite sum). A good
guess is that the explicit form of this hamiltonian will never appear
explicitly in the literature."\foot{But see section (5.3)!}

It is interesting to investigate the  general circumstances permitting the
existence of a
ladder operator in integrable systems.
For classical
integrable systems, the most common recursive structure is the one related to
their bi-hamiltonian character (see e.g. [\ref{A. S. Fokas and I. M. Gelfand,
{\it in} {\rm ``Important Developments in Soliton Theory"}
(A. S. Fokas and V. E.  Zakharov, Eds.), p. 259,  Springer Verlag, 1993;
M. Antonowicz and A. P. Fordy, {\it in} {\rm  ``Soliton Theory:
a Survey of Results"} (A.P. Fordy, Ed.), p. 273, Manchester Univ. Press,
1990.}]).
 Recall that
the existence of {\it two} distinct hamiltonian
structures is
an almost universal signature of integrability is
for classical hamiltonian models.  Take for instance a continuous
system described by a set of fields collectively denoted by
$\varphi$.  The bi-hamiltonian character of the system translates into the
equalities
$$\partial_t\varphi = \{\varphi, H\}_{(1)} = P_1{\delta H\over
\delta \varphi} = \{\varphi, H'\}_{(2)} = P_2{\delta H'\over \delta
\varphi},\eq$$
where $H, H'$ are two distinct hamiltonians, and $P_i$ $(i=1,2)$
denotes the differential operator,
defining the equal-time Poisson bracket $\{~,~\}_{(i)}$
via $$\{\varphi(x), \varphi(y)\}_{(i)} = P_i(x)\delta(x-y).\eqlabel\pois$$
If $H'$ has higher order than  $H$ in terms of an appropriate grading,
it follows that the charge $H_{n+1}$
can be calculated
from the lower order one $H_n$ by
$$P_1^{-1}P_2 {\delta H_n\over \delta
\varphi} = {\delta H_{n+1}\over \delta
\varphi}.\eqlabel\biham$$
The bi-hamiltonian structure furnishes then a recursion
operator,
$P_1^{-1}P_2$, that acts, modulo functional derivatives, as a sort of ladder
operator on the conserved charges.

Constructing the quantum version of a classical integrable  hamiltonian
system amounts to a quantization of  either of the two Poisson structures.
The
point is that the quantization of the system via $P_1$ gives a theory
which, in general
 is quite distinct from the one obtained from the quantization of $P_2$.
Hence, at the quantum level, the bi-hamiltonian character is irremediably
lost.\foot{This
has apparently first been observed in [\ref{B. A. Kuperschmidt and P. Mathieu,
{\it Phys.
Lett.} {\bf B227} (1989), 245.}\refname\KM] in the context of the quantum
Korteweg-de
Vries equation.}  As a consequence,
a priori there are no reasons to expect a recursive scheme
for the calculation of the conserved charges in quantum integrable systems.

As we just argued, the ladder property (\heredi) does not reflect
a sort of accidental bi-hamiltonian structure:
the action of $B$ is clearly not a direct quantum generalization of (\biham).
The classical counterpart of (\heredi)
would be
$$\{B, H_n\}_{(i)} = H_{n+1},\eqlabel\heredicla$$
for some Poisson structure  $\{~, ~\}_{(i)}$
corresponding to the commutator
appearing in (\heredi).
This equation has
a simple, but somewhat unfamiliar
interpretation in soliton theory: it reflects the presence of a
{\it hamiltonian mastersymmetry
of degree 1},
which we denote by
$\tau_1$.
For systems whose evolution equation takes the form
$$\varphi_t = K(\varphi),\eq$$
a mastersymmetry $\tau_1$ is defined by
$$[[\tau_1, K(\varphi)], K(\varphi)]=0,\eq$$
(where the commutators should be understood as  Lie derivatives).
A mastersymmetry is hamiltonian if it can be written as
$$\tau_1= P_i {\delta H_n\over \delta
 \varphi}\eqlabel\hammast$$
for some $H_n$.  If $ P_i$ is a fundamental hamiltonian operator,
meaning that it
cannot be factorized into a product of powers of other hamiltonian
operators,
the existence of a
hamiltonian mastersymmetry of degree 1 is the exception rather than the
rule.\foot{This statement will be made more precise in appendix A, where the
concept of
mastersymmetry is also briefly reviewed. At this point we simply mention that
(\hammast) is not verified for most bi-hamiltonian systems  if
$ P_i$ a fundamental hamiltonian operator.}
But in the
 Landau-Lifshitz model,
which corresponds to a continuous, classical version of the XYZ model,
such a mastersymmetry does exist
[\ref{B. Fuchssteiner, {\it Physica} {\bf 13D} (1984), 387.}\refname\FUC].
This, in a sense,
is the reason for the existence of a ladder operator for the XYZ quantum chain.

In the XYZ context, the
 boost property (\boost) is actually rooted in the Yang-Baxter equation
itself, or more precisely, in the fact that the
 transfer matrix is a product of the  $R$ matrices,
satisfying
  the Yang-Baxter equation [\Tet].  In other words, the Lax operator and the
 $R$ matrix are identical.  In the terminology of [\KBI], such models are said
to be
fundamental.
Somewhat paradoxically, that relation between the boost operator and the
Yang-Baxter
equation
might suggest that
the existence of a quantum ladder operator is a general feature
of integrable quantum chains!  However, since not all integrable spin models
are
fundamental,
 the existence of a ladder operator $B$ satisfying
(\heredi) cannot be the hallmark of quantum integrable chains.
In particular,
no such
operator exists
for the Hubbard model (as shown below); this model is indeed not
fundamental in the above sense.
There
are other examples of integrable chains without ladder operator. For instance,
for the
trigonometric (or hyperbolic) long-range interaction   version of the XXX model
presented
by Haldane and Shastry [\ref{F. D. M. Haldane, {\it Phys. Rev. Lett.}
{\bf 30} (1988), 635; B. S. Shastry, {\it Phys. Rev. Lett.} {\bf 30} (1988),
639.}], no such operator is known.\foot{The
original argument establishing the existence of $B$ for the XYZ model, which is
reviewed in
section (2), breaks down for a chain with long-range interactions.}

It is of great interest to
study the continuous limit of the ladder operator $B$.  Indeed, in appropriate
sectors, ferromagnetic or antiferromagnetic, the XYZ
 model reduces either to the
quantum nonlinear Schr\"odinger equation or the Thirring model (equivalent to
the sine-Gordon equation by bosonization, itself related to the quantum
Korteweg-de Vries equation by a Miura-Feigen-Fuchs transformation).
Although the quantum inverse scattering method furnishes a systematic method
for calculating the conserved charges of these theories, it is again rather
impractical.
Hence, the existence of $B$ in the XYZ model gives a hope that a
ladder operator could exist also in these continuous models. However, as we
will show, this hope does not materialize.


\subsec{An overview of the present results}

Our aim
in this work is to unravel the general pattern of the conserved
integrals.  That amounts to first determining which terms can appear in the
charge of a given order,
and eventually  fixing the values of their coefficients.  Our
strategy is rather straightforward: from the explicit expression of the first
few conserved integrals, we identify the general pattern, which is subsequently
established by means of a recursive argument.  In some favorable circumstances,
all the coefficients can be calculated explicitly. This occurs for two
special cases of the XYZ model: the XXX model
($\a=\b=\c$ in (\xyzham)) and the degenerate XY model ($\c=0$).   The conserved
quantities in the latter case are well-known (the model being equivalent to a
free fermion theory) and have been presented in [\ref{E. V. Gusev,
{\it Theor. Math. Phys.}
{\bf 53} (1983), 1018.}\refname\Gu, \ref{M. Grady, {\it Phys. Rev} {\bf D25}
(1982),
1103.}\refname\Gra, \ref{H. Araki, {\it Comm. Math. Phys. }{\bf 132} (1990),
155.}\refname\Ara, \ref{H. Itoyama and  H. B. Thacker, {\it Nucl. Phys.} {\bf
B320}
(1989), 541.}\refname\IT]. For the infinite XXX model, a general expression for
the  conserved integrals was first reported (without proof)
in  [\ref{ V. V. Anshelevich, {\it Theor. Math. Phys.}
{\bf 43} (1980), 350.}\refname\Ans].  Originally unaware of this work, we have
rediscovered
it in a modified form, presented in [\ref{M. P. Grabowski and P. Mathieu,
{\it Mod. Phys. Lett.} {\bf 9A} (1994), 2197.}\refname\GMa], with
a sketch of the proof.
In this work we give the
details of the
the
proof.
We also extend this construction to the isotropic, $su(M)$ invariant
generalizations  of the XXX chain.
In the general anisotropic case of the $s=1/2$ XYZ model, the
functional form of the integrals of motion is the same as in the isotropic
case.
However, the coefficients become rather complicated polynomials in
the coupling constants. We have been able to determine exactly
only some of these coefficients. No significant simplification of the
XYZ conservation laws has been found in the special XXZ case ($\a=\b$).

In the absence of a ladder operator, as is the case for the Hubbard model, the
general form of the conserved law must first be guessed, with each term
multiplied by an undetermined coefficient. These are then fixed by enforcing
the
commutativity with the hamiltonian.  Fortunately, for the Hubbard model, a
simple pattern soon emerges, in which the building blocks of the conserved
densities are recognized as the XX conserved densities (which is not surprising
since the Hubbard model is essentially two copies of the XX model interacting
along their $z$-component.)  This observation  significantly reduces the number
of a priori undetermined free parameters.  With the
normalization we have chosen for the parameter $U$ (cf. eq. (\hubham)), all
coefficients turn out to be  simply $\pm U^\ell$, with
$\ell$ a non-negative integer.  Furthermore, the different terms in $H_n$
can be grouped into classes, where all the terms of a given class have the
same coefficient and can be obtained from each other by a relative lattice
translation of its components; these classes are in
correspondence with the partitions of the integers $n-2r\geq 0$, where $r$ is
again
a non-negative integer.
The coefficients of the terms
independent of $U$ and those linear in $U$  are found exactly.\foot
{This implies the
knowledge of the explicit form of
all conservation laws of a hypothetical physical model that
would be described by the  Hubbard hamiltonian but with a parameter $U$
satisfying
$U^2=0$, a constraint that does not spoil the integrability of the model.  Such
a constraint can be realized e.g.
by having $U$ of the form $U=\alpha\beta$, with
$\alpha$ and $\beta$ constant Grassmannian parameters.}

Appendices are devoted to the study of some continuous theories related to the
XYZ model.  Appendix A is concerned with a general discussion of
mastersymmetries in
 classical soliton theory, followed by a presentation of the results pertaining
to the
classical Landau-Lifshitz equation.  Although this appendix is mainly composed
with review
material, it contains several
novel observations.  In particular,
it determines
the conditions
under which
classical mastersymmetries and a quantum boost operator can be related.
This provides a way of detecting classical integrable models whose quantum
version will contain a ladder operator.
Next, by a direct analysis presented in the appendix B, we establish the
evanescent
nature of the ladder operator in the continuous limit of the XYZ model, by
disproving the
possible existence of ladder operators both for
the quantum nonlinear Schr\"odinger and the
quantum Korteweg-de Vries equations.

Many of the calculations of the higher order charges in this work, which are
often quite demanding computationally,  were performed
using a set of specialized
``Mathematica"
routines that we developed.


\subsec{Motivation}

What is the motivation for constructing explicitly the conserved charges of the
quantum integrable chains?
First of all, it forces us to look at these systems from a point of view
different from the Bethe ansatz.  New perspectives can obviously lead to
a deeper understanding of these models and more generally of the nature of
quantum integrable systems, for which, after all, relatively little is known.

Another motivation is related to model building.  It is always of interest to
be able to mimic the gross features of a physical problem in terms of an
integrable system.  A simple way of producing new integrable physical models in
the context of an integrable hierarchy, is to take as the basic hamiltonian,
some linear combinations of the conserved charges, choosing the precise
combination that produces the desirable dynamical aspects of the problem under
consideration.  The simplest illustration of this idea is the description of
the
XXZ model in a magnetic field.  It is obtained by adding to the original
hamiltonian, a term proportional to the lowest order conserved charge, namely
the
$z$-component of the total spin, which is multiplied by a constant coefficient
having the physical interpretation of an external magnetic field.  Frahm has
proposed recently a more sophisticated realization of this program, by
considering
a XXZ-type model with basic hamiltonian
$Q_2+cQ_3$
[\ref{H. Frahm, {\it J. Phys. A: Math. Gen.} {\bf 25} (1992), 1417.}].

The knowledge of explicit formulae
for the conserved charges could also be useful in finding exact solutions
for small finite chains (see e.g.
[\ref{K. Fabricius, U. L\"ow, and K.-H. Mutter,
{\it Phys. Rev. B} {\bf 44} (1991), 7476; {\it Zeit. Phys. B} {\bf 91}
(1993), 51.}]).

Our original motivation for this project was to see to what extent information
on the conserved charges of the continuous quantum
integrable systems could be extracted
from those of their lattice regularization.  In particular, if the
explicit formulae for all the XXX
conserved charges are known, does it mean that all the conserved charges of
the corresponding continuous theory are also known?
However, the continuous limit of the higher order charges is
not straightforward.
The problem is essentially that
products of operators in the
continuous theory have to be regularized in some way (e.g.
via normal order)
and no clear print of this regularization appears to be encoded in the
lattice version.
A full discussion of these questions is deferred to a subsequent work.


\def \ms {{\ominus}}
\def \ps{{\oplus}}
\def \pt{{\otimes}}
\def \mt {{\oslash}}
\def \ee {{~~}}
\def \calC{{\cal {C}}}
\def \calS{{\cal {S}}}
\def \calF{{\cal {F}}}
\def \calP{{\cal {P}}}
\def \calA{{\cal {A}}}
\def \calB{{\cal {B}}}
\def \sumC{{\sum_{\calC\in\calC^{(n,k)}}}}
\def \hal{{\frac {1} {2}}}

\def \mult {{\cdot}}
\def \pr{{^\prime}}

\def \ksi{{\xi}}

\def \r#1{{}}
\def \V{{\bf {V}}}
\def \Bbb#1{{\bf #1}}

\def \a{{\la_x}}
\def \b{{\la_y}}
\def \c {{\la_z}}
\def \S{\sigmaB }
\def \sb{\sigmaB}
\def \Sh {{\hat \S}}
\def \sh {{\hat \s}}
\def \St {{\tilde \S}}
\def \st {{\tilde \s}}
\def \ft {{\tilde f}}
\def \F {{\tilde F}}

\def \slashh{\widehat}

\def \Q {{Q }}

\newsec{The Yang-Baxter equation, commuting transfer matrices and the
 boost operator}

We first briefly review the relation between the conserved charges of a quantum
spin chain with
 the commuting family transfer matrices and the Yang-Baxter equation.
Since we do not consider  in this work
non-trivial boundary effects,
the following discussion is confined to infinite
chains or finite chains with periodic boundary conditions.

 Consider a quantum chain with some
spin-like variables
$S^a_i$,
$a=1,...,d$,  defined at site
$i$ and valued in some
Lie algebra, which thereby fixes the defining commutation relation of
the model.  The dynamics is governed by a hamiltonian $H$ containing only
nearest-neighbor
interaction terms.  The canonical equation of motion
$${d {\bf S}_i\over dt} = [{\bf S}_i,H]\eq$$
is supposed to be equivalent to the compatibility condition for the linear
system
$$\eqalign{
{d {\bf S}_i(t)\over dt} =& U_i(\la){\bf S}_i,\cr
{\bf S}_{i+1}(t) =& L_i(\la){\bf S}_i,\cr}\eq$$
which reads
$${d L_i\over dt} = U_{i+1}L_i-L_iU_i.\eq$$
$U$ and $L$ are $d\times d$ matrices depending upon a complex spectral
parameter $\la$.
Finally, the Lax operator $L$ is assumed to satisfy the following
relation:
$$\Rc(\la-\mu)( L_i(\la)\otimes L_i(\mu)) =(L_i(\mu)\otimes
L_i(\la))\Rc(\la-\mu),\eqlabel\rllrel$$ where $\Rc$ is a $d^2\times d^2$ matrix
with
normalization $\Rc(0) =I$.  The above tensor product lives in the space
$V_1\otimes V_2$
where $V_1\simeq V_2\simeq V$ is the
$d$-dimensional space on which the $d\times d$ quantum matrix operators
$L_i(\la)$ act.  To
indicate that $\Rc$ acts on $V_1\otimes V_2$, we rewrite it as $\Rc_{V_1V_2}$
or
$\Rc_{12}$. Equation (\rllrel) implies as a compatibility condition
the famous Yang-Baxter equation:
$$\Rc_{23}(\la-\mu) \Rc_{12}(\la) \Rc_{23}(\mu) =
\Rc_{12}(\mu) \Rc_{23}(\la)
\Rc_{12}(\la-\mu).\eqlabel\yba$$
where $V_3\simeq V$.
 It is usually written in terms of the variable $R$ related to $\Rc$ by
$$\Rc = PR,\eqlabel\rrc$$
where $P$ is a transposition operator, i.e.
$$P_{i,j}R_{jk}P_{i,j} = R_{ik}.\eq$$
The relation (\yba) directly implies that the transfer matrix $T(\la)$
$$T(\la) = \Tr_{V_0} L_N(\la)...L_1(\la)\eq$$
commutes with $T(\mu)$. Expanding the transfer matrix as
$$\ln ~T(\la) =2i  \sum_{n=0}^\infty {\la^n\over n!}Q_{n+1},\eqlabel\defq$$
one obtains a family of mutually commuting quantities $\{Q_n\}$:
$$[Q_n,Q_m]=0.\eq$$

The  matrix elements
of the  operator $L_i(\la)$ are functions of the
quantum variables ${\bf S}_i$, themselves $d\times d$ matrices. Matrix entries
of $L_i$
act then on a $d$-dimensional space, called the quantum space and denoted by
$V_0$.
Because $V_0\simeq V$, we can set
$$L_i(\la) = R_{i0}\eq$$
(up to a normalization factor). In that case, (\rllrel) becomes
$$R_{12}(\la-\mu) R_{10}(\la) R_{20}(\mu) =
R_{20}(\mu) R_{10}(\la)
R_{12}(\la-\mu),\eqlabel\ybb$$
which, thanks to (\rrc), is the same as (\yba).

Spin models for which the Lax operator is the same as the $R$ matrix are called
fundamental [\KBI].  In such cases, the quantum hamiltonian can be expressed
directly in terms of the $R$ matrix. Indeed, the transfer matrix is then a
product of
$R$ matrices, one for each site:
$$T(\la) = \Tr_{V_0} R_{N0}(\la)...R_{10}(\la),\eq$$
so that
$$\eqalign{
{dT(\la)\over d\la}|_{{}_{\la=0}} = &\sum_i  \Tr_{V_0} P_{N,0}...P_{i+1,0}
{dR_{i0}(\la)\over
d\la}|_{{}_{\la=0}}~(P_{i,0}P_{i,0})P_{i-1,0}..P_{1,0}\cr
 =&\sum_i {d\Rc_{ii+1}(\la)\over
d\la}|_{{}_{\la=0}}~ \Tr_{V_0} P_{N,0}...P_{1,0}\cr}\eq$$
(in the first line we used the fact that $P_{i,0}^2=I$).  Up to a constant,
this is simply
the quantum chain hamiltonian
$$H = -T(0)^{-1}{dT(\la)\over d\la}|_{{}_{\la=0}} = -\sum_i
{dR_{ii+1}(\la)\over
d\la}|_{{}_{\la=0}}~P_{i,i+1} = \sum_i h_{i,i+1}.\eqlabel\hdefi$$

 We next briefly
review the relation between the Yang-Baxter equation and the boost operator
[\Tet].  The starting point is (\ybb) with $1\to k,~2\to k+1$ and
$\la=\nu+\mu$.  The
derivative of this expression with respect to $\nu$ evaluated at $\nu=0$ yields
$${dR_{0k}(\mu)\over
d\mu}R_{0k+1}(\mu)-R_{0k}{dR_{0k+1}(\mu)\over
d\mu} = [h_{k,k+1},R_{0k}R_{0k+1}],\eq$$
where $h_{k,k+1}$ is defined in (\hdefi).  This result is then multiplied from
the left by
$\prod_{n<k}R_{0n}(\mu)$ and from the right by $\prod_{n>k+1}R_{0n}(\mu)$:
$$\eqalign{
(\prod_{n<k}R_{0n}(\mu)){dR_{0k}(\mu)\over
d\mu}(\prod_{n>k}R_{0n}(\mu)) - &
(\prod_{n<k+1}R_{0n}(\mu)){dR_{0k+1}(\mu)\over
d\mu}(\prod_{n>k+1}R_{0n}(\mu))\cr
&\qquad=[h_{k,k+1},T(\mu)].\cr}\eq$$
Multiplication by $k$ followed by a summation over $k$ from $-\infty$ to
$\infty$ leads
then to (\boost),
with $B$ defined by (\bovsh).   From (\defq), this directly implies
$$[B, Q_n] = Q_{n+1}.\eqlabel\qndef$$
For finite chains with periodic boundary conditions, the argument is the same
except that
the coefficients are understood to be defined modulo $N$, the number of sites.

The logarithm in (\defq) ensures the locality of the charges $Q_n$,
i.e. that
$Q_n$ does not
contain interactions between spins at distances greater then $n$ lattice units.
As shown by L\"uscher [\Lush] for the XYZ model, these
charges can be put in the form:
$$Q_n=\sum_{\{ i_1, \dots, i_{n-1} \} } G_{n-1}^T(i_1, \dots,
i_{n-1}),\eqlabel\GT$$
where the summation is over ordered subsets $\{i_1,\dots,i_{n-1}\}$ of
the chain, and $G^T$ is a translation covariant and
totally symmetric function, with  the locality property:
$$G_n^T(i_1,\dots,i_n)=0,\quad {\rm for}\quad |i_n-i_1|\ge n .\eqlabel\locpro$$
For the infinite XXX chain, further properties of the conserved charges,
including their
completeness, have been proved in [\ref{ D. Babbit, L. Thomas, {\it J. Math.
Anal.}
 {\bf 72}
(1979), 305}\refname\Babbi].\foot{To our knowledge, no such completeness
proof exists for the anisotropic case. }

In the rest of the paper we will reserve the notation $\{Q_n\}$ for the
charges defined by (\defq). Note that a rescaling of the spectral
parameter $\la\to\alpha^{-1}\la$ results in a multiplicative
redefinition of the $Q_n$'s : $Q_n\to\alpha^{n-1} Q_n$.
For the XYZ model it is convenient to use the freedom to rescale the
spectral parameter so that  the logarithmic derivative of the transfer matrix
becomes:
$$
 -T^{-1}(0) \dot T(\la)|_{\la=0} =
{1 \over {2 i}} Q_2=
{1 \over {2 i}}
\sumL
[\a\s^x_j\s^x_{j+1}+\b\s^y_j\s^y_{j+1}+\c\s^z_j\s^z_{j+1}].
\eqlabel\Tnorm$$
With this normalization, the boost operator for the XYZ model reads:
$$B={1 \over {2 i}}\sumL j
[\a\s^x_j\s^x_{j+1}+\b\s^y_j\s^y_{j+1}+\c\s^z_j\s^z_{j+1}].\eq $$
The structure of the conservation laws is however more transparent in another
basis,
which may be obtained by taking an appropriate
linear combinations of $Q_n$'s. Such a basis will be denoted $\{H_n\}$
and chosen so that  $H_n = {1\over {(n-2)!}} Q_n+$ a linear combination of
lower order $Q_i$'s.


\newsec{The conserved charges of the XYZ model: notation and generalities}

\subsec{Notation}
We begin by introducing the necessary
notation.
 We consider  either a finite spin lattice with
periodic boundary conditions ($\Lambda=\{1,\dots, N\}$, with $N+1\equiv 1$)  or
an infinite
one  ($\Lambda={\Bbb Z}$).  Our constructions  applies  equally well to
both cases, provided that
addition in $\Lambda$ is understood modulo $N$ for the finite chain.
We now introduce two general objects that enter naturally in the construction
of the
conserved charges: clusters and patterns.

A sequence of $n$
lattice sites ${\calC=\{{i_1}, ..., {i_n}\}}$,
with $i_1<i_2<...<i_n$, will be called  a {\it cluster} of order $n$.  A
disordered cluster  will refer to a sequence with the ordering condition
relaxed.\foot{This
definition of a cluster differs slightly from the one that
we used previously in [\GMa].}
A cluster starting at $i_1$ and ending at $i_n$, and containing  non-adjacent
sites has $k=i_n-i_1+1-n$ {\it holes} (sites in between  $i_1$ to  $i_n$
that are not included in $\calC $); $k=0$ for a cluster containing only
contiguous spins.
The set of all clusters of $\Lambda$ of
order $n$ with $k$ holes  will be  denoted as  ${\calC}^{(n,k)}$.
 For instance, $\calC^{(4,2)}$ contains
$\{1,2,3,6\}$, $\{1,2,5,6\}$, $\{1,4,5,6\}$,
$\{1,2,4,6\}$, $\{1,3,4,6\}$, $\{1,3,5,6\}$ and all their translations.

Let $\calS^{(n)}$ denote the set of all sequences of $n$ basis spin
matrices  of the form
\eqn\spseq{{\calS=\{\s_{i_1}^{a_1}, ..., \s_{i_n}^{a_n}\},}}
where $a_i=x,y,z$,
and  $i_1<i_2<...<i_n$. Clearly, such a sequence is completely specified by
the cluster $\{i_1,\dots,i_n\}$ and by the assignment of group
indices, $\{a_1,\dots,a_n\}$, which we call a {\it pattern}.
We denote by $\calP^{(n)}$ the set of all
$n$-spin {\it patterns}, i.e.  sequences of $n$ elements
taking values in the set $\{x,y,z\}$.
There exist natural projections, assigning to a
spin-sequence $\calS\in\calS^{(n)}$ its corresponding pattern or cluster:
$$p:\calS^{(n)}\to \calP^{(n)} \text{and} p\pr : \calS^{(n)}\to \bigcup_k
{\calC}^{(n,k)}\eq$$ given by:
$$p(\calS)=\{a_1, a_2,\dots,a_n\}, \quad\quad
p\pr (\calS )=\{i_1, i_2, \dots,i_n\}.\eq$$
 We will write $\calC^\pi$ to denote
the sequence $\calS$ with  cluster $C$ and  pattern $\pi$.

Next  we  construct the $n$-linear spin polynomials $\ft_n$, defined
on basis sequences in $\calS^{(n)}$, which will be
the building blocks of our construction.
First we define two different rescaling of the spin
variables:\foot{Although these
rescalings involve square roots of the coupling constants,
the resulting expressions for charges will
contain only integer powers of
the couplings.}
\eqn\Sresc{\eqalign{ \sh^a_j&=\sqrt{\la_a}~ \s^a_j ,\cr
\st^a_j&=
{\sqrt{ {\a\b\c} \over {\la_a} }}~ \s^a_j.} }
In these variables, the XYZ hamiltonian reads:
$$ H_2=\sumL \sh^a_j \sh^a_{j+1}. \eq $$
Then, for any sequence of spins $\calS\in \calS^{(n)} $, we define
$\V_m(\calS)=\{V_m^x, V_m^y, V_m^z\}(\calS)$ (for $m<n$) as the
vector product of the first $m$ spins of the sequence, where the leftmost
spin factor appears
with a hat and all others with a tilde, with parentheses nested toward the
left,
i.e.:
 \eqn\jj {\eqalign{&\V_1(\calS) ={\bf \Sh}_{i_1},\cr
&\V_2(\calS)={\bf \Sh}_{i_1}\times {\bf \St}_{i_2},\cr
&\V_3(\calS)=({\bf \Sh}_{i_1}\times {\bf \St}_{i_2})\times {\bf \St}_{i_3},
\cr &\dots \cr
&\V_{m}(\calS)=\V_{m-1}(\calS) \times {\bf \St}_{i_{m}}.}}
We also denote by $a_m(\pi)$ the `` direction" of ${\bf V}_m(\calC^\pi)$, i.e:
$$ a_m(\pi)=(\dots(a_1\times a_2)\times a_3)\dots\times a_m).\eqlabel\api$$
Then,
we construct scalar $n$-linear polynomials, defined in the space of
spin sequences, from the  scalar product of ${\bf V}_{n-1}(\calS)$ and ${\bf
\Sh}_{i_n}$,
multiplied by an appropriate constant $g(\calS)$:
\eqn\ftn {\ft_n(\calS)=g(\calS) (\V_{n-1}(\calS)\mult {\bf \Sh}_{i_n}),}
with $\ft_0(\calS)=\ft_1(\calS)=0$.
The coefficient $g(\calS)$ is a  polynomial in $\a, \b, \c$, determined by
the pattern and the
positions of the holes in the cluster:

\n (i)- If $\calS$ contains only contiguous spins,
$g(\calS)=1$. In this case, \eqn\ftnbis{\eqalign {\ft_n(\calS)&=
((\dots ((\sh^{a_1}_{i_1}
\times \st^{a_2}_{i_2})\times
\st^{a_3}_{i_3})\times \dots ) \st^{a_{n-1}}_{i_{n-1}})\cdot  \sh^{a_n}_{i_n},
\cr
&=\epsilon(\pi) \quad
\sh^{a_1}_{i_1}
\st^{a_2}_{i_2}\dots
\st^{a_3}_{i_3} \st^{a_{n-1}}_{i_{n-1}}\sh^{a_n}_{i_n}, }}
where
$$\epsilon(\pi)=\sgn[ a_{n-1}(\pi)] \delta_{a_{n-1}(\pi),a_{n} }\eq$$
with the convention $\sgn(0)=0$.  Hence, $\epsilon(\pi)$ can be either 0 or
$\pm 1$.
For example, we have:
\eqn\ff{\eqalign{ &\ft_2( \{\s^x_i,\s^x_{i+1}\})=
\a \;\s^x_{i}\s^x_{i+1}, \cr
&\ft_3(\{\s^x_i,\s^y_{i+1},\s^z_{i+2}\})=\a \b \;\s^x_{i} \s^y_{i+1}
\s^z_{i+2},\;
\cr &\ft_3(\{\s^x_i,\s^y_{i+1},\s^y_{i+2}\})=0 ,\cr
&\ft_4(\{\s_i^x,\s_{i+1}^y, \s^y_{i+2},\s^x_{i+3}\})=-
\a^2\c \;\s^{x}_{i} \s^y_{i+1}\s^y_{i+2}\s^x_{i+3}.}}

\n (ii)- For clusters containing holes, the coefficient $g(\calS)$ is given by
\eqn\gofS{
 g(\calS)=\prod_{j=i_1+1;\atop j\notin\calC}^{i_{n-1}} 
{  {\a\b\c} \over {\la_{|a_{L(j)}(\pi)|}}  }
, }
where $L(j)$ is the number of sites in $\calC$ to the left of site $j$ and
$a_{L(j)}(\pi)$
has been defined above in (\api).
For example,
\eqn\ff{\eqalign{ &\ft_2( \{\s^x_i, \s^x_{i+2}\})=
(\b\c)\a  \;\s^x_{i}\s^x_{i+2}, \cr
&\ft_3(\{\s^x_i,\s^y_{i+2},\s^z_{i+3}\})=(\b\c) \a \b \;\s^x_{i} \s^y_{i+2}
\s^z_{i+3},\; \cr
&\ft_3(\{\s_i^x,\s_{i+1}^y, \s^y_{i+3},\s^x_{i+4}\})=-
(\a\b) \a^2\c\; \s^x_{i} \s^y_{j}\s^y_{k}\s^x_{l},
}}
(where the coefficient $g(\calS)$  is indicated in parentheses).

Finally, we define the linear
spaces $\calF_{n,k}$ spanned by the
translation invariant local polynomials built from $\ft_n$'s:\foot
{$\calF_{n,k}$ is a linear subspace of dimension $3 \cdot 2^{n-2}$
of the space of observables; since the product of
two local polynomials is not local, it is not an algebra.}
\eqn\spaceF{ \calF_{n,k} =
\sum_{\pi \in \calP^{(n)} } \alpha(\pi)
\sum_{\calC\in \calC^{n,k}} \ft_n(\calC^{\pi}), }
where $\alpha(\pi)$ are some real-valued coefficients.

In the isotropic case, i.e. for the XXX model,
all the coupling dependent coefficients
become one, which results in a significant  simplification.
We can then define the polynomials $f_n$ whose arguments are clusters in
 $\calC^{n,k}$, as
follows:
\eqn\fnXXX{ f_n(\calC)=\sum_{\pi\in \calP^{(n)} }\ft_n(\calC^{\pi}) .}
In particular, omitting the argument of the $f_n$'s for brevity, we have
\eqn\ff{\eqalign{ &f_0=f_1=0,\cr&f_2={\bf \S}_{i_1}\mult {\bf \S}_{i_2}, \cr
&f_3=({\bf \S}_{i_1} \times {\bf\S}_{i_2})\mult {\bf\S}_{i_3},\cr
 &f_4=(({\bf\S}_{i_1}\times
{\bf\S}_{i_2})\times {\bf\S}_{i_3})\mult {\bf\S}_{i_4}.}}

The multilinear polynomials $f_n$'s and $\ft_n$'s satisfy an interesting
property which is that the dot product
can be placed at an arbitrary position, provided that parentheses to its left
(right)
are nested toward the left (right), e.g:
\eqn\exnest{\eqalign{f_5 &=((( {\bf\S}_{i_1}\times {\bf\S}_{i_2})\times
\sb_{i_3})\times \sb_{i_4})
\mult \sb_{i_5}= (( \sb_{i_1}\times \sb_{i_2})\times \sb_{i_3})\cdot (\sb_{i_4}
\times \sb_{i_5})\cr &= \sb_{i_1} \cdot ( \sb_{i_2}\times (\sb_{i_3}\times
(\sb_{i_4}
\times \sb_{i_5}))).  }}
This is a direct consequence of the familiar vector identity:
\eqn\famid{ ( {\bf A}\times {\bf B})\cdot {\bf C}={\bf A}\cdot
({\bf B}\times {\bf C}).   }

\subsec{The structure of the XYZ conservation laws}

The conserved charges of the XYZ model have the general form:
\eqn\Hxyz{ Q_n=  \sum _{k=0}^{[n/2]-1} \sum_{\ell=0}^{k} {\tilde
{F}}^n_{n-2k,\ell},}
where
 the square bracket stands for integer part, and
$\tilde F^n_{n-2k,\ell}$ is an element of ${\cal F}_{n-2k,\ell}$. More
precisely,
\eqn\Hform{ \tilde F^n_{n-2k,\ell}=\sum_{\pi\in \calP^{(n-2k)}}
\theta^n_{n-2k,\ell}(\pi) \sum_{
\calC \in {\calC^{(n-2k,\ell)} }} \tilde f_{n-2k} (\cal C^\pi),}
where $\theta^n_{n-2k,\ell}$ is a polynomial of degree $2(k-l)$ in the
coupling constants $\a,\b,\c$.
It is convenient to think of $Q_n$ in terms of a triangle,
consisting of the operators $\tilde F^n_{n-2k,\ell}$:
\eqn\xyztr{\eqalign{&\F^n_{n,0}\cr \F^n_{n-2,1}\quad\quad &\F^n_{n-2,0}\cr
\F^n_{n-4,2}\quad\quad \F^n_{n-4,1}\quad\quad &\F^n_{n-4,0}\cr
\F^n_{n-6,3}\quad\quad \F^n_{n-6,2}\quad\quad \F^n_{n-6,1}\quad\quad
&\F^n_{n-6,0}\cr
\F^n_{n-8,4}\quad\quad \F^n_{n-8,3}\quad\quad \F^n_{n-8,2}\quad\quad
\F^n_{n-8,1}\quad\quad &\F^n_{n-8,0}\cr
\dots \quad \quad \quad \quad \quad &\quad
\quad\quad
}}
The bottom edge of the triangle is formed by
the sequence $\{\F^n_{2,k}\}_{k=0, \dots, [n/2]-1}$ if $n$ is even, or
$\{\F^n_{3,k}\}_{k=0, \dots, [n/2]-1}$ if $n$ is odd.
Note that linear combinations of the form
\eqn\lcom{H_n=\sum_{k=0}^{[n/2]-1}
p_{n-2k}(\a,\b,\c)
Q_{n-2k} ,}
 where
$p_{n-2k}$ is an arbitrary
 polynomial of degree $n-2k$ in the coupling constants
preserve the form of the triangle, redefining only
the polynomials $\theta(\pi)$.

The problem is how to determine the coefficients $\theta(\pi)$.
In section (5), where we prove \Hxyz,
we give a recursive method to calculate $\F^n_{n-2k,\ell}$.
Some of the coefficients are easy to calculate: e.g.
one finds  that $\theta^n_{n-2k,k}(\pi)=(n-2)!$ for all patterns $\pi$ (thus
$\F^n_{n-2k,k}\sim \sum_{\pi\in \calP^{(n-2k)}}\sum_{\calC\in
\calC^{(n-2k,k)}}\ft_n(\calC^\pi)$).
This fixes the coefficients on the left edge of the triangle.
Unfortunately, finding all the coefficients in the
general anisotropic XYZ case is a very difficult task and we have
not succeeded in disentangling the equations for the $\theta$'s apart for those
on the left edge.
However, there are two particular cases in
which all the coefficients are relatively simple and can be
determined exactly, namely when $\c=0$ or in the isotropic limit where
$\a=\b=\c$.
In the XY model, which is analyzed in
section (6), the triangle degenerates into its right edge.  On the other hand,
for the XXX
model, treated in detail in the next section, all the coupling-dependent
coefficients
become constants.  In that case, by an appropriate
transformation of the type \lcom,  one can eliminate all the terms on the
right edge, except for $\F^n_{n,0}$.  This decoupling provides enough
simplification in the
recursion relations for the coefficients to allow for an exact solution.

For the XXZ model, the recursion formulae are somewhat simplified, being
dependent only on a single parameter $\c/\a$. However, this
simplification is not as dramatic as in the XXX case, and
it does not seem to permit a solution in closed form for all the
coefficients.

\newsec{The explicit form of the conservation laws in the XXX case}

\subsec{The general formula}

For the isotropic XXX
model the only quantities that enter in the construction of the
 local integrals of motion
are
\eqn\Fnk{ F_{n,k}=\sum_{\calC \in \calC^{(n,k)}} f_n(\calC).}
Quite remarkably, the
linear combination of $F_{n,k}$'s that builds up
$H_n$ is encoded in a simple tree,
in which the vertices are labeled by the $F_{n,k}$'s.
This tree
 is displayed in Fig. 1.

\vskip 0.7cm
\vbox{
$$ {\epsfxsize= 10.0cm \epsfbox{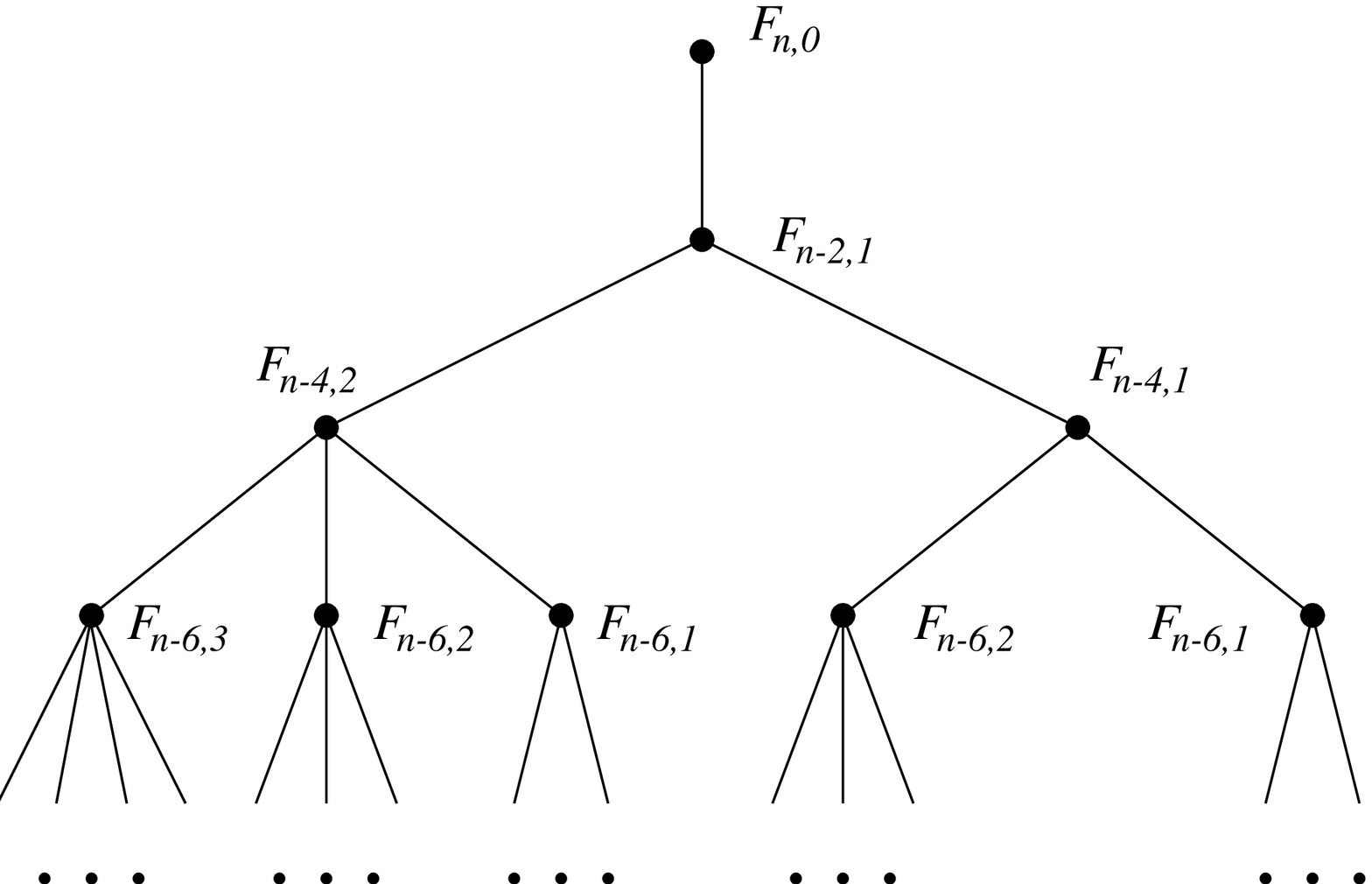}}\ \ \  $$
\vskip 0.8cm
Fig. 1.
The tree structure corresponding to $H_n$. The tree stops with the terms
$F_{2,\ell}$ ($F_{3,\ell}$) when $n$ is even (odd).
Note that the trees corresponding to $H_{2m}$ and $H_{2m+1}$ have identical
structure, differing  only in labeling of their vertices.
}
\vskip 1.0cm

The sum of all the
vertices of the tree, with each vertex contributing with unit weight,
gives $H_n$.
Summing up all the vertices of the $H_n$ tree leads then to the following
expression:
\eqn\Hndef{ H_n= F_{n,0}+ \sum _{k=1}^{[n/2]-1} \sum_{\ell=1}^{k}
\alpha_{k,\ell} F_{n-2k,\ell},}
where
the coefficients $\alpha_{k,\ell}$ are defined  recursively by the
relation:
\eqn\alphadef{ \alpha_{k+1,\ell}=\sum_{m=\ell-1}^k \alpha_{k,m},}
with $\alpha_{1,1}=1$ and $\alpha_{k,0}=0$.
Notice that $\alpha_{k,1}=\alpha_{k,2}$ for $k\ge 2$.
These coefficients turn out to be  generalized Catalan numbers, as can be seen
by rewriting the
recurrence relation \alphadef\
in the form:
\eqn\recalpha{ \alpha_{k,\ell}=\alpha_{k-1,\ell-1} +\alpha_{k,\ell+1},}
with the convention   $\alpha_{k,\ell}=0$ if $\ell>k$.
This is the defining relation for the
 generalized Catalan numbers,
${ \alpha_{k,\ell}=C_{2k-l-1, \ell},}$
 with $C_{n,m}$ given by
\eqn\Catalrel{ C_{n,m}=\pmatrix {n-1 \cr p}-
\pmatrix {n-1 \cr p-2},}
where
$ \pmatrix{ a \cr b}$ are the binomial coefficients, with
$p=[(n-m+1)/2]$,
$m+n$ odd and
$m<n+2$.
In particular, $\alpha_{k,1}=C_{2k,1}$ reproduce the familiar  sequence of
Catalan numbers:
$\{1,1,2,5,14,42,132,429,1430,4862\dots\}$.
 The tree in Fig. 1 is known as a Catalan tree.

The first few conserved charges are:
\eqn\Hnex {\eqalign{ &H_2=F_{2,0} , \cr 
&H_3=F_{3,0},\cr
&H_4=F_{4,0}+F_{2,1},\cr
&H_5=F_{5,0}+F_{3,1},\cr
&H_6=F_{6,0}+F_{4,1}+F_{2,2}+ F_{2,1},\cr
&H_7=F_{7,0}+F_{5,1}+F_{3,2}+ F_{3,1}, \cr
&H_8=F_{8,0}+F_{6,1}+F_{4,2}+ F_{4,1}+ F_{2,3}+ 2 F_{2,2}+2 F_{2,1},\cr
&H_9=F_{9,0}+F_{7,1}+F_{5,1}+F_{5,2}+ F_{3,3} + 2 F_{3,2}+2 F_{3,1},\cr
&H_{10}=F_{10,0}+F_{8,1}+F_{6,2}+ F_{6,1}+ F_{4,3}+ 2 F_{4,2}+2 F_{4,1}\cr
&\qquad~ +F_{2,4}+ 3 F_{2,3}+5 F_{2,2} +5 F_{2,1}.}}

We note that \Hndef\ can be put in the form \Hxyz\ with
$F^{n}_{n,0}=F_{n,0}$, $F^n_{n-2k,0}=0$ for $0<k<[n/2]-1$,
$F^n_{n-2k,\ell}=\alpha_{k,\ell}F_{n-2k,l}$.

The above construction gives
$N-1$ charges $\{ H_2,\dots H_N \} $
for the XXX chain of length $N$ with periodic boundary conditions.
We can add to the set $\{H_n\}$ 
any of the three components of the total spin,
${ H_1^a=\sum_{j\in \Lambda} S^a_j }$.
As the $F_{n,k}$'s are invariant under global spin rotation,
$[H_1^a, H_n]=0$.\foot{Thus the charges \Hndef\
survive in the presence of a magnetic field
coupled to any of the  components of the total spin.}
For the infinite isotropic Heisenberg chain we have then an infinite
sequence of independent local charges $\{ H_1^a, H_2,...,   H_n,... \}$.

\subsec{Details of the proof}

We now present the direct algebraic  proof of the mutual commutativity of all
the $\{ H_n \}
$'s.
We will first show that
${[ H_2, H_n]=0}$,
by evaluating the commutator:
\eqn\comlink{ [H_2, f_n(\calC)]=\sum_{j\in\Lambda} [\S_j \cdot\S_{j+1},
f_n(\calC)].}
The proof is rather lengthy and some of its steps are quite tedious.
It is divided into nine parts.


\subsubsec{a)  Three useful identities}

To proceed, we need a few simple identities for commutators
in the $su(2)$ tensor product space.
Let ${\bf A}=\{\s^x_j,\s^y_j,\s^z_j\}$
and ${\bf B}=
\{\s^x_k,\s^y_k,\s^z_k\}$
denote  the basis of two different ($j\ne k$)  $su(2)$ factors, and let
${\bf L}=\{L^1,L^2,L^3\}$ and ${\bf R}=\{R^1,R^2,R^3\}$
be arbitrary tensor products commuting with all components of both ${\bf A}$
and ${\bf B}$. The
following identities hold:
\eqn\idone{[ {\bf A}\cdot {\bf B}, ({\bf B}\times {\bf L})\cdot {\bf R}]=
-2 i (( {\bf A}\times {\bf B})\times {\bf L})\cdot {\bf R} ,}
\eqn\idtwo{[ {\bf A}\cdot {\bf B}, ({\bf A}\times {\bf B})\cdot {\bf R}]=
4 i ( {\bf B}\cdot {\bf R}) - 4 i ({\bf A}\cdot {\bf R}) ,}
\eqn\idthr{[ {\bf A}\cdot {\bf B}, (({\bf L}\times {\bf A})\times {\bf B})
\cdot {\bf R}]=
2 i ( {\bf L}\times {\bf B})\cdot {\bf R}
-2 i ( {\bf L}\times {\bf A})\cdot {\bf R}
 .}
These identities are simple consequences of the basic $su(2)$
multiplication rule for the components of ${\bf A, B}$:
\eqn\bsutwo{ A^a A^b=\delta^{ab}+ i \epsilon_{abc} A^c.}

\subsubsec{b) Commuting  $\sb_i \sb_{i+1}$ with $f_n(\calC)$}

The  above identities are helpful in evaluating the commutators of the basic
links $\sb_i \sb_{i+1}$ with $f_n(\calC)$.
It is remarkable that all such commutators
are of the form
$f_{n\pm 1}(\calC^\prime)$, where the cluster $\calC^\prime$
can be  obtained by applying
 a number of simple transformations
to the cluster
$\calC=\{{i_1},\dots,{i_n}\}$:
\eqn\rulescom{\eqalign{
{}_{i_1-1} \calC &\equiv \{{i_1-1}, {i_1},\dots,{i_n}\},\cr
\calC{}_{i_n+1} &\equiv \{{i_1},\dots,{i_n}, {i_n+1}\},\cr
\calC_{\slashh {i_k}} &\equiv \{ {i_1}, \dots, {i_{k-1}},{i_{k+1}},\dots,
{i_n}\},\cr
\calC_{ {i_j}\to {i_k} {i_\ell} } &\equiv \{ {i_1},\dots,
{i_{j-1}}, ({i_k}, {i_\ell}),{i_{j+1}},\dots,{i_n}\}.\cr}}
Note that the last operation is  defined only if ${i_k}, {i_\ell}$ are not
in $\calC_{ \slashh {i_j}}$; the brackets indicate the change in nesting
of vector products when the expression appears as an argument of
the polynomial $f_{n+1}$.  As a result, this fourth
operation produces a disordered
cluster, unless $i_j$ is at the extremity, i.e. $j=1$ or $j=n$.  Although this
may not be completely clear at this point, it will be illustrated
shortly by an example.  On the other hand, the first three transformations
always yield
ordered clusters.

For $\calC\in\calC^{(n,k)}$ the terms arising in \comlink\ correspond to
clusters
with $(n\pr,k\pr)=(n+1,k),\;(n+1,k-1), (n-1,k), \; (n-1,k-1)$.
More precisely, using the above identities, the different terms in
\comlink\ are found to be:
$$\eqalign{
 [\S_{i_1-1} \cdot\S_{i_1}, f_n(\calC)] &= - 2 i f_{n+1} ({}_{i_1-1} \calC),\cr
[\S_{i_n-1} \cdot\S_{i_n}, f_n(\calC)] &= 2 i f_{n+1} (\calC {}_{i_n+1}),\cr
[\S_{i_1} \cdot\S_{i_1+1}, f_n(\calC)] &=4 i f_{n-1}( \calC_{\slashh {i_1}})
- 4 i f_{n-1}(
\calC_{\slashh {i_1+1}})\text{if} i_2=i_1+1\cr
&=2 i
 f_{n+1}( {}_{i_1} {}_{i_1+1} \calC_{\slashh {i_1}})\text{if} i_2\not=i_1+1,\cr
[\S_{i_n-1} \cdot\S_{i_n}, f_n(\calC)] &=4 i f_{n-1}( \calC_{\slashh {i_n-1}})
- 4 i f_{n-1}(
\calC_{\slashh {i_n}})\text{if} i_{n-1}=i_n-1\cr
&=- 2 i
 f_{n+1}(\calC_{\slashh {i_n}}{}_{i_n-1} {}_{i_n})\text{if}
i_{n-1}\not=i_n-1,\cr
[\S_{i_j} \cdot\S_{i_j+1}, f_n(\calC)] &=2 i
 f_{n-1}(\calC_{\slashh {i_j}})-2i
f_{n-1}( \calC_{\slashh{i_{j+1}}}) \text{if} i_j,i_{j+1}\in \calC\cr
&= -2 i
 f_{n+1}( \calC_{{i_j}\to
{i_j}{i_j+1}}) \text{if} i_j\in
\calC,~,i_{j+1}\not\in\calC,\cr}\eqlabel\comlik$$
where in the last case, it is assumed that $j\not=1,n-1$.

The last term in \comlik\ contains contributions from disordered clusters.
For example, consider the cluster
$\calC=\{1,2,4,5\}$ and the commutator of
$$f_4(\calC)=
\S_1\cdot(\S_2\times(\S_4\times\S_5))=(\S_2\times(\S_4\times\S_5))\cdot\S_1
.\eq$$ with
$\S_2\cdot\S_3$.  Taking for $f_4$ the second expression above allows us to use
\idone\
directly, with the result
$$[\S_2\cdot\S_3,(\S_2\times(\S_4\times\S_5))\cdot\S_1] =
-2i((\S_3\times\S_2)\times(\S_4\times\S_5))\cdot\S_1\eq$$
We see clearly that the nesting of the spins has been modified (that is the
parentheses
cannot be all nested toward one side or the other).   In other words, the
result cannot
be written in terms of $f_5(\{1,2,3,4,5\})$.  As this calculation illustrates,
the special
property \famid\ makes the direct application of the identities \idone\
-\idthr\
straightforward.

\subsubsec{c) Summation over links inside a cluster}

Adding up all these terms gives:
\eqn\comHfnC{  [ H_2, f_n(\calC)]= a_{n+1,k}(\calC) +
b_{n-1,k+1}(\calC)+d_{n+1,k-1}(\calC)+e_{n-1,k}(\calC)+r(\calC), }
where the various quantities appearing in this expression are defined by:
$$\eqalign{
a_{n+1,k}(\calC) &= - 2 i f_{n+1} ({}_{i_1-1} \calC)+ 2 i f_{n+1} (\calC
{}_{i_n+1})
,\cr
 b_{n-1,k+1}(\calC)&= -4 i  f_{n-1}( \calC_{\slashh {i_2}}) \delta_{i_1+1,i_2}
+4 i f_{n-1}(\calC_{\slashh {i_{n-1}}})\delta _{i_{n-1}+1,i_{n}} \cr
 &~~ + 2 i \sum_{j=2}^{n-2}
[ f_{n-1}(\calC_{\slashh {i_j}})-
f_{n-1}( \calC_{\slashh{i_{j+1}}}) ]
 \delta_{i_{j+1},i_j+1}
,\cr
d_{n+1,k-1}(\calC)&=2 i
 f_{n+1}( {}_{i_1} {}_{i_1+1} \calC_{\slashh {i_1}}) (1-\delta_{i_1,i_2-1})\cr
&~~~- 2 i
 f_{n+1}(\calC_{\slashh {i_n}}{}_{i_n-1} {}_{i_n})
(1-\delta_{i_n, i_{n-1}+1}),\cr
 e_{n-1,k}(\calC)&=4 i f_{n-1}( \calC_{\slashh {i_1}}) \delta_{i_1+1,i_2}-
4 i f_{n-1}( \calC_{\slashh {i_n}})\delta_{i_{n-1}+1,i_n},\cr
 r(\calC)&=-2 i
\sum_{j=2}^{n-2} f_{n+1}( \calC_{{i_j}\to
{i_j}{i_j+1}})
 (1-\delta_{i_{j+1}, i_j+1}).\cr
&~~~+2 i
\sum_{j=3}^{n-1} f_{n+1}(\calC_{{i_j}\to {i_j-1}{i_j} })
 (1-\delta_{i_{j-1}, i_j-1})\cr}\eqlabel\abcder$$
Caution: in the above formulae, one should distinguish carefully between
sites ${i_j+1}$ and ${i_{j+1}}$.

\subsubsec{d) Summation over clusters}

With the exception of $r(\calC)$, all of the terms on
the right hand-side of \comHfnC\ involve only regular nested product of the
full set of
spin variables. Fortunately, when summing up over all clusters in
$\calC^{(n,k)}$, the
unwanted contributions from improperly-nested spin sequences
cancel:
$$\sum_{\calC\in{\calC^{(n,k)}} } r(\calC)=0.\eq$$
This can be seen as follows.  $r$-type terms arise only from clusters
containing holes.
Consider a particular cluster $\calC'$ having a hole at site $j$ but not at
site $j+1$.
Let us write its associated polynomial $f$ as
$$f(\calC') =  ({\bf L}\times \S_{j+1})\cdot {\bf R},\eq$$
where ${\bf L}$ stands for the contribution of all the
spins at the left of the site $j$
and ${\bf R}$ for that of the spins at the right of $j+1$.  The contribution of
this
cluster to the commutator with $\S_j\S_{j+1}$ is
$$[\S_j\S_{j+1},({\bf L}\times \S_{j+1})\cdot {\bf R}] =- 2i({\bf L}\times
(\S_j\times\S_{j+1}))\cdot {\bf R}.\eqlabel\comm$$
In the summation over all possible clusters having the same number of spins and
same number
of holes, we will encounter another cluster $\calC''$, which differs from
$\calC'$ only in
that the hole which was in position $j$ now appears in position $j+1$.  The $f$
polynomial
of this cluster being
$$f(\calC'') =  ({\bf L}\times \S_{j})\cdot {\bf R},\eq$$
its contribution to the commutator with $\S_j\S_{j+1}$ is
$$[\S_j\S_{j+1},({\bf L}\times \S_{j})\cdot {\bf R}] =- 2i({\bf L}\times
(\S_{j+1}\times\S_{j}))\cdot {\bf R},\eqlabel\comma$$
which is exactly the negative of (\comm).  Hence, in the summation over all
clusters of a
fixed type, all the contributing $r$-terms cancel two by two.

It then follows that the commutator
$[H_2, F_{n,k}]$ can be expressed as a linear combination of terms
of the type $F_{n\pr,k\pr}$. Equivalently, we may interpret this commutator
as a transformation of the set of all clusters into itself.
Now, with an argument similar to the one used to show
the cancellation of the $r$ terms, we can prove that the sum over the
$e$-type terms
 also adds up to zero:
$$\sum_{\calC\in{\calC^{(n,k)}}} e_{n-1,k}(\calC)=0.\eq$$
Let us illustrate this calculation  by a detailed example.
Take $N=7,~ n=4$ and $k=1$.
The different clusters $\calC^{(4,1)}$ are
$$\eqalign{
\{\calC^{(4,1)}\}_{N=7} =& \{\{1,2,4,5\},\{2,3,5,6\},\{3,4,6,7\},\{1,3,4,5\}
,\{2,4,5,6\},\cr
&\quad\{3,5,6,7\},\{1,2,3,5\},\{2,3,4,6\},\{3,4,5,7\}\}.\cr}\eq$$
The $e_{3,1}$ terms for the first three clusters are
$$\eqalign{
4i[f_3(\{2,4,5\})-&f_3(\{1,2,4\})+f_3(\{3,5,6\})-
f_3(\{2,3,5\})\cr
&+f_3(\{4,6,7\})-f_3(\{3,4,6\})],\cr}\eq$$
those for the following three clusters are
$$4i[-f_3(\{1,3,4\})-f_3(\{2,4,5\})-f_3(\{3,5,6\})],\eq$$
and for the last three, these are
$$4i[f_3(\{2,3,5\})+f_3(\{3,4,6\})+f_3(\{4,5,7\})].\eq$$
Adding up all these contributions, we see that terms cancel in pairs.

Therefore,
\eqn\coHFnk{ [ H_2, F_{n,k}]=A_{n+1,k}+D_{n+1,k-1}+ B_{n-1,k+1} ,}
where
$$\eqalign{
A_{n+1,k}=&
\sumC a_{n+1,k}(\calC),\cr
 B_{n-1,k+1}=&\sumC b_{n-1,k+1}(\calC),\cr
D_{n+1,k-1}=&\sumC d_{n+1,k+1}(\calC).\cr} \eq$$

\subsubsec{e) The special case where  $n=N$}

In the case where $n=N$, the calculation of \comlink\ is slightly
different, as a new type of
term
appears in the commutator of the hamiltonian and the
highest order term in $H_N$, i.e.
$${ F_{N,0}=\sum_{j=1}^N f_N(\calC_j)},\eq$$
where $$\calC_j=\{ j,j+1\dots,{j+N-1}\}.\eq$$
This commutator reads:
\eqn\HNofC{ [H_2,f_N(\calC_j)]=g_j+e_{N-1,1}(\calC_j)+ b_{N-1,1}(\calC_j),}
where
\eqn\gj{\eqalign{ g_j&=[ \S_{j+N-1}\cdot\S_j, f_N(\calC_j)]
\cr&=2 i \epsilon^{ka_0b_0}\epsilon^{b_0a_1b_2}\dots\epsilon^{b_{N-3}a_{N-2}k}
(
\s^{a_0}_{j+1} \s^{a_1}_{j+2}\dots\s^{a_{N-2}}_{j+N-1} -
\s^{a_0}_{j}\s^{a_1}_{j+1}\dots \s^{a_{N-2}}_{j+N-2}). }}
It is obvious that
\eqn\sgi{\sum_{j=1}^N g_j=0.}
Summing up over all $\calC_j$ in \HNofC\ gives then
\eqn\HFnzer{ [ H_2,F_{N,0}]=B_{N-1,1}.}

\subsubsec{f) Conditions for the commutativity of $H_n$ with $H_2$}

A simple consequence of \coHFnk\ and \HFnzer, is that  the
commutation with the hamiltonian changes the order of the clusters in $F_{n,k}$
from $n$ to
$n\pm1$.  Therefore, to prove that the charge
$H_n$ of the form \Hndef\ commutes with the hamiltonian,
it is sufficient to show,
for any $n\le N$, that:

\n (i) $[ H_2, F_{n,0}]$ does not contain terms of order $n+1$,

\n (ii) $[ H_2, F_{n,k}+\sum_{\ell=1}^{k+1} F_{n-2,\ell}]$ does not contain
terms of order
$n-1$.

{\noindent  {
The first assertion is easily verified  for $n<N$
since ${ A_{n+1,0}=0}$, by translational symmetry, that is
$$ A_{n+1,0}=2 i \sum_{j\in \Lambda} [-f_{n+1}(\{j-1,j,...,j+n\}) +
f_{n+1}(\{j,j+1,...,j+n+1\})] = 0.\eq$$
 For $n=N$, (i) immediately follows from
\HFnzer.
The second assertion is more difficult to prove. It  is equivalent to:
\eqn\comtwo{ B_{n-1,k+1}+\sum_{\ell=1}^{k+1} (A_{n-1,\ell}+ D_{n-1,\ell-1}
)=0.}
As the sum above contains terms corresponding to clusters of order $n-1$, with
hole numbers ranging from 0 to $k+1$, it can  vanish if and
only if all terms with a given number of holes cancel, i.e.
\eqn\BplusA{ B_{n-1,k+1} +A_{n-1,k+1}=0,}
\eqn\AplusD{ A_{n-1,\ell} +D_{n-1,\ell}=0, \;\; (1\le \ell \le k)}
\eqn\Dnminone{ D_{n-1,0}=0.}
\AplusD\ and \Dnminone\ are direct consequences of (\abcder).  In fact, they
follow
immediately once we notice that $D_{n-1,\ell}$ can be rewritten as
$$D_{n-1,\ell} = -\sumC a_{n+1,\ell}(\calC)\eq$$
The proof of
\BplusA\
is much more tedious.

\subsubsec{g) A resummation trick}

In order to demonstrate \BplusA, one first has to rewrite the $B$ term in a way
that can be compared with the form of $A$.  $B$ naturally decomposes into two
parts;
$$ B_{n-1,k+1}=  B_{n-1,k+1}^{(1)}+ B_{n-1,k+1}^{(2)},\eq$$
with
$$ B_{n-1,k+1}^{(1)} =  -4 i\sum_{\calC\in \calC^{(n-1,k+1)}}  [ f_{n-1}(
\calC_{\slashh {i_2}}) \delta_{i_1+1,i_2} - f_{n-1}(\calC_{\slashh
{i_{n-1}}})\delta
_{i_{n-1}+1,i_{n}}], \eq$$
and
$$ B_{n-1,k+1}^{(2)} =2 i\sum_{\calC\in \calC^{(n-1,k+1)}} \sum_{j=2}^{n-2}
[ f_{n-1}(\calC_{\slashh {i_j}})-
f_{n-1}( \calC_{\slashh{i_{j+1}}}) ]
 \delta_{i_{j+1},i_j+1}.\eqlabel\btwo$$
Both parts vanish identically for $n=2,3$.  For $n\geq4$, the set of clusters
$\calC^{(n-1,k+1)}$ can be decomposed into the following four classes:
$$\eqalign{
\calC_1=\{i_1,i_{1+1},K_0,i_{n}-1,i_{n}\}~&\sim~\bullet\bullet
K_0\bullet\bullet\cr
\calC_2=\{i_1,i_{1+1},K_0,i_{n-1}<i_{n}-1,i_{n}\}~&\sim~\bullet\bullet
K_1\circ\bullet\cr
\calC_3=\{i_1,i_{2}>i_1+1,K_0,i_{n}-1,i_{n}\}~&\sim~\bullet \circ
K'_1\bullet\bullet\cr
\calC_4=\{i_1,i_{2}>i_1+1,K_0,i_{n-1}<i_{n}-1,i_{n}\}~&\sim~\bullet\circ
K_2\circ
\bullet\cr}\eqlabel\cuscla$$ Here $K_p$ (with or without prime) stands for an
element of
$\calC^{(n-5+p,~k+1-p)}$.   On the rhs we have introduced a symbolic
description of the
cluster in which the first two entries refer to the sites $i_1$ and $i_1+1$
($\bullet$ means
that the corresponding site is included in the cluster and $\circ $ means that
it is not) and the last two give the information relative to the sites $i_n-1$
and $i_n$.  Everything in between is specified by the $K_i$'s. Note that the
clusters of the type $\calC_4$ do not contribute to $
B_{n-1,k+1}^{(1)}$ due to the constraint encoded in the delta function.  With
the above
schematic description for the clusters, we have:
$$\eqalign{
 B_{n-1,k+1}^{(1)} =  -4 i\sum_{K_p\in \calC^{(n-5+p,k+1-p)}}  [&
f_{n-1}(\bullet\circ K_0\bullet\bullet) - f_{n-1}(\bullet\bullet
K_0\circ\bullet)\cr
+&f_{n-1}(\bullet\circ K_1\circ \bullet)
- f_{n-1}(\bullet\circ K'_1\circ \bullet)].\cr} \eq$$
Since the summation of terms containing
 $K_1$ and $K'_1$ runs over identical sets, their contributions
cancel out, leaving us with
$$ B_{n-1,k+1}^{(1)} =  -4 i\sum_{K_0\in \calC^{(n-5,k+1)}}  [
f_{n-1}(\bullet \circ K_0\bullet\bullet) - f_{n-1}(\bullet\bullet
K_0\circ\bullet)].\eqlabel\bun$$  We now turn to
$B_{n-1,k+1}^{(2)}$.  The crucial step in the present argument is the rewriting
of (\btwo)
in the form
$$ B_{n-1,k+1}^{(2)} =2 i[Pr\sum_{\calC\in \calC^{(n-1,k+1)}} \sum_{j=2}^{n-2}
 f_{n-1}(\calC_{\slashh {i_j}})-Pr\sum_{\calC\in \calC^{(n-1,k+1)}}
\sum_{j=2}^{n-2}  f_{n-1}( \calC_{\slashh{i_{j+1}}})].\eqlabel\btwoo$$
Notice first that the constraint $i_{j+1}=i_j+1$ has been relaxed.  The symbol
$Pr$
indicates a projection operation whose action amounts to rescale the
multiplicity of a
given cluster to one.  The rationale for this projection is as follows.  In
(\btwo), the
clusters $\calC_{\slashh {i_j}}$ for a fixed $j$ are all distinct due to the
constraint
$i_{j+1}=i_j+1$. But when the constraint is relaxed, this is no longer so and
it is then
necessary to rule out repetitions.  Here is an example.  Translational
invariance being
not an issue here, consider a subset of $\calC^{(n,k)}$, consisting of
clusters with fixed extremities
(and this is
what the tilde denotes below).  Let $n=4,k=1,i_1=1,i_4=5$:
$$\eqalign{
\{\tilde{\calC}\}  =& \{\{1,2,3,5\},\{1,2,4,5\},\{1,3,4,5\}\},\cr
\{\tilde{\calC}_{\slashh {i_2}}\}_{i_{3}=i_2+1}
=&\{\{1,3,5\},\{1,4,5\}\},\cr
\{\tilde{\calC}_{\slashh {i_3}}\}_{i_{3}=i_2+1}   =
&\{\{1,2,5\},\{1,3,5\}\},\cr}
\eq$$
so that
$$ \sum_{ \tilde{\calC}}[f_{3}(\tilde{\calC}_{\slashh {i_2}})-
f_{3}( \tilde{\calC}_{\slashh{i_{3}}})]\delta_{i_{3},i_2+1} =
 f_3( \{1,4,5\})-f_3( \{1,2,5\}).\eq$$
On the other hand, since
$$\eqalign{
\{ \tilde{\calC}_{\slashh {i_2}}\}
=&\{\{1,3,5\},\{1,4,5\},\{1,4,5\}\},\cr
\{\tilde{\calC}_{\slashh {i_3}}\}   =
&\{\{1,2,5\},\{1,2,5\},\{1,3,5\}\},\cr}
\eq$$
we see that
$$ \sum_{\tilde{\calC}}[f_{3}(\tilde{\calC}_{\slashh {i_2}})-
f_{3}( \tilde{\calC}_{\slashh{i_{3}}})] = 2[f_3( \{1,4,5\})-f_3(
\{1,2,5\})].\eq$$
By working out another simple example, with $n=5,k=2$, the reader will
convince himself that the projection has to be done for each sum separately
before
evaluating their difference.

We now prove the equivalence between (\btwo) and (\btwoo).
Clearly (\btwoo) incorporates all the terms in (\btwo), which are those
associated to
clusters  containing both $i_j$ and $i_j+1$.  The extra terms in (\btwoo) turn
out to
cancel two by two: the contribution of a positive term, with $i_j\in \calC$ and
$i_j+1\not\in \calC$ is canceled by the contribution of a negative term
associated to a
cluster $\calC'$  which differs from $\calC$ only in that $i_j+1\in \calC'$ and
$i_j\not\in \calC'$. Schematically, if
$$\eqalign{
\calC~\sim~& K\bullet\circ K'\qquad(\bullet\text{at site}j),\cr
\calC'~\sim~& K\circ\bullet K'\qquad(\bullet\text{at site}j+1),\cr}\eq$$
then $$\calC_{\slashh {i_j}} = \calC'_{\slashh {i_{j+1}}}.\eq$$

Having established (\btwoo), we now rewrite it in the form
$$ B_{n-1,k+1}^{(2)} =2 i[Pr\sum_{\calC\in \calC^{(n-1,k+1)}}
 f_{n-1}(\calC_{\slashh {i_2}})-Pr\sum_{\calC\in \calC^{(n-1,k+1)}}
 f_{n-1}( \calC_{\slashh{i_{n-1}}})].\eqlabel\btwooo$$
The sum over clusters can again be separated into sums over the four classes
(\cuscla), with
the result:
$$\eqalign{
 B_{n-1,k+1}^{(2)} =  2 i~Pr\sum_{K_i\in \calC^{(n-5+i,k+1-i)}}  [&
f_{n-1}(\bullet\circ K_0\bullet\bullet) + f_{n-1}(\bullet\circ
K_1\circ\bullet)\cr
+&f_{n-1}(\bullet\circ {K'_1}_{\slashh {i_2}}\bullet \bullet)
+ f_{n-1}(\bullet\circ {K_2}_{\slashh {i_2}}\circ \bullet)]\cr
-2 i~Pr\sum_{K_i\in \calC^{(n-5+i,k+1-i)}}  [&
f_{n-1}(\bullet\bullet K_0\circ\bullet) + f_{n-1}(\bullet\bullet {K_1}_{\slashh
{i_{n-1}}}\circ\bullet)\cr +&f_{n-1}(\bullet\circ {K'_1}\circ \bullet)
+ f_{n-1}(\bullet\circ {K_2}_{\slashh {i_{n-1}}}\circ \bullet)].} \eq$$
The second and the seventh terms cancel and similarly for the fourth and the
eight ones.  The
contribution of the third term being already included in that of the first one,
it can be
ignored (recall that due to the projection factor, multiplicities are
irrelevant).
Similarly, the sixth term can be dropped.  With these simplifications, $
B_{n-1,k+1}^{(2)}$
take the form
$$ B_{n-1,k+1}^{(2)} =  2 i\sum_{K_0\in \calC^{(n-5,k+1)}}  [
f_{n-1}(\bullet \circ K_0\bullet\bullet) - f_{n-1}(\bullet\bullet
K_0\circ\bullet)],\eqlabel\buna$$
the projection operator being now unnecessary.
Adding (\bun) and (\buna) yields
$$ B_{n-1,k+1} =  -2 i\sum_{K_0\in \calC^{(n-5,k+1)}}  [
f_{n-1}(\bullet \circ K_0\bullet\bullet) - f_{n-1}(\bullet\bullet
K_0\circ\bullet)].\eqlabel\buns$$
The trivial modification
$$\eqalign{
 B_{n-1,k+1} =  -2 i\sum_{K_0, K_{-1}}  [&
f_{n-1}(\bullet \circ K_0\bullet\bullet) +f_{n-1}(\bullet \bullet
K_{-1}\bullet\bullet)\cr
-& f_{n-1}(\bullet\bullet
K_0\circ\bullet)-f_{n-1}(\bullet \bullet K_{-1}\bullet\bullet)],\cr}
\eqlabel\bunss$$
allows us to reach the form
$$\eqalign{ B_{n-1,k+1} = &-2i\sum_{\calC\in \calC^{(n-1,k+1)}}[
f_{n-1}(\calC~\bullet)-f_{n-1}(\bullet~ \calC)]\cr
= &-2i\sum_{\calC\in \calC^{(n-1,k+1)}}[
f_{n-1}(\calC_{\slashh {i_n+1}})-f_{n-1}({}_{\slashh {i_1-1}} \calC)]\cr
=&-A_{n-1,k+1},}\eq$$
which thereby proves the desired result.

\subsubsec{h) Mutual commutativity of the $H_n$'s and relation with the $Q_n$
basis}

Having established the commutativity of the $H_n$'s with the
hamiltonian, the final step in the proof of integrability
is  to show that all these conservation laws  commute among themselves.
For this, we
will
proceed indirectly by first expressing the charges $H_n$ in the $\{Q_n\}$ basis
defined
recursively by
$$[B,Q_n] =Q_{n+1}\eq$$
with $Q_2=H_2$, and then use the fact that the $Q_n$ mutually commute, being
directly related to the commuting transfer matrices.  We first evaluate the
commutator of
$H_n$ ($n\ge2$) with the boost operator.  The result is:
\eqn\coBHn{ [ B, H_n]=\sum_{k=0}^{\max(1,[n/2]-1)} \beta^{(n)}_k H_{n+1-2k}, }
where.
$$\eqalign{
\beta^{(n)}_0=&n-1,\cr
\beta^{(n>2)}_1=&5-3n,\cr
\beta^{(n)}_{1<\ell < [n/2]}=&-(n-2 \ell-1) \alpha_{\ell,1} \cr}
\eqlabel\betacoef $$
The simplest way of obtaining this formula is to compare
the coefficients of the terms $F_{n-2k+1,0}$ on both sides.
With this result, it is simple to obtain the expression for the charges $Q_n$
in terms of
the
$\{H_n\}$'s.  It is clear from \coBHn\ that
for even (odd) $n$, $Q_n$ can be
expressed as a linear combination of the $H_m$ with even (odd) $m\le n$:
\eqn\QnHn{ \Q_n=\sum_{p=0}^{[n/2]-1}\gamma^{(n)}_p H_{n-2p}.}
The coefficients $\gamma$ can be determined from a simple recurrence
relation:\foot{Notice
that the formulae for $\gamma$ reflect our choice of
normalization in (\Tnorm); they need to be modified in another
normalization,
to take into account a rescaling of the $Q_n$'s,
corresponding to a  rescaling of the spectral parameter.}
\eqn\gammarec{\gamma_{\ell}^{(n+1)}=\sum_{p,m\ge 0 \atop p+m=\ell}
\gamma_p^{(n)}\beta_{m}^
{(n-2p)}, }
with $\gamma^{(2)}_p=\delta_{p,0}$.
For example, up to additive constants,
\eqn\QninHn{\eqalign{ & Q_4=2 H_4 - 4 H_2,\cr
&Q_5=6 H_5-18 H_3,\cr
&Q_6=24 H_6- 96 H_4 + 72 H_2.\cr}}
Therefore, we see that $\{H_n\}$ represents a basis of the space of the
conservation laws of the XXX chain, which can be obtained taking linear
combinations of $\{Q_n\}$'s.
Mutual commutativity
of $H_n$'s now follows trivially from that of the $Q_n$'s.
\subsubsec{i) An alternative finale}

It is also possible to prove the involutive nature of $H_n$ without
using
the transfer matrix formalism
at all,
by applying an inductive argument using \coBHn\ and the Jacobi identity.
We sketch this  argument below.
First, we note that \coBHn\ can be rewritten in the form
\eqn\coBHnnew{ H_{n+1}={1\over{ (n-1)}} [B, H_n] +R_n,}
where $R_n$ is a linear combination of the charges $H_{m<n}$. Then,
assuming that $[H_n, H_m]=0$ for all $n,m<n_0$, we prove that
$[H_{n_0+1}, H_k]=0$, for $k<n_0$. For $k=2$, the proof is the calculation
given above in sections (b) to (g). For $k=3$, the commutativity of
$H_{n_0+1}$ and $H_3$ can be established using the Jacobi identity and the fact
that $[H_{n_0+2}, H_2]=0$. Similarly, one may successively show that
$[H_{n_0+1},
H_{k>3}]=0$ using the Jacobi identity and the
 relations $[H_{n_0+\ell}, H_2]=0$ for $\ell=1,\dots,k-1$.
The above method constitutes an alternative purely algebraic proof of the
integrability of the XXX model.

\newsec{The XYZ model revisited}

\subsec{Proof  of the
 general pattern for the charges}

We now present the proof that the charges of the XYZ model have the form
\Hxyz.
We will again proceed directly, calculating
$[ B, \ft_n(\calS)] $, for an arbitrary sequence $\calS=\calC^\pi$ with
cluster $\calC=\{i_1, \dots,i_n\}$ and pattern $\pi=\{a_1,\dots,a_n\}$.
 This commutator contains only
terms of the form $\ft_{n\pm 1}(\calS\pr)$,
where $\calS\pr$ can be obtained from $\calS$ by a few simple
transformations, given below:
\eqn\bHxyz{\eqalign{
\s^{a_0}_{i_1-1} \calS &\equiv \{{i_1-1}, {i_1},\dots,{i_n}\}
^{\{ a_0, a_0\times a_1, \dots, a_n\} },\cr
\calS\s^{a_{n+1}}_{i_n+1} &\equiv \{{i_1},\dots,{i_n}, {i_n+1}\}^
{\{ a_1,\dots, a_{n+1} \times a_n, a_{n+1}\} },\cr
\calS_{\slashh \s_{i_k}^{a_k}}
&\equiv \{ {i_1}, \dots, {i_{k-1}},{i_{k+1}},\dots,
{i_n}\}^
{\{ a_1, \dots, a_{k-2},a_k \times a_{k-1}, a_{k+1},\dots,a_n\} } ,\cr
{}_{\slashh \s_{i_k}^{a_k}}\calS &\equiv \{ {i_1}, \dots,
{i_{k-1}},{i_{k+1}},\dots,
{i_n}\}^
{\{ a_1, \dots, a_{k-1},a_k \times a_{k+1},a_{k+2},\dots,a_n\} } ,\cr
\calS_{ \s_{i_j}^{a_j}\to \s_{i_k}^{a_k} \s_{i_\ell}^{a_\ell} }
&\equiv \{ {i_1},\dots,
{i_{j-1}}, ({i_k}, {i_\ell}),{i_{j+1}},\dots,{i_n}\}
^{\{ a_1,\dots,a_{j-1},(a_k,a_\ell),a_{j+1},\dots,a_n\} }.\cr}}
 The following conventions will be adopted:
$$\eqalign{
\{a_1,\dots, 0,\dots,a_n\}&=0,\cr
\{a_1,\dots,-a_k,\dots,a_n\}&=-\{a_1,\dots,a_k,
\dots,a_n\},\cr}\eq$$ and $$\eqalign{
\ft_n(C^0)=&0,\cr
\ft_n(\calC^{-\pi})=&-\ft_n(\calC^\pi).\cr}\eq$$

The basic commutators of the form
$[ j\s_j\s_{j+1}, \ft_n(\calS)]$
may be evaluated similarly as in the XXX case.
Summing up all these terms yields:
\eqn\comHfxyz{ [ B, \ft_n(\calS)]= b^+(\calS) +
b^{-+}(\calS)+b^{+-}(\calS)+b^-(\calS),}
where
$$b^{\alpha}(\calS)=\sum_{e=x,y,z} b^\alpha_e(\calS), \quad\quad
\alpha\in\{+,-+,+-,-\},\eq$$ and
$$\eqalign{
 b^{+}_e(\calS)=& -  (i_1-1) \ft_{n+1} (\s^{e}_{i_1-1} \calS)+  i_n \ft_{n+1}
(\calS \s^{e}_{i_n+1}),\cr
 b_e^{-+}(\calS)= & -2 i_1 \ft_{n-1}
( \calS_{\slashh {\s^e_{i_2}}}) \delta_{i_1+1,i_2}
+2 i_{n-1}\ft_{n-1}({}_{\slashh {\s^e_{i_{n-1}}}}\calS)
\delta _{i_{n-1}+1,i_{n}} \cr
 &~~ +  \sum_{j=2}^{n-2}
i_j [  \ft_{n-1}({}_{\slashh {\s^e_{i_j}}}\calS)-
\ft_{n-1}( \calS_{\slashh{\s^e_{i_{j+1}}}}) ]
 \delta_{i_{j+1},i_j+1},\cr
 b_e^{+-}(\calS)= &
i_1 (\lambda_e)^2 \ft_{n+1}
( \calS_{\s_{i_1}^{a_1}\to \s^{e\times a_1}_{i_1+1} \s^e_{i_1}})
(1-\delta_{i_1,i_2-1})\cr
&~~-
 (i_n-1) (\lambda_e)^2 \ft_{n+1}
( \calS_{\s^{a_n}_{i_n}\to \s^e_{i_n-1}\s^{e\times{a_n}}_{i_n}})
(1-\delta_{i_n, i_{n-1}+1})\cr
&~~+\sum_{j=2}^{n-2} i_j (\lambda_e)^2 \ft_{n-1}
( \calS_{\s^{a_j}_{i_j}\to \s^{e\times a_j}_{i_j}\s^e_{i_j+1}})
 (1-\delta_{i_{j+1}, i_j+1})\cr
&~~+
\sum_{j=3}^{n-1} (i_j-1) (\lambda_e)^2 \ft_{n+1}
(\calS_{\s^{a_j}_{i_j}\to \s^e_{i_j-1}\s^{e\times a_j}_{i_j}})
 (1-\delta_{i_{j-1}, i_j-1}),\cr
 b_e^{-}(\calS)=& 2 i_1 (\lambda_e)^2\ft_{n-1}
( {}_{\slashh {\s^e_{i_1}}}\calS) \delta_{i_1+1,i_2}-
2 i_{n-1} (\lambda_e)^2 \ft_{n-1}( \calS_{\slashh {\s^e_{i_n}}})
\delta_{i_{n-1}+1,i_n}.\cr}\eqlabel\bcof$$

With the exception of the third and fourth
terms in $b^{+-}(\calS)$,
all the terms in \comHfxyz\
 contain only expressions corresponding to ordered sequences.
Fortunately, thanks to the factors arising from the
coefficients $g(\calS)$, the contributions due to disordered clusters
will cancel\foot{This can be seen by an
argument similar to the one used in
section (4.2.d) to demonstrate the cancellation of the $r$-terms in
the XXX case.}
 in summation over all
possible clusters for a given pattern $\pi$.
Therefore, the commutator $[B, \tilde F^n_{m,k}]$ can be
expressed as a linear combination of the terms of the type
$\ft(\calS\pr)$, with $\calS\pr$ being an ordered sequence.

A simple inductive step finishes the proof. Assume that the $n$-th
conserved charge is of the form \Hxyz.  As we have just shown, the
$n+1$-th charge $Q_{n+1}$, given by $[B, Q_n]$, is
a linear combination of the polynomials $\ft(\calS)$.
Now, because the commutator $[B, Q_n]$ is translation
invariant,
it must in fact be a linear combination of
elements of the spaces $\calF_{n+1-2k,l}$. Therefore, the charge $Q_{n+1}$
is again necessarily of the form \Hxyz.

\subsec{Recursion relations for the $\tilde F$'s}

Using the operations (\bcof)  (which naturally extend to any
space $\calF_{n,m}$), one can derive recursion relations for the $\tilde F$'s.
It is convenient to represent this recursion as a transformation
of the triangle corresponding to $Q_n$ into a new triangle corresponding to
$Q_{n+1}$ (cf. Fig. 2.).

\vskip 0.7cm
\vbox{
$$ {\epsfxsize= 6cm \epsfbox{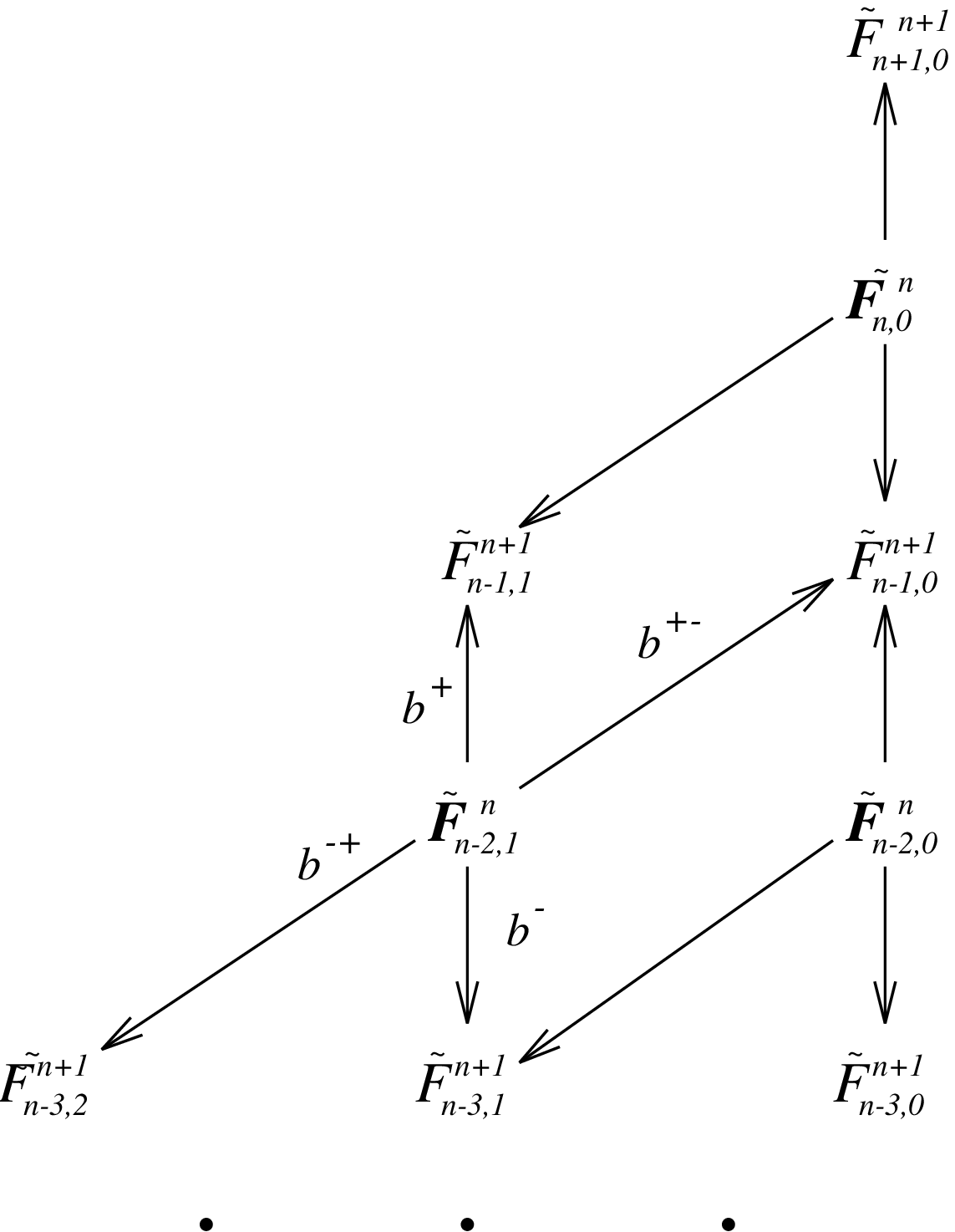}}\ \ \  $$
\vskip 0.8cm
Fig. 2.
 The  action of the boost
operator on the triangle  representing $Q_n$ for the XYZ model.
}
\vskip 1cm

It is clear that the boost operator
does not modify the structure of the triangle
 if $n$ is even
(i.e. the triangles corresponding to $Q_{2m}$ and $Q_{2m+1}$ differ
only in labeling of points). For $n$ odd,
the action of the boost  produces a
new strip at the bottom of the triangle
(therefore adding $[(n+1)/2]$ new points).
The left edge of the triangle representing $Q_{n+1}$ is given by:
\eqn\led{
\F^{n+1}_{n+1-2k,0<k<[(n+1)/2]}=
b^{-+}(\F^n_{n-2(k-1),k-1})+b^+(\F^n_{n-2k,k}),}
with
$$ \F^{n+1}_{n+1,0}=b^+(\F^n_{n,0}).\eq$$ The right edge of the $Q_{n+1}$
triangle is:
\eqn\rghted{ \F^{n+1}_{n+1-2k,0}=b^{-}(\F_{n-2(k-1),0}^n) +
b^+(\F^n_{n-2k,0})+b^{+-}(\F^n_{n-2k,1}). }
The interior of the triangle can be obtained from:
$$\eqalign{ \F^{n+1}_{n+1-2k,0<\ell<k}= &
b^-(\F^n_{n-2(k-1),\ell})+
b^{-+}(\F^n_{n-2(k-1),\ell-1})\cr&~+
b^+(\F^n_{n-2k,\ell})+
b^{+-}(\F^n_{n-2k,\ell+1}).}\eq$$
Finally, for $n$ odd, the bottom edge of the triangle $Q_{n+1}$ is:
\eqn\newbott{ \F^{n+1}_{2,0<\ell<[(n+1)/2]} =
b^-(\F^n_{3,\ell})+ b^{-+}(\F^n_{3,\ell-1}).}

These relations provide a recursive way to calculate the conserved charges.
As in the isotropic case,
the basis $\{Q_n\}$ contains terms
proportional to lower order charges; it is more convenient to
express the results
in another basis, denoted $\{H_n\}$, obtained by  taking appropriate linear
combinations of the form \lcom, in which lower order contributions are
subtracted.


\subsec{Explicit form of the first few conserved charges}

We present below several of the lowest-order conservation laws of the XYZ
model.
The first non-trivial charge beyond the hamiltonian is:
\def \Fh{{\tilde F}}
\eqn\hxyzt{
H_3=\Fh^3_{3,0}=\sum_{j\in \Lambda} (\Sh_j\times \St_{j+1})\cdot \Sh_{j+2}.}
The four-spin charge is:
\eqn\hxyzf{H_4=\Fh^4_{4,0}+\Fh^4_{2,1}+\Fh^4_{2,0},}
where
$$\eqalign{
\Fh^4_{4,0}=&\sum_{j\in\Lambda}
((\Sh_j\times \St_{j+1})\times\St_{j+2})\cdot \Sh_{j+3},\cr
 \Fh^4_{2,1}=&
\sum_{j\in\Lambda}\sum_{a\in\{x,y,z\}} ({ {\a\b\c} \over {\lambda_a}})
\sh^a_j \sh^a_{j+2},\cr
\Fh^4_{2,0}=&\sum_{j\in\Lambda}\sum_{a\in\{x,y,z\}} (\lambda_a)^2
\sh^a_j \sh^a_{j+1}.\cr}\eq$$
The five-spin charge corresponds to a triangle with the same structure as
that of $H_4$, namely:
\eqn\hxyzfif{H_5=\Fh^5_{5,0}+\Fh^5_{3,1}+\Fh^5_{3,0},}
with
$$\eqalign{ \Fh^5_{5,0}=&\sum_{j\in \Lambda}
(((\Sh_{j} \times \St_{j+1})\times
\St_{j+2})\times \St_{j+3})\cdot
\Sh_{j+4},\cr
\Fh^5_{3,1}=&
\sum_{j\in\Lambda}\sum_{a\in\{x,y,z\}}
({ {\a\b\c} \over {\lambda_a}}) \epsilon^{abc}(
\sh^a_j \st^b_{j+2}\sh^c_{j+3}
+\sh^b_j \st^c_{j+1}\sh^a_{j+3}),\cr
\Fh^5_{3,0}=&
- \sum_{j\in\Lambda}
\sum_{ a\ne b \ne c}
 (\lambda_b)^2 \epsilon^{abc}
\sh^a_j \st^b_{j+1}\sh^c_{j+2}.\cr}\eq$$
The six-spin ``monster'' is:
\eqn\hxyzs{
H_6=\Fh^6_{6,0}+\Fh^6_{4,1}+\Fh^6_{2,2}+\Fh^6_{4,0}+\Fh^6_{2,1}+\Fh^6_{2,0},}
with
$$\eqalign{
\Fh^6_{6,0}=& \sum_{j\in\Lambda} (((\Sh_{j} \times \St_{j+1})\times
\St_{j+2})\times \St_{j+3})\times \St_{j+4})\cdot\St_{j+5},\cr
\Fh^6_{4,1}=&
\sum_{j\in\Lambda} \sum_{a\ne b}[
({ {\a\b\c} \over {\lambda_a}})
(\sh_j^a\st_{j+2}^b\st^a_{j+3}\sh^b_{j+4}-
\sh_j^a\st_{j+2}^b\st^b_{j+3}\sh^a_{j+4}\cr
 &~~+\sh_j^b\st_{j+1}^a\st^b_{j+2}\sh^a_{j+4}-
\sh_j^a\st_{j+1}^b\st^b_{j+2}\sh^a_{j+4}) \cr &~~+
\lambda_a\lambda_b(
\sh^a_j\st^b_{j+1}\st^a_{j+3}\sh^b_{j+4}-
\sh^a_j\st^b_{j+1}\st^b_{j+3}\sh^a_{j+4})] ,\cr
\Fh^6_{2,2}=&
\sum_{j\in\Lambda} \sum_{a\in\{x,y,z\}}
({ {\a\b\c} \over {\lambda_a}})^2 \sh^a_j\sh^a_{j+3},\cr
\Fh^6_{4,0}= &\sum_{j\in\Lambda} \sum_{a\ne b}
(\lambda_b^2-\lambda_a^2)
\sh^a_j\st^b_{j+1}\st^b_{j+3}\sh^a_{j+4},\cr
\Fh^6_{2,1}=&
\sum_{j\in\Lambda} \sum_{a\in\{x,y,z\}}
\lambda_a^2({\a\b\c \over {\lambda_a}}) \sh^a_j\sh^a_{j+2},\cr
\Fh^6_{2,0}=&
\sumL
\sum_{ a\ne b \ne c}
(\lambda_a^2 +\lambda_b^2) (\lambda_a^2 +\lambda_c^2)
 \sh^a_j\sh^a_{j+1}.}\eq$$

The logarithmic derivatives of the transfer matrix can be expressed as
linear combinations of the $H_n$'s:
$$\eqalign{
Q_3 =& H_3,\cr
Q_4 =& 2 H_4 -2
p^2
H_2,\cr
Q_5 =& 6 H_5 -4 p^2 H_3,\cr
Q_6 =& 24 H_6 -32 p^2 H_4 + 8 p^4 H_2. } \eqlabel\Qnxyz$$
with $p^2=\a^2+\b^2+\c^2$.
In the isotropic limit ($\a=\b=\c=1$), the XYZ charges reduce to
linear combinations of the XXX charges given by \Hndef,
which we denote below as $H_n^{XXX}$ to avoid misunderstanding:
$$\eqalign{
H_3 =& H_3^{XXX},\cr
H_4 =& H_4^{XXX}+H_2^{XXX},\cr
H_5 =& H_5^{XXX}-H_3^{XXX},\cr
H_6 =& H_6^{XXX}+4 H_2^{XXX}.}\eq $$
{}From (\Qnxyz) we then recover the relations \QninHn.


\newsec{The conserved charges in the XY model}

\subsec{Generalities}

 For completeness, we present in this section the results for the
charges for the XY model:
$$H_{XY}=\sumL \a \s^x_j\s^x_{j+1}+
\b \s^y_j\s^y_{j+1}. \eqlabel\hamxy$$
As is well known, the XY spin chain is equivalent to a
free fermion theory; the charges for this system have thus a particularly
simple form, and they have been discussed by many authors
[\Gu-\IT].  Although the following results are not new,
the derivation given here is different from the approach of [\Gu-\IT].
An interesting  feature of the XY case is that there exist two independent
families of  conservation laws.
Both
families persist when the model is perturbed by a perpendicular (i.e. in the
$z$-direction) magnetic field.

Since the XY model is just a special case of the XYZ hamiltonian,
we can apply to it the apparatus developed in sections (3) and (5).
The general formula \Hxyz\ greatly simplifies for the XY case. First,
note that since $\c=0$, the coefficient $g(\calS)$ vanishes for
any sequence containing holes. Furthermore, $\ft_n(\calS)=0$ for sequences
with $a_1=z$ or $a_n=z$ or $a_2,\dots,a_{n-1}=x,y$. Thus,
$\ft_n(\calS)$ can be non-vanishing only for the sequences of the form:
$$\calS_{n,j}^{\alpha\beta}
=\{\s^\alpha_j,\s^z_{j+1},\dots, \s^z_{j+n-2},\s^\beta_{j+n-1}\}, \eq$$
with $\alpha,\beta\in\{x,y\}$.
Observe further that
$$\ft_n(\calS_{n,j}^{\alpha\beta})=\epsilon ~e_n^{\alpha\beta},\eq$$
where
$$e_{n,j}^{\alpha\beta}=
\s^\alpha_j\s^z_{j+1}\dots \s^z_{j+n-2}\s^\beta_{j+n-1},\eq$$
and the coefficient $\epsilon$ assumes values $\pm1$ or $0$.
It vanishes when $\alpha\ne\beta$ if $n$ is even, or $\alpha=\beta$,
if $n$ is odd.
We also introduce the notation
$$e_n^{\alpha\beta}=\sumL e_{n,j}^{\alpha\beta}\eq$$
The charges \Hxyz\ for the XY model take thus the simple form:
$$ Q_n=\sum_{k=0}^{[n/2]-1} \F^n_{n-2k,0},\eq$$
with $\F^n_{n-2k,0}\in \calF_{n-2k,0}$.
 This shows that the triangle corresponding to $Q_n$ collapses to its right
edge.
Note that the spaces $\calF_{n-2k,0}$ are two-dimensional.
A convenient basis is given by
$\{e_{n-2k}^{xx},e_{n-2k}^{yy}\}$ for $n$ even, and
$\{e_{n-2k}^{xy},e_{n-2k}^{yx}\}$ for $n$ odd. As we will shortly see,
the family $\{Q_n\}$ contains only half of the integrals of
motion of the XY model.

\subsec{The XX case}

Let us consider first the special case of the XX model with $\a=\b$.
One finds two conserved quantities for each $n$, which define
two families:
\eqn\hxypl{ \eqalign{ H^{(+)}_n &=
 e_n^{xx} +e_n^{yy}\quad\quad n~{\rm even}, \cr
&=
 e_n^{xy} -e_n^{yx}\quad\quad n~{\rm odd},
}}and
\eqn\hxym{ \eqalign{ H^{(-)}_n&=
 e_n^{xy}-e_n^{yx}\quad\quad n~{\rm even}, \cr &=
e_n^{xx}+e_n^{yy}\quad\quad n~{\rm odd}.  }}
In particular, there are two two-spin hamiltonians;
$$H_2^{(+)}=\sum_{j\in\Lambda}
\s^x_j \s^x_{j+1}+ \s^y_j \s^y_{j+1} ,\eq $$
$$H_2^{(-)}=\sum_{j\in\Lambda} \s^x_j \s^y_{j+1}- \s^y_j \s^x_{j+1}. \eq $$
$H_2^{(+)}$ is just the XX hamiltonian, while $H_2^{(-)}$ is a
special case of the Dzyaloshinski-Moriya interaction
[\ref{I. E. Dzyaloshinski, {\it J. Phys. Chem. Solids} {\bf 4} (1958), 241;
T. Moriya, {\it Phys. Rev. Lett} {\bf 4} (1969), 228.}].

Commutativity of $H_n^{(\pm)}$ with both hamiltonians $H_2^{(\pm)}$
can be easily verified directly by a
short calculation.
Furthermore, the boost operator,
$$B={1 \over {2 i}}\sumL j
[\s^x_j\s^x_{j+1}+\s^y_j\s^y_{j+1}],\eq $$
has ladder properties:
\eqn\bhpl{[B, H_n^{(+)}]=(n-1)(-1)^{n-1} (H_{n+1}^{(+)} +H_{n-1}^{(+)}),}
(valid for all $n$ with the understanding that $H_1^{(+)}=0$).
Since
the XX hamiltonian
$H_2^{(+)}$
is proportional to
to the first logarithmic derivative of the transfer matrix $Q_2$,
it follows that
$H^{(+)}_n $ can be expressed as a linear combination of the
logarithmic derivatives of $T$.  This shows the mutual
commutativity of the $H_n^{(+)}$'s.
Note that
the family $\{H_n^{(+)}\}$
can be obtained from the family $\{Q_n\}$ by a transformation of the
form  \lcom.
Clearly, the
triangle corresponding to $H_n^{(+)}$ degenerates to a single point.
In terms of the
associated triangles, this transformation corresponds then
to redefining to zero all the points on the right edge,
except for the vertex $F_{n,0}^n$.

It is clear from \bhpl\ that
for even (odd) $n$, $Q_n$ can be
expressed as a linear combination of the $H^{(+)}_m$ with even (odd) $m\le n$:
\eqn\QnHn{ \Q_n=\sum_{p=0}^{[n/2]-1}\gamma^{(n)}_p H^{(+)}_{n-2p}.}
The coefficients $\gamma$ satisfy the recurrence relation
\gammarec, where the coefficients $\beta$ are now:
 $\beta^{(n)}_0=\beta^{(n)}_1=(-1)^{(n-1)} (n-1)$.
For example, up to additive constants,
\eqn\QHXX{\eqalign{ & Q_3=-H_3^{(+)},\cr
&Q_4=2 H^{(+)}_4 - 2 H^{(+)}_2,\cr
&Q_5=6 H^{(+)}_5+ 8 H^{(+)}_3,\cr
&Q_6=24 H^{(+)}_6+40 H^{(+)}_4 + 16 H^{(+)}_2.\cr}}
\QHXX\ can be also obtained as a special case from the formulae
\hxyzt-\hxyzs\ for the general XYZ model.

In contrast,
the charges of the second family
$H_n^{(-)}$ are not of the general form \Hxyz.
Remarkably, the boost operator associated to $H_2^{(+)}$
acts as a ladder operator for the second
family too
(as first noticed in [\ref{E. Barouch and  B. Fuchssteiner, {\it Stud. Appl.
Math.}
 {\bf 73} (1985),
221.}\refname\BarFuch
]):
$$ [B, H_{n}^{(-)}]= ({n-1})(-1)^{n}( H_{n+1}^{(-)}+H_{n-1}^{(-)}),
\eqlabel\bhmin$$
with $H_1^{(-)}=-2 \sumL \s^z_j$. Using (\bhmin), \bhpl, together with the
Jacobi identity one can prove by an inductive argument
similar to the one in section (4.2),
that the charges of the second
family commute among themselves and with those of the
first family.

Note that under
a spin rotation by $\pi/2$
around the $z$-axis but only on even sites, i.e.:
$$\eqalign{
\s_{2j}^x\to\s_{2j}^y,\quad~\quad \s^y_{2j}\to -\s^x_{2j}\cr
\s_{2j+1}^x\to\s_{2j+1}^x,\quad\quad \s^y_{2j+1}\to\s^y_{2j+1}\cr}\eq$$
the two-spin charge $H_2^{(-)}$ transforms into the staggered XX hamiltonian:
$$ H_2^{(-)}\to \sum_{j\in\Lambda} (-1)^j
(\s^x_j \s^x_{j+1}+ \s^y_j \s^y_{j+1}).\eq$$

Both
$H_n^{(+)}$ and $H_n^{(-)}$ behave as  scalars under a global spin
rotation around the $z$-axis.
Indeed, it is simple to check that all members of both families commute with
the generator of such rotation, that is the $z$ component of the
total spin,
$$S^z=\sum_{j\in\Lambda} \s_j^z.\eq$$
Therefore, both families survive when the model is perturbed by a
magnetic field term $h S^z$.  On the other hand, while $H_n^{(+)}$ is invariant
under global parity
transformations, i.e. $\s_j^a\to -\s_j^a$, $H_n^{(-)}$ changes sign and for
this reason we qualify it as a
pseudoscalar.


\subsec{The general XYh case}

We turn now to the analysis of the general XY model
incorporating the magnetic field term.  The defining hamiltonian is thus
$$H_{XY}+hS^z =\sum_{j\in\Lambda}
\a\s^x_j \s^x_{j+1}+\b
\s^y_j \s^y_{j+1} + h\s^z_j .\eqlabel\HXYh$$
As in the XX case, we look for the charges in terms
of the basis $e_n^{\alpha\beta}$.
We denote by $\calA_n$ the span of
$\{e_n^{xx}, e_n^{yy}\}$ for $n$ even,
and the span of
$\{e_n^{xy}, e_n^{yx}\}$ for $n$ odd (i.e. $\calA_n=\calF_{n,0}$).
 Similarly,
$\calB_n$ will denote the span of
$\{e_n^{xy}, e_n^{yx}\}$ for $n$ even, and
that of $\{e_n^{xx}, e_n^{yy}\}$ for $n$  odd.
In addition, $\calA_1=0$, $\calB_1={\rm span}(S^z)$.
Let $\calA=\bigoplus_{n=1}^N \calA_n$ and $\calB=\bigoplus_{n=1}^N \calB_n$.
We now introduce two derivations on $\calA\oplus\calB$ given by:
$$\delta_{XY}(P)={1 \over {2i}}[H_{XY}, P],\eq$$
$$\delta_{Z}(P)={h\over{2i}}[S^z, P],\eq$$
for $P\in \calA\oplus\calB$.
The action of these operators can be visualized on the diagram in
Fig. 3.

\vskip 0.5cm
\vbox{
$$ {\epsfxsize= 4cm \epsfbox{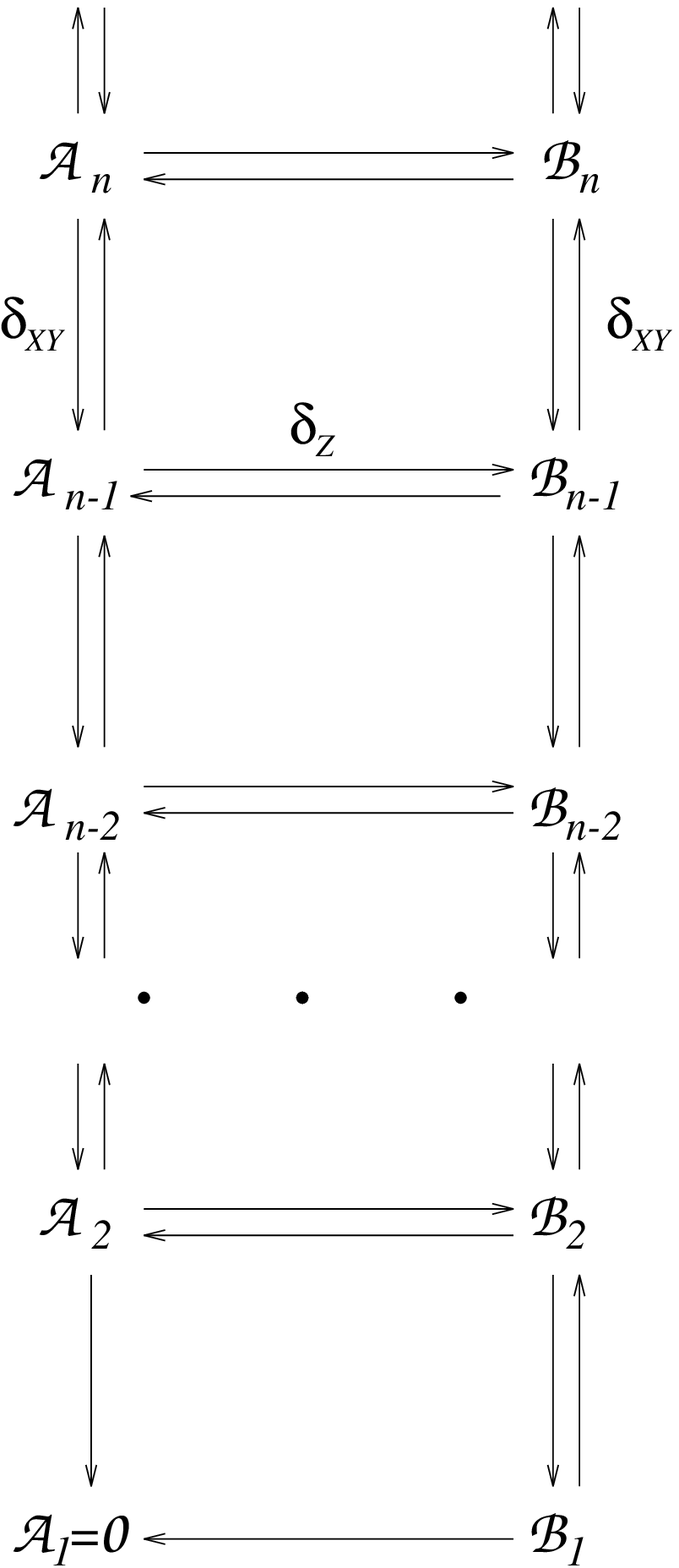}}\ \ \  $$
\vskip 0.7cm
Fig. 3.
The action of the derivations $\delta_{XY}, \delta_Z$ in the
space
$\calA\oplus\calB$ of the XY model.}
\vskip 1cm

The derivation $\delta_Z$ acts ``horizontally," i.e.
$\delta_Z: \calA_n \to \calB_n$,
$\delta_Z: \calB_n \to \calA_n$,
while $\delta_{XY}$ acts ``vertically," i.e.
$\delta_{XY}: \calA_n\to \calA_{n-1}\oplus\calA_{n+1}$, and
$\delta_{XY}: \calB_n\to \calB_{n-1}\oplus\calB_{n+1}$.
The action of $\delta_{XY}$ on an arbitrary $X\in \calA\oplus\calB$ can be
written as:
$$\delta_{XY}(X)=u(X)+d(X),\eq$$
where $u:\calA_n\to \calA_{n+1}$, $u:\calB_n\to \calB_{n+1}$,
and $d:\calA_n\to \calA_{n-1}$, $d:\calB_n\to \calB_{n-1}$.
A simple calculation yields:
\eqn\uddz{\eqalign{
u(e_n^{xx})=&-\b(e_{n+1}^{yx}+e_{n+1}^{xy}),\cr
u(e_n^{yy})=&\a(e_{n+1}^{yx}+e_{n+1}^{xy}),\cr
u(e_n^{xy})=&u(e_n^{yx})=\a e_{n+1}^{xx}-\b e_{n+1}^{yy},\cr }}
and
\eqn\dd{\eqalign{
d(e_n^{xx})=&-\a(e_{n-1}^{yx}+e_{n-1}^{xy}),\cr
d(e_n^{yy})=&\b(e_{n-1}^{yx}+e_{n-1}^{xy}),\cr
d(e_n^{xy})=&d(e_n^{yx})=\b e_{n-1}^{xx}-\a e_{n-1}^{yy}.\cr }}
Similarly,
\eqn\zzz{\eqalign{
\delta_Z(e_n^{xx})=&h(e_{n}^{yx}+e_{n}^{xy}),\cr
\delta_Z(e_n^{yy})=&-h(e_{n}^{yx}+e_{n}^{xy}),\cr
\delta_Z(e_n^{xy})=&\delta_Z(e_n^{yx})=h(e_{n}^{yy}-e_{n}^{xx}).}}

The kernel of $u$ in the space $\calA_n\oplus\calB_n$
is thus two dimensional (it is the span of $e_n^{xy}-e_n^{yx}$
and $\a e_n^{xx}+\b e_n^{yy}$),
allowing for the existence of two families of
integrals of motion.  The explicit expressions for
half of the XX charges are not affected by the
anisotropic deformation of the hamiltonian:
it is simple to check, using \uddz, that
$$H_n^{(1)}=e_n^{xy}-e_n^{yx}\eq$$
commutes with the hamiltonian, for any
$n$.  In order to find explicit expressions for the remaining
charges, we write them in  the form
\eqn\hxyform{H_n^{(2)}=
\a e_n^{xx}+\b e_n^{yy}+\sum_{k=1}^{[n/2]}
 X_{n-2k} +\sum_{k=1}^{[(n+1)/2]} Y_{n+1-2k},}
with $X_{n-2k}\in \calA_{n-2k}( \calB_{n-2k})$
and  $Y_{n+1-2k}\in \calB_{n+1-2k}( \calA_{n+1-2k})$
if $n$ is even (odd).
The condition
$[H_{XY}+h S^z, H_n^{(2)}]=0$ gives then the following system of equations:
$$
 \eqalign{ u(Y_{n-1})+ \delta_Z(\a e_n^{xx}+\b e_n^{yy})&=0,\cr
u(X_{n-2k})+\delta_Z(Y_{n-2k+1})+d(X_{n-2k+2})&=0,
\text{for}k=1,\dots,[(n+1)/2],\cr}\eqlabel\XYcon$$
which allow for a recursive determination of the $X,Y$'s: using
\uddz,
$Y_{n-1}$ is fixed by the first equation; the second one, with $k=1$,
determines
$X_{n-2}$; then, the case $k=2$ leads to
$Y_{n-3}$, etc. Note that since the kernel of $u$ in $\calA_m\oplus\calB_m$
is two-dimensional, (\XYcon) does not fix the terms
$X_{n-2k}$ and $Y_{n+1-2k}$ uniquely, but only up to
an admixture of lower order integrals of motion.
In solving (\XYcon) one may thus impose additional
constraints. One possibility is to demand that
$X_{m}$ and $Y_{m}$ be orthogonal to the kernel of $u$ in
$\calA_m\oplus\calB_m$. This yields
$$H_n^{(2)}=\a e_n^{xx} +\b e_n^{yy} +\sum_{k=1}^{n-1}
c_k (\b e_{n-k}^{xx}-\a e_{n-k}^{yy}),\eq$$
where the coefficients $c_k$ satisfy $$c_{k}={1\over{(\a^2+\b^2)}}
(h (\a+\b) c_{k-1}-2\a\b c_{k-2}),\eq$$
with
$$c_1={h (\a-\b)\over (\a^2+\b^2)}~,\text{and} c_0={( \b^2-\a^2)\over (
\a^2+\b^2 )}\eq$$
 Alternatively, one may obtain a simpler solution by
demanding that the number of
terms $X_{n-2k}$ and $Y_{n+1-2k}$ in \hxyform\ be minimal.
This leads to following expressions, that have first been written in [\Gu]:
$$ H_n^{(2)}= \a e_n^{xx}+\b e_n^{yy} +Y_{n-1}+X_{n-2},\eq$$
where
\eqn\XYeq{\eqalign{
Y_{n-1}=&-h (e_{n-1}^{xx}+e_{n-1}^{yy}),\cr
X_{n-2}=&~\a e_{n-2}^{yy}+\b e_{n-2}^{xx}.}}

As in the XX case, it can be shown that
the boost acts as a ladder operator.
The recursive application of the boost,
 starting from the hamiltonian (\HXYh),
generates a family of scalar conserved charges.
For $h=0$ the $n$-th order  charge of this family
coincides  with the $n$-th
logarithmic derivative (\qndef) of the XYZ transfer matrix, evaluated at
$\c=0$.
The recursive application of the boost operator, starting from
$H_2^{(-)}$ produces a family of pseudoscalar conservation
laws.\foot{Note
that for $h=0$ the charges of second family cannot be expressed
in terms of the logarithmic derivatives of the XYZ transfer matrix, in the
parametrization adopted here. Observe
however that both families can be obtained from the XYh transfer matrix,
written down in
[\ref{E. Barouch, {\it Stud. Appl. Math.}
 {\bf 70} (1984),
151.}\refname\Bar].}

To end this section, let us mention that the degenerate case
$\b=0$ of the XYh model (which could be called the Xh model)
is known as the Ising chain.
It is evident that this system inherits from the XYh model
two distinct families of conserved charges.

\subsec{Higher order ladder operators}

An interesting feature of the XYh model is the existence of higher order
ladder operators [\Ara, \BarFuch].
For simplicity we focus again on the XX case.
The first moment of $H_n^{(+)}$
acts then as a ladder operator
whose application to a 
charge of order $m$ generates a
charge of order $m+n-1$.
For example,
$$B^{(+)}_3={1\over 2i}\sumL j[ e_{3,j}^{xy}-e_{3,j}^{yx}] ,\eq$$
acts in the following way:
$$ \eqalign{ [ B^{(+)}_3, H_2^{(+)}]&= H^{(+)}_4- H_2^{(+)}, \cr
 [ B_3^{(+)}, H_{n>2}^{(+)}]&= (n-1) (H^{(+)}_{n+2}- H_{n-2}^{(+)}) .}\eq$$
The first moments of the pseudoscalar charges transform
scalar (pseudoscalar)  charges into pseudoscalar (scalar) ones,
e.g.
$$B_2^{(-)}={1 \over {2 i}}\sumL j [\s^x_j\s^y_{j+1}-\s^y_j\s^x_{j+1}],\eq $$
acts as follows:
$$[B_2^{(-)}, H_n^{(\pm)}]= (n-1)( H_{n+1}^{(\mp)} -
 H_{n-1}^{(\mp)}).\eq$$
Since the commutator of two ladder operators has again the ladder property,
any two such operators
can be used to generate an infinite family of ladder operators.
No such higher order ladder operators
are known  in the general nondegenerate case of the XYZ model.
In particular, the first moment of the density of the three-spin
XYZ charge does not have the ladder property.


\newsec{The $su(M)$ invariant spin chain}

\subsec{Formulation of the model}

In this section we consider the
isotropic $su(M)$ version of the
 XXX model
in the fundamental representation.
The hamiltonian of this system reads [\ref{B. Sutherland, {\it Phys. Rev.}
{\bf 12} (1975), 3795.}\refname\Suth]:
\eqn\hsun{
H_2=\sum_{j\in\Lambda} {t^a_j t^a_{j+1} }, }
where $t^a_j$, $a=1,..., M^2-1$, are the  $su(M)$ generators in the fundamental
representation, acting non-trivially only on the
$j-th$ factor of the Hilbert space $\bigotimes_j \CC^N$.
For $M=2$ \hsun\ is simply  the hamiltonian of the spin-1/2 XXX model.
It is convenient to choose the normalization  so that $t^a$ are the
$su(M)$ Gell-Mann matrices:
\eqn\sungm{\eqalign{ [t^a,t^b]=&2  i f^{abc} t^c, \cr
t^a t^b +t^b t^a=&4 \delta_{ab}/M+2d^{abc}t^c,}}
where $f^{abc}$ are the structure constants of $su(M)$, and
$d^{abc}$ is a completely symmetric tensor, non-trivial for all $M>2$.

\subsec{Expressions for the conserved charges}

Remarkably, the conserved charges for general $M$ have the same
structure as in the  $M=2$ case (the ordinary $s=1/2$ XXX
model). The expressions for the general case can be obtained from \Hndef\ by
a simple substitution:
$$\s^a\to t^a~\qquad\epsilon^{abc}\to f^{abc}.\eq$$
For example,
\eqn\hsunthr{H_3=\sumL f^{abc} t^a_j t^b_{j+1} t^c_{j+2},}
\eqn\hsunfour{H_4=\sumL [f^{abp}f^{pcd} t^a_j t^b_{j+1} t^c_{j+2}
t^d_{j+3} + t^a_j t^a_{j+2}].}
As for $M=2$, we can define
the spin polynomials $f_n(\calC)$,
for an arbitrary cluster $\calC=\{j_1,j_2,\dots, j_n\}$, as follows:
\eqn\fnsun{f_n(\calC)=f^{a_1a_2b_2} f^{b_2a_3 b_3}\dots
f^{b_{n-3}a_{n-2}b_{n-2}}
f^{b_{n-2}a_{n-1}a_n}
t^{a_1}_{j_1}
t^{a_2}_{j_2} \dots
t^{a_n}_{j_n} .}
The general expression for the $su(M)$ XXX conserved charges can then be
written in a universal form, (corresponding to the Catalan tree in
Fig. 1.), valid for all $M$:
$${ H_n= F_{n,0}+ \sum _{k=1}^{[n/2]-1} \sum_{\ell=1}^{k}
\alpha_{k,\ell} F_{n-2k,\ell},}\eq$$
with
$$ F_{n,k}=\sumC f_n(\calC).\eq$$

The proof that the $H_n$'s form a commuting
family parallels the analogous proof in the
 $su(2)$ case. First, we show
that the identities \idone-\idthr\ remain valid for $su(M)$, but now with
${\bf A}=\{t^1_j,\dots,t^{M^2-1}_j\}$ and
${\bf B}=\{t^1_k,\dots,t^{M^2-1}_k\}$, (with $j\ne k$)
 and with ${\bf A}\times {\bf B}$
understood as a vector with
$M^2-1$ components given by
$$({\bf A}\times {\bf B})^c=f^{abc}  A^a B^b.\eq$$
The demonstration of
\idone-\idthr\ is elementary and uses several identities for
the tensors $f^{abc}$ and $d^{abc}$:\foot{The proofs of
the first three identities can be found in
[\ref{A. J. MacFarlane, A. Sudbery and P. H. Weisz,
{\it Comm. Math. Phys.} {\bf 11} (1968), 77.}]. The fourth identity can be
proved using the first two equalities as well as two additional
well known identities:
$$\eqalign{ f^{adc} f^{cbe} +f^{bdc} f^{ace}+ f^{edc}f^{abc}=0,\cr
f^{adc} d^{cbe} +f^{bdc} d^{ace}+ f^{edc}d^{abc}=0. }$$}
\eqn\fid{\eqalign{
f^{abc}f^{cde}f^{ega}&=-(M/2) f^{bdg},\cr
f^{abc}f^{cde}d^{ega}&=-(M/2) d^{bdg},\cr
f^{abc}f^{abd}&=M\delta^{cd},\cr
f^{abc}f^{cde}f^{egh}d^{hai}&=-f^{adc}f^{cbe}f^{eih}d^{hag}.
}}
The rest of the proof is exactly the same as in the $su(2)$ case:
from  \idone-\idthr\ one derives \comHfnC\ and proves first $[H_2,H_{n}]=0$
and then the mutual commutativity of all the $H_n$'s.

\subsec{Relation to higher spin $su(2)$ models}

The $su(M)$ XXX system can be also regarded as a $su(2)$ spin-$s$ chain,
with $s=(M-1)/2$. The hamiltonian \hsun\
can be rewritten in terms of the variables
$S_j^a$, which are the
spin-$s$ $su(2)$ generators, acting non-trivially only at site $j$:
\eqn\hsunsutw{ H=\sum_{j\in\Lambda} G_M (x_j),}
where $G_M$ is a polynomial of order $M^2-1$,
and $x_j=S^a_j S^a_{j+1}$.
In particular, the expressions for the first few
polynomials $G_M$
(obtained  by expressing the matrices $t^a$ in terms of the standard spin-$s$
matrices) read:
\eqn\sunpol{ \eqalign{
G_3(x)=&x+x^2,\cr
G_4(x)=&x-{4\over {81}} (11 x^2 +4 x^3), \cr
G_5(x)=&x+{1\over{90}}(13 x^2 -6 x^3 -x^4),\cr
G_6(x)=&x+{8\over {178435}}(14911 x^2+636 x^3 -360 x^4- 32 x^5),\cr
G_7(x)=&x+{1\over {14094}}(-6417 x^2- 1713 x^3 - 8 x^4 +19 x^5+ x^6).
}}
It is convenient to use the basis provided by the projection operators
$P^j_{s}$, defined by
$$ P^j_{s}(x)=\prod_{ l=0 \atop {l\ne j}}^{2s}{ x-y_j \over {y_l - y_j}},
\eq$$
where $j=0,\dots 2s$, and
$$y_l=\hal [2 l(l+1) -s(s+1)].\eq$$
Acting on the states of the tensor product of two spins,
$P^j_s$ projects  onto the states with total spin $j$. In terms of this basis,
modulo constants,
$$G_M(x)=\sum_{j=0}^{2s} (-1)^j P^j_{(M-1)/2}(x).\eq$$
Similarly, the density of a XXX conserved charge of order $n$ in
terms of the $su(2)$ variables
becomes a polynomial in $n$ (matrix) variables $x_j,\dots, x_{j+n-1}$.

\newsec{Equivalent representations of charges in spin chains}

 \subsec {$su(2)$ spin chains in the Weyl representation}

It is well known that
 the $su(2)$ generators can be
represented as differential operators acting in the space of functions of
two complex variables $u, v$ satisfying $|u|^2+|v|^2=1$:
\eqn\Weyl{\eqalign{ \s^z&=u\partial_u -v\partial_v,\cr
\s^x&=u\partial_v+v\partial_u,\cr
\s^y&=i( v\partial_u-u\partial_v).}}
This is often called the Weyl representation.
The spin-$s$ representation can be obtained by the restriction of
the space of functions to polynomials
of degree $2s$. States of the spin-$s$ chain with $N$ sites can be
therefore represented by polynomials of degree $2s$ in $2N$ variables
${u_i,v_i}$ ($i=1,\dots,N$).
The conserved charges $H_n$ become commuting differential operators
of degree $2n$ in the variables $u_i,v_i,\partial_{u_i},\partial_{v_i}$.
For example, the XXZ hamiltonian in the Weyl representation is:
\eqn\htwoWeyl{\eqalign{H_2=&
\sumL [ (2-\c) (u_j v_{j+1}\partial_{v_j}\partial_{u_{j+1}}+
v_{j}u_{j+1}\partial_{u_j}\partial_{v_{j+1}} )\cr &
+ \c (u_j u_{j+1} \partial_{u_j}\partial_{u_{j+1}}+
v_j v_{j+1} \partial_{v_j}\partial_{v_{j+1}})].}}
Higher charges can be obtained in the same manner,
but the resulting expressions do not seem  particularly transparent.

Observe that in deriving the explicit form \Hndef\ of
the charges $H_n$
for the $s=1/2$ XXX chain
one makes use of both the $su(2)$ commutator and the
anticommutator $\s^a\s^b+\s^b\s^a=\delta^{ab}$.
In the Weyl representation, this anticommutator is not encoded in \Weyl,
but results from the restriction of the representation space
to multilinear polynomials. It is therefore clear that the charges
\Hndef\ cease to commute for spin chains with $s>1/2$.
Establishing the explicit form of the charges in this case is a
much more difficult task.

\subsec {Representations of the XXX charges in terms of braids}

 For both the $s=1/2$ and the $su(M)$  XXX models,
 the hamiltonian can be written, modulo constants,  as:
 \eqn\hamexch{H= \sum_{j\in\Lambda}  P_{j,j+1},}
 where $P_{j,k}$ are operators exchanging particles at
 sites $j$ and $k$.
 The spin polynomials $f_n(C)$,  where $\calC=\{i_1,i_2,\dots,i_{n-1},i_n\}$
can
 be rewritten in terms of the exchange operators as follows:
  \eqn\exchf{ f_n(C)=[...[P_{i_1,i_2},P_{i_2,i_3}],
 P_{i_3,i_4}],P_{i_4,i_5}],....,P_{i_{n-1},i_n}].}
 This permits to express the  higher-order hamiltonians for the  XXX
 system as a sum of products of the exchange operators.

 Let $b_i=P_{i,i+1}$ be a nearest-neighbor exchange operator. It is simple
to check
that the
 $b_i$'s satisfy the defining algebra of the braid group $B_N$,
 (where $N=|\Lambda|$):
 \eqn\brgral{\eqalign{b_i b_{i+1} b_i&= b_{i+1} b_i b_{i+1} , \cr
 b_i b_j&=b_j b_i, \quad |i-j|>1,}}
and an additional constraint
$$ b_i^2=1. \eqlabel\brconst$$
Note that any binary exchange (not necessarily nearest-neighbor)
can be realized as a product of nearest-neighbor exchanges.
 Therefore, it follows that the spin polynomials \exchf\ can
 be expressed in terms of the generators of $B_N$.
  For example, if $\calC$ contains only adjacent spins,
 $$f_n(C)=[...[b_1,b_2],b_3],b_4],....,b_n].\eq$$
 Hence the XXX charges can be interpreted in terms of
 equivalence classes (with respect to (\brconst)) of braid polynomials (see
Fig. 4. for an example).

\vskip 0.7cm
\vbox{$$ {\epsfxsize= 9.1  cm \epsfbox{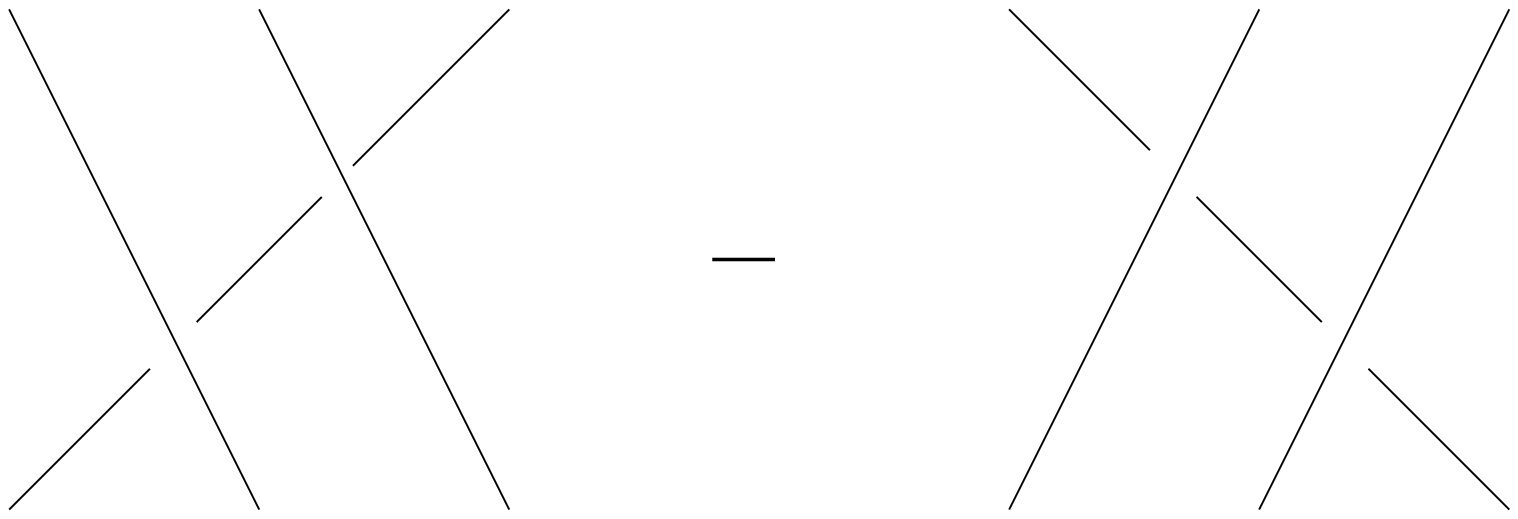}}\ \ \  $$
\vskip 0.5cm
Fig. 4.
A braid polynomial corresponding to the three-spin XXX charge
}
\vskip 1cm


\newsec{Structure of conserved charges in the Hubbard model}

\subsec{Introduction}

The Hubbard model (for which original references can be found in [\ref{A.
Montorsi,
Ed., {\rm ``The Hubbard Model,"} World Scientific, 1992.}]) is usually defined
in terms
of fermionic operators, by the hamiltonian
$$ H= -2\sum_{j,~s=\ua, \da} (a^\dg_{j,s} a_{j+1,s} + a^\dg_{j+1,s} a_{j,s}) +
4U\sumL (n_{j,\ua}-1/2)( n_{j,\da}-1/2),\eq$$
where
$$n_{j,s} = a^\dg_{j,s} a_{j,s},\eq$$
and $U$ is some coupling constant.  The sum over $j$ runs from 1 to $N$, the
number of sites.  The operators $a^\dg_{j,s}$ and $ a_{j,s}$ are respectively
the
creation and the  annihilation operator of an electron of spin $s$ at site $j$,
subject to the anti-commutation relation:
$$a^\dg_{j,s} a_{k,s'} + a_{k,s'} a^\dg_{j,s} = \delta_{j,k}\delta_{s,s'},
\eqlabel \anticomf$$
and
 $$
a^\dg_{j,s}a^\dg_{k,s'} = -a^\dg_{k,s'}a^\dg_{j,s},\qquad a_{j,s}
a_{k,s'}=-a_{k,s'}
a_{j,s}.\eqlabel\anticoms$$
The hamiltonian can be rewritten in terms of spin variables by means of the
standard Jordan-Wigner transformation:
$$\eqalign{&a_{j,\ua} = (\s_1^z\s_2^z...\s_{j-1}^z)~\s_j^-,\cr
&a_{j,\da} =
(\s_1^z\s_2^z...\s_N^z)(\t_1^z\t_2^z...\t_{j-1}^z)~\t_j^-.\cr}\eqlabel\JW$$
$\{\s_j^a\}$ and $\{\t_j^a\}$, $a\in\{x,y,x\}$, $j\in\Lambda$,
 are two independent sets of Pauli
matrices:
$$[\s_j^i,\t_k^\ell] = 0,\eq$$
and the normalization used for $\s_j^\pm$ is
$$\s_j^\pm = \hf(\s_j^x \pm i \s_j^y).\eq$$
In this way, the hamiltonian $H=H_2$ is easily seen to be equivalent to
(\hubham).
Notice that for $U=0$, the Hubbard model
reduces to two uncoupled XX models and that for
$\t_j^\pm = 0$  and $ \t_j^z = h$ for all $j$, it reduces to the XX model in
an external magnetic field.

 The first non-trivial conservation law was obtained by
Shastry and found to be [\Shas]:
$$\eqalign{H_3 =
&\sumL [(\s_j^x\s_{j+1}^z\s_{j+2}^y-\s_{j}^y\s_{j+1}^z\s_{j+2}^x)\cr  &~- U
(\s_j^x\s_{j+1}^y-\s_{j}^y\s_{j+1}^x)(\t_{j}^z+\t_{j+1}^z)] +
[\s\lra\t].\cr}
\eq$$ It is simple to verify that it indeed commutes with $H_2$.

To conclude this
short introduction of
the Hubbard model, we point out that there
exists another equivalent way to represent it: redefining the spin
variables as
$\s^a_j\to S^a_{2j}$ and
$\t^a_j\to S^a_{2j+1}$, we obtain a lattice with $2N$ sites with
next-to-nearest neighbor interactions and bond alternations:
$${H_2=\sum_{j} S^+_{j}S^-_{j+2} + S^-_{j}S^+_{j+2}+
    U [1+(-)^j]S^z_{j} S^z_{j+1}.}\eqlabel\Hubeq$$


\subsec{The non-existence of a ladder operator}

We first demonstrate that for the Hubbard model there is no ladder operator $B$
whose commutator with $H_2$ would produce $H_3$.
The most general possible expression for $B$ is [\FF]:
$$B= \sumL  [~jb_j + c_i],\eq$$
where $b_j$ and $c_j$ are translation invariant densities, at most bilinear in
(nearest-neighbor) spin variables.  Enforcing the symmetry $\s\lra \t$ and
requiring $B$
to reproduce the known ladder operator in the two limiting cases $U\ra 0$ and
$\t^z_j\ra
h$ gives
$$\eqalign{b_j=&\s^x_j\s^x_{j+1}+\s^y_j\s^y_{j+1}+\t^x_j\t^x_{j+1}+
\t^y_j\t^y_{j+1}+{U\over 2}(\s^z_j\t^z_{j}+\s^z_{j+1}\t^z_{j+1}),\cr
c_j=&0.\cr}\eq$$
The resulting $B$ is thus the first moment of the hamiltonian, with the density
of the
latter  symmetrized with respect to sites $j$ and $j+1$:
$$B= \sumL ~j[\s^x_j\s^x_{j+1}+\s^y_j\s^y_{j+1}+\t^x_j\t^x_{j+1}+
\t^y_j\t^y_{j+1}+{U\over 2}(\s^z_j\t^z_{j}+\s^z_{j+1}\t^z_{j+1})]
.\eqlabel\boosca$$ This
form of $B$ is in fact generic to all the cases where it exists.

With this candidate ladder operator, a simple
calculation yields:
$$\eqalign{[B,H_2] = -2i&\sumL
[\s_j^x\s_{j+1}^z\s_{j+2}^y-\s_{j}^y\s_{j+1}^z\s_{j+2}^x \cr&~~ -{U\over 2}
(\s_j^x\s_{j+1}^y-\s_{j}^y\s_{j+1}^x)(\t_{j+1}^z +\t_{j}^z)]
+[\s\lra\t].\cr}\eq$$
The sum has the same form as the above $H_3$, except that the coefficient of
the $U$
term is here $-1/2$ instead of $-1$.  This particular value of the relative
coefficient
between the
$U$ independent term and the linear one is of course crucial for the
commutativity of $H_3$ with $H_2$.  In other words,
$[H_2,[B,H_2]]$ does not vanish.
This finishes the proof of   the
non-existence of a ladder operator for the Hubbard model.


\subsec{Higher order charges}

As already indicated in the introduction, higher order
charges must be calculated by brute force methods and they become rather
complicated.  For instance $H_4$ reads:\foot{One
could have hoped that the above operator $B$ would still be of some
use, at
least for generating the correct terms in $H_4$ (despite the fact that the
coefficients
cannot be expected to be right) in the commutator
$[B,H_3]$.  But some terms in $H_4$ are not produced in this way.}
$$\eqalign{H_4 = &\sumL
[(\s_j^x\s_{j+1}^z\s_{j+2}^z\s_{j+3}^x+
\s_{j}^y\s_{j+1}^z\s_{j+2}^z\s_{j+3}^y)\cr
 &~- U
(\s_j^x\s_{j+1}^z\s_{j+2}^x+\s_{j}^y\s_{j+1}^z\s_{j+2}^y)
(\t_{j}^z+\t_{j+1}^z+\t_{j+2}^z)\cr
&~- {U\over 2}
(\s_j^x\s_{j+1}^y-\s_{j}^y\s_{j+1}^x)(\t_j^x\t_{j+1}^y-\t_{j}^y\t_{j+1}^x)
\cr &~-U
(\s_j^x\s_{j+1}^y-\s_{j}^y\s_{j+1}^x)(\t_{j+1}^x\t_{j+2}^y-
\t_{j+1}^y\t_{j+2}^x)\cr
&~-U(\s_j^ z\t_{j+1}^z) -{U\over 2}(\s_j^ z\t_{j}^z) \cr
&~+U^2(\s^x_j\s^x_{j+1}+\s^y_j\s^y_{j+1})(\t_{j}^z\t_{j+1}^z + 1)]\cr
&~+ [\s\lra\t].\cr}\eq$$
  This charge has also been
found in [\ref{H. Grosse, {\it Lett. Math. Phys.} {\bf 18} (1989), 151. },
\Zhou].  Observe
that these integrals of motion can be written much more compactly in
terms of the tensor products of the densities of the XX charges.
It is convenient to  use a notation which
differs slightly from the one used in section (6); namely, we define:
$$\eqalign{h^{(+)}_{n,j} =& e_{n,j}^{xx}+e_{n,j}^{yy}
=\s_j^x\s_{j+1}^z\s_{j+2}^z...\s_{j+n-1}^z\s_{j+n}^x+
\s_j^y\s_{j+1}^z\s_{j+2}^z...\s_{j+n-1}^z\s_{j+n}^y,\cr
h^{(-)}_{n,j} = & e_{n,j}^{xy}-e_{n,j}^{yx}=
\s_j^x\s_{j+1}^z\s_{j+2}^z...\s_{j+n-1}^z\s_{j+n}^y-
\s_j^y\s_{j+1}^z\s_{j+2}^z...\s_{j+n-1}^z\s_{j+n}^x,}\eqlabel\defha$$
for $n>1$, and
$$h^{(+)}_{1,j} = -\s_j^z~,\qquad h^{(-)}_{1,j}=0,\eqlabel\defhb$$ and
similarly
$$g^{(\pm)}_{n,j}(\t)= h^{(\pm)}_{n,j}(\s\ra\t).\eqlabel\defg$$
With this  notation, the first three charges take the
form:
$$\eqalign{
H_2 = & \sumL [ h^{(+)}_{2,j}+g^{(+)}_{2,j}+U~ h^{(+)}_{1,j}g^{(+)}_{1,j}],\cr
H_3 = &\sumL [ h^{(-)}_{3,j}+g^{(-)}_{3,j} + U~\{
h^{(-)}_{2,j}(g^{(+)}_{1,j}+ g^{(+)}_{1,j+1} )+ g^{(-)}_{2,j}
(h^{(+)}_{1,j}+ h^{(+)}_{1,j+1} )\}],\cr
H_4 = &\sumL [ h^{(+)}_{4,j}+g^{(+)}_{4,j} + U~\{
h^{(+)}_{3,j}(g^{(+)}_{1,j}+ g^{(+)}_{1,j+1}+ g^{(+)}_{1,j+2} ),\cr&~~+
g^{(+)}_{3,j}
(h^{(+)}_{1,j}+ h^{(+)}_{1,j+1} + h^{(+)}_{1,j+2} )
-h^{(-)}_{2,j}(g^{(-)}_{2,j-1}+g^{(-)}_{2,j}
+g^{(-)}_{2,j+1})\cr&~~ -
h^{(+)}_{1,j}(g^{(+)}_{1,j-1}+g^{(+)}_{1,j}+g^{(+)}_{1,j+1})\}\cr
&~~+U^2~\{h^{(+)}_{2,j}(g^{(+)}_{1,j}g^{(+)}_{1,j+1}+1)+g^{(+)}_{2,j}
(h^{(+)}_{1,j}h^{(+)}_{1,j+1}+1)\}]
.\cr}
\eq$$
We have also obtained the explicit form of $H_5$ which is
$$\eqalign{H_5 = &\sumL [ h^{(-)}_{5,j} + U~\{
h^{(-)}_{4,j}(g^{(+)}_{1,j}+ g^{(+)}_{1,j+1}+ g^{(+)}_{1,j+2}
+ g^{(+)}_{1,j+3} )\cr
&~~+h^{(+)}_{3,j}(g^{(-)}_{2,j-1}+
g^{(-)}_{2,j}+g^{(-)}_{2,j+1}+g^{(-)}_{2,j+2})\cr&~~
-h^{(-)}_{2,j}(g^{(+)}_{1,j-1}+g^{(+)}_{1,j}+g^{(+)}_{1,j+1}+
g^{(+)}_{1,j+2})\}\cr&~~
+U^2~\{h^{(-)}_{3,j}(g^{(+)}_{1,j}g^{(+)}_{1,j+1}+g^{(+)}_{1,j}g^{(+)}_{1,j+2}+
g^{(+)}_{1,j+1}g^{(+)}_{1,j+2})\cr&~~
+h^{(+)}_{2,j}(g^{(-)}_{2,j-1}g^{(+)}_{1,j+1}
+g^{(-)}_{2,j+1}g^{(+)}_{1,j})\}\cr&~~-U^3~\{
h^{(-)}_{2,j}(g^{(+)}_{1,j}+ g^{(+)}_{1,j+1} )\}] + [h\lra g].\cr}\eq$$

We now present a number of simple observations concerning the structure of
the conserved charges, indicated by the above results.

\n 1- $H_n$ is composed of sum of terms of the form
$$h^{(\ep)}_{\ell_1}
h^{(\ep')}_{\ell_2}...g^{(\ep'')}_{\ell_{p-1}}g^{(\ep''')}_{\ell_p}, \qquad
\eqlabel\formt$$
with $\ell_i\geq 1$, $i=1,\dots,p$, where $p$ is some integer
and $\ep, \ep',... = \pm$. A  parity of (\formt) is  given by
$$\ep \ep'...\ep''\ep''' = (-)^n.\eq$$
Furthermore, these terms must all be scalars, which means that
in the product (\formt) pseudoscalar factors must occur in pairs.\foot{A
pseudoscalar changes sign when the
direction of the $z$-axis is reversed.  $h^{(+)}_n$ is scalar (pseudoscalar)
for
$n$ even (odd), while $h^{(-)}_n$ is scalar (pseudoscalar) for $n$
odd (even).  The same holds for $g$.}

\n 2- With the normalization we have chosen for $H_2$, the relative
coefficients of the
various components of
$H_n$ are
$\pm$ times some powers of $U$.

\n 3- The term independent of $U$ in $H_n$ is given by
$$H_n(U=0)= \sumL [h^{(\ep)}_{n,j}+g^{(\ep)}_{n,j}],\eq$$
with $\ep = (-)^n$. Terms depending upon $U$ involve products
of $h$ and $g$ factors of
the form (\formt)
 with
$$\sum_{i=1}^p \ell_i = n-2r,\eq$$ where
$r$ is a non-negative integer.  These terms can be therefore
grouped into classes,
associated to the partitions of $n-2r$, for $r=0,\dots,[n/2]-1$.

\n 4- The terms in $H_n$ associated to partitions of $n$ (those for
which
$r=0$) have a simple pattern. The number of parts $p$ in the
partition is related to the power of
$U$ in the coefficient of the corresponding term in $H_n$. This power is simply
 $p-1$.
For example, the terms with $r=0$ which are linear in $U$
are obtained from all possible decomposition of $n$ into two parts of distinct
`colors'
(the
$\s$ and
$\t$ factors), e.g.
$$5=4+(1)=1+(4)=3+(2)=2+(3),\eq$$
(where terms referring to $g$ factors are written in parentheses).  Similarly
the terms of
order $U^2$ are associated with partitions of $n$ into three parts of two
colors. For
instance, the terms appearing with a coefficient of absolute value $U^2$ in
$H_5$ are
in correspondence with the partitions
$$5=3+(1+1)= 1+1+(3) = 2+(2+1) = 2+1+(2).\eq$$
  But not all
partitions into more than two parts are eligible (e.g., the partition
$5=(1)+2+2$ is
absent).  Some, but not all, are eliminated by the scalar and/or the parity
conditions stated above. For instance, $5=1+1+1+1+1$, with any arrangement of
parentheses, is excluded by the parity and the scalar condition.

\n 5-  The above correspondence is not one-to-one
since more than one term is associated to a given partition.
Consider a class of terms corresponding to a certain partition.
Terms within such class are related to
each other by a relative translation of their components.  For the first
two-component
partition, i.e., $n=n-1+(1) $, the different terms in the class correspond
to the $n-1$ positions of $g^{(+)}_{1,k}$, $k=j,...,j+n-1$, relative to the
$h_{n-1,j}$ factor.  For the other two-component partitions,
there are also $n-1$ terms in the class, with the difference that now the first
term in the class $n=n-k+(k)$ is $g_{k, j-k+1}$ and the other terms are
obtained by translating this $g$ factor by one unit with respect to
$h_{n-k,j}$, up to the site $j+n-k$.  The different terms in classes labeled by
partitions into more parts can be described similarly.

It is more difficult to find a
simple pattern for the terms associated to partitions of $n-2r, ~r>1$ because
they are
not universal, in the sense that they can be modified by the addition of a
linear
combination of the lower order charges $H_{n-2r}$.  Nevertheless,
the terms linear in $U$ are described by the explicit formula presented below
in section (9.5).


\subsec{A diagrammatic description of the conserved charges}

Formulated in words, some of the above rules may look a bit complicated.
However they
have a very simple diagrammatic description.
To the densities $h_{n,j}^{(\pm)}$,
$g_{n,j}^{(\pm)}$, we will assign sequences of $n$ symbols (corresponding
to $n$ successive sites of the chain starting at site $j$), as
follows:
$$\eqalign{
h_{n,j}^{(+)}=& \ps\ps\dots\ps\ps, \cr
h_{n,j}^{(-)}=& \ms\ms\dots\ms\ms, \cr
g_{n,j}^{(+)}=& \pt\pt\dots\pt\pt,\cr
g_{n,j}^{(-)}=& \mt\mt\dots\mt\mt .}$$
The product of two or more
densities will be represented by writing the corresponding
sequences of symbols, appropriately shifted, in successive rows, e.g:
$$h_{2,j}^{(-)} g_{2,j+1}^{(+)} =\matrix{ \ms\ms\ee\cr\ee\pt\pt} ~~~.$$
It is understood that the leftmost symbol in the top row is at site $j$.
In this notation, (omiting the sums over $j$), the charges
read:
$$ H_2= \matrix{\ps\ps}\quad +\quad \matrix{\pt\pt}\quad +\quad
         U \matrix{\ps\cr\pt}, $$
$$ H_3= \matrix{\ms\ms\ms}\quad +\quad
         U( \matrix{\ms\ms\cr \pt\ee }+
        \matrix{ \ms\ms\cr \ee\pt })+ (\s \lra \tau),$$
$$\eqalign{ H_4=& \matrix{\ps\ps\ps\ps}\quad
+\quad
         U(~\matrix{\ps\ps\ps&\cr\pt\ee\ee&}+
          \matrix{\ps\ps\ps&\cr\ee\pt\ee&}+
          \matrix{\ps\ps\ps&\cr\ee\ee\pt& }~)
\cr
\quad +\quad
         &U(~\matrix{&\ps\ee\ee\cr&\pt\pt\pt}+
          \matrix{&\ee\ps\ee\cr&\pt\pt\pt}+
         \matrix{&\ee\ee\ps\cr&\pt\pt\pt}~)
\cr
 \quad - \quad
 & U(~\matrix{&\ee\ms\ms\cr&\mt\mt\ee} +
 \matrix{&\ms\ms\cr&\mt\mt} +
 \matrix{&\ms\ms\ee\cr&\ee\mt\mt})
\cr
\quad - \quad
& U~ (~ \matrix{\ee\ps\cr\pt\ee} +
 \matrix{\ps\cr\pt} +
 \matrix{\ps\ee\cr\ee\pt}~)+
U^2 ( 
\matrix{&\ps\ps\cr&\pt\ee\cr&\ee\pt})+
U^2 ( 
\matrix{&\ee\ps\cr&\ps\ee\cr&\pt\pt})
\quad - \quad
U^3 ( \matrix{\ps\cr\pt})
,}$$
$$\eqalign{ H_5=& \matrix{\ms\ms\ms\ms\ms}\quad +\quad 
         U~( ~ \matrix{\ms\ms\ms\ms\cr\pt\ee\ee\ee }+ 
          \matrix{\ms\ms\ms\ms\cr\ee\pt\ee\ee }+
          \matrix{\ms\ms\ms\ms\cr\ee\ee\pt\ee }+
          \matrix{\ms\ms\ms\ms\cr\ee\ee\ee\pt})
\cr
\quad + \quad
         & U~(~\matrix{\ee\ps\ps\ps\cr\mt\mt\ee\ee }+
          \matrix{\ps\ps\ps\cr\mt\mt\ee}+
          \matrix{\ps\ps\ps\cr\ee\mt\mt}+
          \matrix{\ps\ps\ps\ee\cr\ee\ee\mt\mt}~)~
\cr
\quad -\quad
         & U~(~\matrix{\ee\ms\ms\cr\pt\ee\ee }+
          \matrix{\ms\ms\cr\pt\ee }+
          \matrix{\ms\ms\cr\ee\pt}+
          \matrix{\ms\ms\ee\cr\ee\ee\pt})
\cr
\quad + \quad
& U^2~(~ \matrix{\ms\ms\ms\cr\pt\ee\ee\cr\ee\pt\ee}+
   \matrix{\ms\ms\ms\cr\pt\ee\ee\cr\ee\ee\pt}+
   \matrix{\ms\ms\ms\cr\ee\pt\ee\cr\ee\ee\pt})
\quad + \quad
U^2~(~\matrix{\ee\ps\ps\cr\mt\mt\ee\cr\ee\ee\pt} +
 \matrix{\ps\ps\ee\cr\ee\mt\mt\cr\pt\ee\ee} )
\cr
\quad -\quad
& U^3~(~ \matrix{\ms\ms\cr\pt\ee} +
\matrix{\ms\ms\cr\ee\pt}~)
\quad + \quad
 (\s \lra \tau).}$$


\subsec{The explicit form of the term linear in $U$}

We now give the explicit expression of the term linear in $U$ in the charge
$H_n$:
$$\eqalign{ H_n =& ~\sumL [h^{(-)^n}_{n,j}+g^{(-)^n}_{n,j}
\cr &~~+
U~\sum_{k=0}^{[n/2]-1}~\sum_{m=1}^{n-2k-1}~\sum_{\ell=0}^{n-2}
(-)^{n+k+m(n-m)+1}~h^{(-)^{n+m+1}}_{n-m-2k,j}~
g^{(-)^{m+1}}_{m,j-m-k+\ell+1}]\cr &~~+
{\cal O}(U^2)\cr
\equiv &~H_n^{(0)}+ U~H_n^{(1)} + {\cal
O}(U^2).}\eqlabel\hublin$$
We now explain the origin
of the three internal summations for the linear term.  The first one, over $k$,
keeps
track of all the integers $n-2k$ whose two-term partitions are to be
considered.   The
second summation takes care of all the possible ways the integer $n-2k$ can be
separated
into two parts.  Finally, the sum over $\ell$ generates the different terms of
a
given class, that is the different translations of the $g$ factor with respect
to
$h$, whose position is kept fixed. Although it may not be manifest, this
expression is
symmetric with respect to the interchange of $h$ and $g$.

To prove this result we have to show that, up to terms of order $U^2$,
(\hublin) commutes with $H_2$.  The proof  is based on the following
general commutation
relations:
$$\eqalign{  [H_2^{(0)}, h^{(\pm)}_{n,j}] =&~[\sum_{i\in \Lambda}
h^{(+)}_{2,i},
h^{(\pm)}_{n,j}] \cr=&~
\pm 2i\{h^{(\mp)}_{n+1,j-1}+h^{(\mp)}_{n+1,j}+
\delta(h^{(\mp)}_{n-1,j+1}-h^{(\mp)}_{n-1,j})\},}\eqlabel\commua$$
and
$$\eqalign{  [H_2^{(1)}, h^{(\pm)}_{n,j}] =&~[\sum_{i\in\Lambda}
h^{(+)}_{1,i}g^{(+)}_{1,i},
h^{(\pm)}_{n,j}] \cr=&~
\pm 2i\{h^{(\mp)}_{n,j}g^{(+)}_{1,j}-h^{(\mp)}_{n,j}g^{(+)}_{1,j+n-1}
\}.}\eqlabel\commub$$
In the first expression, we have introduced an operator $\delta$ whose action
on
$h^{(\pm)}_{n,i}$ or
$g^{(\pm)}_{n,i}$ is defined to be multiplication by
2 if $n=1$ and 1 otherwise:
$$\delta  h^{(+)}_{1,j}= 2h^{(+)}_{1,j}~,\qquad\delta  h^{(\pm)}_{n\not=1,j}=
h^{(\pm)}_{n,j}.\eq$$
Exactly the same results applies for $h\ra g$.  Notice that both type of
commutators change
the parity of the density on which it acts.

These commutators are very easily
demonstrated
using the definitions (\defha), (\defhb) and (\defg).

We now proceed to the evaluation of $[H_2^{(0)}, H_n^{(1)}]$.  Using
(\commua), one gets
$$\eqalign{
&[H_2^{(0)}, H_n^{(1)}] =
2i\sumL~\sum_{k=0}^{[n/2]-1}~\sum_{m=1}^{n-2k-1}~
\sum_{\ell=0}^{n-2} (-)^{n+k+m(n-m)+1}\cr
&~~~\times\{(-)^{n+m+1}[h^{(-)^{n+m}}_{n-m-2k+1,j-1}-h^{(-)^{n+m}}_{n-m-2k+1,j}
	+\delta h^{(-)^{n+m}}_{n-m-2k-1,j+1}\cr&~~~~~-\delta
h^{(-)^{n+m}}_{n-m-2k-1,j}]
{}~g^{(-)^{m+1}}_{m,j-m-k+\ell+1}\cr
&~~~~+(-)^{m+1}h^{(-)^{n+m+1}}_{n-m-2k,j}[g^{(-)^{m}}_{m+1,j-m-k+\ell}-
g^{(-)^{m}}_{m+1,j-m-k+\ell+1}\cr&~~~~~
+\delta g^{(-)^{m}}_{m-1,j-m-k+\ell+2}-\delta g^{(-)^{m}}_{m-1,j-m-k+\ell+1}]
\}.}\eq$$
We first relabel the $j$ summation index in each term in order to have all $h$
factors at
site $j$ (e.g. $j-1\ra j$ in the first term, etc.).  Next, in the first,
second,
seventh and eighth terms, we relabel $m$ as: $m-1\ra m$.  We then evaluate the
 summation over
$\ell$. All terms cancel two by two except for those at the boundaries
of the $\ell$ interval.  For the summation over $m$, a similar situation
holds, except
that now two values at both extremities of the interval contribute due to the
$\delta$
factors which are responsible for partial cancellations.  Most of the
resulting terms can
again be canceled if in half of them we reshuffle the $k$ index as: $k+1\ra
k$.  Summing
over
$k$ yields then
$$[H_2^{(0)}, H_n^{(1)}] = 2i\sumL(-)^{n+1}\{ h^{(-)^{n+1}}_{n,j}(
g^{(+)}_{1,j}-g^{(+)}_{1,j+n-1})+  g^{(-)^{n+1}}_{n,j}(
h^{(+)}_{1,j}-h^{(+)}_{1,j+n-1})\},\eq$$
which is exactly the same as $-[H_2^{(1)}, H_n^{(0)}]$.  This shows that, up
to terms
of order $U^2$, $[H_2, H_n]$ indeed vanishes.

\newsec{Concluding remarks}

The main concern of the present
 work is to exhibit the algebraic structure
of conservation laws in integrable spin chains.
For the XYZ model, this has led to a description in terms of
simple multilinear polynomials in spin variables. The terms
describing a charge $H_n$ can be grouped into
classes corresponding to different types of clusters with
prescribed number of spins and holes. The coefficients
of the different terms satisfy recursion relations. In the general
anisotropic case, we have not been able to solve these relations.
In two special cases however, for the XXX  and XY models,
explicit expressions for all the
conserved charges can be found in closed form.
The structure of charges that emerges for the Hubbard model
is organized
in terms of  tensor products of
densities of two XX models.
However, in the absence of a recursive structure
we were able to prove this only for the zero-order and linear terms
in the model's free parameter.

For the XXX and XY models, the explicit construction of the charges
provides an alternative proof of integrability, independent of the
Bethe Ansatz or the transfer matrix formalism. Admittedly, this
pedestrian proof is quite tedious and it does not even adress
the issue of finding the eigenstates and eigenvalues of
the hamiltonian. On the other hand, the proof for the
Heisenberg model
 can be generalized in a completely straightforward way for the
 $su(M)$ invariant XXX chain (which required the use
of the nested Bethe Ansatz).  One may therefore hope that there
are other types of systems in which the unorthodox direct
approach could be effectively used to prove integrability.

One expects to find a similar pattern of conserved charges
in integrable chains with non-trivial boundary conditions.
In particular, for the XYZ model with nonzero boundary terms in the
hamiltonian, the construction of higher order conserved charges
can be achieved by modifying the densities of the
XYZ higher order charges only near the boundary; in the ``bulk,"
the densities of the conserved charges are not affected by boundary
effects. A similar reasoning can be made for integrable  chains
with impurities.
It is also of interest to find explicitly the
deformation of higher order
conserved charges
in integrable models with a general quantum group symmetry.

Another very interesting problem is to determine the
structure of conservation laws in integrable chains with
long-range interactions.   The explicit expressions for
the charges $H_3$ and $H_4$ in the Haldane-Shastry model [\ref{F. D.
Haldane, Z. N. C. Ha, J. C. Talstra, D. Bernard
and V. Pasquier, {\it Phys. Rev. Lett.} {\bf 69} (1992), 2021.}\refname\HHTBP]
are analogous
to the corresponding formulae for
the Heisenberg chain. Indeed, one can regard the
Heisenberg model as a limiting case of a long-range
model. It is thus natural to expect the structure
of the charges in the models of the Haldane-Shastry type
to be similar to the XXX case.

In a different perspective,
the structure of the higher conserved charges for the models
analyzed here
may provide new insights into the problem of testing
integrability for  general spin chains.
The integrability of a quantum chain is
usually demonstrated rather indirectly by
showing that the model can be solved by the coordinate Bethe
ansatz,
or that the
hamiltonian can be derived from a commuting family of transfer matrices
related to the
Yang-Baxter equation.  But these  are only sufficient
conditions for integrability.  Moreover,
testing these sufficient conditions is often not
easy and systematic. Alternatively, one may
look directly at the existence of higher order charges,
with a general structure similar as in the basic models
considered in this work.
A heuristic integrability
test based on this approach will be considered in a
forthcoming publication.

For higher spin chains, anisotropic $su(M)$ chains, or
models with underlying algebra different from $su(M)$,
the determination of the explicit form of the conservation laws
remains a difficult challenge. As we have stressed in the
introduction, the problem is not only with
the actual computation - here one may be helped, to  a certain
extent, by effective computer algebra programs - but in
finding a pattern in the huge
amounts of data that emerge out of these
calculations.
In our view, integrability reflects itself into such structural patterns.


\appendix{A}{Mastersymmetries and hamiltonian structures in classical soliton
theory}

\subsec{Symmetries and mastersymmetries}

We consider continuous systems subject to the evolution equation
$$\vp_t = K_1(\vp),\eqlabel\evol$$
where $\vp$ collectively describes the independent fields under consideration.
 $K_1$
stands for a vector field whose evaluation at the point $\vp$ yields the
scalar quantity $K_1(\vp)$.  The Lie derivative in the direction of $K_1$,
noted
$L_{K_1}$, is the same as the time derivative along the integral curves of the
above
evolution equation.  Thus, the time derivative of a generic tensor field $A$,
that can
depend explicitly upon time, is simply $\partial_t A +L_{K_1}A$.  $A$ is said
to be
invariant with respect to (\evol) if its time derivative vanishes,
$$\partial_t A +L_{K_1}A = 0.\eqlabel\Avf$$
A  vector field satisfying  (\Avf) is called a symmetry of (\evol).
The Lie derivative of
a vector field is defined as
$$ L_{K}A \equiv [K, A]_L = A'[K] - K'[A],\eq$$
where $K'(\vp)[A]$ denotes the Fr\'echet derivative of $K$ at the point $\vp$
in the
direction of
$A$:
$$K'[A] = {\partial \over \partial\ep} K(\vp + \ep A)|_{\ep = 0}.\eq$$
If (\evol) is one member of an integrable hierarchy, then there exists an
infinite
number of time independent vector fields $K_m$ all commuting together:
$$[K_n, K_m]_L = 0.\eq$$
All $K_{m>1}$ are symmetries of (\evol). A mastersymmetry is defined to be a
vector
field $\tau$ whose commutator with any $K_n$ lies in the commutant of $K_n$
[\ref{A. S. Fokas and B. Fuchssteiner, {\it Phys. Lett.} {\bf 86A} (1981),
341;
W. Oevel and B. Fuchssteiner, {\it Phys. Lett.} {\bf 88A} (1981), 323;
B. Fuchssteiner, {\it Prog. Theor. Phys.} {\bf 70} (1983), 1508.},
\ref{W. Oevel, {\it in} {\rm  ``Topics
in Soliton Theory and Exactly Solvable Nonlinear Equations"} (M. Ablowitz et
al, Eds.),
World Scientific, 1986, and {\it Master symmetries: weak action/angle
structure for
hamiltonian and non-hamiltonian systems}, Paderborn preprint (1985).
}\refname\WO, \ref{
A. S.Fokas, {\it Stu. Appl. Math}. {\bf 77} (1987), 253.}\refname\Fok] :
$$[[\tau, K_n],K_n]_L = 0, \quad \forall ~n.\eq$$

\subsec{Hamiltonian structures in integrable systems}

The evolution equation (\evol) is said to be hamiltonian
[\ref{L. A. Dickey, {\rm ``Soliton
Equations and Hamiltonian Systems,"} World Scientific, 1991.},\ref{ L. D.
Faddeev and L. A.
Takhtajan, {\rm ``Hamiltonian Methods in the Theory of Solitons,"}
 Springer Verlag,
1987.}\refname\FTa]
if it can be
written in the form
$$\vp_t = P{\delta H\over \delta \vp} = \{\vp, H\},\eq$$
where the Poisson bracket is defined in terms of the differential operator $P$
by
$$\{\varphi(x), \varphi(y)\} = P(x)\delta(x-y)\eq$$
and
$${\delta H\over \delta \vp}= {\delta \over \delta \vp}\int h~dx  =
\sum_{k\geq 0}
(-\partial_x)^k{\partial h\over \partial(\partial_x^k\vp)}.\eq$$
A (matrix) differential operator $P$ defining a Poisson bracket, satisfying
the
usual requirements of antisymmetry and the
Jacobi identity, is said to be hamiltonian.
Take for instance the KdV equation
$$u_t = u_{xxx} + 6uu_x.\eq$$
It can be written as a hamiltonian system in two different ways
 [\ref{F. Magri, {\it J. Math. Phys.} {\bf 18} (1977), 1405.}]:
$$\eqalign{u_t =&P_1 {\delta H_3\over \delta u} =
\partial_x{\delta \over \delta u}\int
(u^3-\hal~ u_x^2)dx\cr
=&P_2 {\delta H_2\over \delta u} = \hal
(\partial_x^3+4u\partial_x+2u_x){\delta \over \delta
u}~\int u^2dx.\cr}\eq$$ We stress that $P_1$ and $P_2$ define two distinct
Poisson brackets (cf. (\pois)). (Notice also that here the subscript for the
conserved
charges stands for half the usual degree of the density: deg $u$ = 2~deg
$\partial_x$ =2.)  Any
equation in the KdV hierarchy can be written in the form
$$u_t = P_1 {\delta H_{n+1}\over \delta u} =P_2 {\delta H_n\over \delta
u}.\eq$$
This translates into the L\'enard scheme (\biham) for calculating the
conserved
charges.  Such a system is
said to be bi-hamiltonian\foot{More precisely, it is also
required that any linear combination of $P_1$ and $P_2$ must be hamiltonian.}
and it is
characterized to a large extent by its recursion operator
$$R= P_2P_1^{-1}\eq$$
(the adjoint of the
 L\'enard recursion operator).  It is not difficult to show that if both
$P_1$ and
$P_2$ are hamiltonian, then the infinite sequence of operators
$$P_m \equiv R^{m-2} P_2\eq$$
are also hamiltonian.

\subsec{Hamiltonian mastersymmetries for the KdV and NLS equations}

A mastersymmetry not only maps (via $L_\tau$) symmetries into new symmetries,
but also
any invariants (conservation laws, hamiltonian operators, mastersymmetries)
into other
invariants of the same type.  In particular, there is an infinite number of
mastersymmetries.

An almost universal symmetry in integrable systems is the scaling symmetry and
this can
be used as a  convenient structural organizing tool for the description of
mastersymmetries [\WO, \Fok]. Let
$\tau_0$ be the time-independent part of the vector field associated with this
symmetry
(an example is given below). The infinite set of mastersymmetries can be
denoted by
$\{\tau_n\}$
 where in terms of the scaling, $\tau_n$ has degree $n$.
The mastersymmetries satisfy the algebra
$$[\tau_n,\tau_m]_L = c(m-n)\tau_{n+m},\eqlabel\masalg$$
where $n,m\ge 0$ and $c$ is some constant.
  We also have
$$[\tau_n,K_m]_L = a_{n,m}K_{n+m}\eq$$
(with $a_{n,m}=~$constant). These relations show that the whole structure of
the
hierarchy can be extracted out of a degree 1 mastersymmetry $\tau_1$.
Actually, it is
natural to expect that for an evolution equation of the form (\evol),
the existence of a degree 1 mastersymmetry is sufficient to ensure the
existence of an
infinite number
of mastersymmetries, a definite characteristic of an
integrable system.\foot{This point of view has been mainly advocated by
Fuchssteiner.  A related conjecture is that integrability is ensured by the
existence of
one non-Lie point symmetry [\Fok] (section 4 of this reference explains the
connection
between these concepts).  For integrability tests based on symmetries, see
[\ref{A. B.
Shabat and A. V. Mikahilov, {\it in}
{\rm  `` Important Developments in Soliton Theory"}
(A. S. Fokas and V. E. Zakharov, Eds.), p. 355,
Springer Verlag, 1993.}].
}  In terms of
$\tau_1$, the conserved charges can be generated recursively by
$$L_{\tau_1}H_m \equiv H_m'[\tau_1] = H_{m+1}.\eq$$

A mastersymmetry is called hamiltonian if it can be written in the form
$$\tau  = P {\delta B\over \delta \vp},\eq$$
for some hamiltonian operator $P$ and some integral $B$.  For bi-hamiltonian
systems with
scaling symmetry, all mastersymmetries appear to be hamiltonian with respect
to an
appropriate
$P_m$.

Again, the KdV equation offers the simplest illustration of this feature.  We
first
note that this equation is invariant under the following scaling
$$u(x,t)\ra \ep^2u(\ep x, \ep^3 t).\eq$$
This symmetry is generated by the time-dependent vector field
$$\Gamma =2u+ xu_x + 3tu_t,\eq$$
whose time-dependent part is the mastersymmetry $\tau_0$.  It is not difficult
to check
that
$$L_{\tau_0}R = 2 R.\eq$$ An infinite number of mastersymmetries can be
generated out of
$\tau_0$ by the recursion operator $R$ [\WO]:
$$\tau_n = R^n \tau_0 = (\partial_x^2+4u+2u_x\partial_x^{-1})^n
(2u+xu_x).\eq$$
 These $\tau_n$ satisfy the algebra (\masalg) with $c=2$. In particular
$$\tau_1 = x(u_{xxx} + 6uu_x )+8u^2+4u_{xx} +2u_x(\partial_x^{-1}u).\eq$$
Then, denoting
$$B= \hal~\int x u~dx,\eq$$
all $\tau_n$ are seen to be hamiltonian, each with respect to a different
hamiltonian
operator [\ref{W. Oevel, private communication.}]:
$$\eqalign{
\tau_{-1} &= P_1 {\delta B\over \delta u}, \cr
\tau_{0} &= P_2 {\delta B\over \delta u} ,\cr
\tau_{1} &= P_3 {\delta B\over \delta u} ,\cr
&\qquad...\cr
\tau_{n} &= P_{n+2} {\delta B\over \delta u}
.\cr}\eq$$
Since
$$L_{\tau_n} H_m = H_m'[\tau_n] = \int \tau_n {\delta H_m\over \delta u} ~dx =
H_{n+m},\eq$$
the above relations can be reexpressed
in the form
$$\{B, H_n\}_{m} = H_{n+m-2}\eqlabel\recumas$$
(where
$\{~,~\}_{m}$ is defined in terms of $P_m$).  This shows
that a KdV raising ladder operator (mapping $H_n$ to $H_{n+1}$ through some
Poisson
bracket) is necessarily defined in terms of a composite hamiltonian operator,
namely
$$P_3 = P_2 P_1^{-1}P_2.\eq$$
Equivalently, $\tau_1$ is hamiltonian, but with respect to the composite
operator $P_3$.
With respect to $P_2$, $B$ is not a ladder operator.
On the other hand, the first moment of the hamiltonian $H_2$,
that is $\hal~\int
xu^2~dx$, cannot serve as a ladder operator with respect to $P_2$ because
$$P_2{\delta \over \delta u}\hal~\int xu^2~dx = P_2 (xu)\eq$$
is not a mastersymmetry.

For the nonlinear Schr\"odinger equation, the analogue of (\recumas) also
holds, where
now
$$B = \int x\phi^*\phi~dx,\eq$$
and the Poisson bracket defined by $P_1$ is
$$\{\phi^*(x), \phi(y)\} = \delta(x-y),\eq$$
whose quantization yields the usual form of the quantum Schr\"odinger
equation.  This
$B$ acts as a lowering ladder operator:
$$\{B, H_n\}_{1} = H_{n-1}.\eqlabel\recusch$$

For the
quantum version of these theories, the implication of these results is clear.
The quantum nonlinear Schr\"odinger equation is obtained from the
quantization of the first hamiltonian structure while the quantum KdV equation
is obtained
by the quantization of the classical KdV second hamiltonian structure.  As
already
stressed in the introduction, for quantum systems the bi-hamiltonian property
is lost.
The fact  that for both cases no classical ladder operator can be defined in
terms of
the hamiltonian structure appropriate to its quantum version, strongly
suggests that no
quantum ladder operators exists for the corresponding quantum theories.  This
conclusion
is supported by a direct analysis presented in appendix B.

\subsec{Hamiltonian mastersymmetry of the Landau-Lifshitz equation}

The situation described above for the KdV case applies to most bi-hamiltonian
systems. However, there exist exceptional systems for which the
mastersymmetry of degree 1
is hamiltonian with respect to one of the two basic hamiltonian operators
 $P_{1,2}$.  This
is exactly what happens for the Landau-Lifshitz equation, to which we now
turn.

The canonical equation of motion defined by the XYZ hamiltonian
$$H = \sumL \la_a\s^a_j\s^a_{j+1},\eq$$
is
$$
{d\s^a_j\over dt} =[\s^a_j, H] =
2i\la_b\ep_{abc}\s^c_j(\s^b_{j-1}+\s^b_{j+1}).\eqlabel\canqu$$
Its continuous classical limit is obtained from the substitutions [\ref{E. K.
Sklyanin,
LOMI preprint 1979.}]:
$$\eqalign{\s^a_i\ra S^a(x),\qquad&\qquad\s^a_{i+1}\ra S^a(x+\Delta),\cr
\la_a\ra 1+{\Delta^2\over 2}J^a ,\qquad&\qquad t\ra -i\Delta^{-2} t,\cr}\eq$$
with $\Delta$ being the lattice spacing,  and it reads
$${dS^a\over dt} = 2\ep_{abc}S^c(J^b S^b+S^b_{xx}).\eqlabel\LLeq$$
(\LLeq) is called the Landau-Lifshitz equation.  It can be written in
hamiltonian form by
means of the following Poisson bracket
$$ \{S^a(x) ,S^b(y) \} =  2\ep_{abc}S^c(x) \delta(x-y),\eq$$
and the hamiltonian
$$H = \hal~\int (-S^a_xS^a_x+ J^a S^aS^a)~dx.\eq$$
It follows from the field equation that $S^aS^a$ is time independent; we can
thus set
$$S^aS^a = 1.\eq$$
To proceed, we introduce a more compact notation:
$$S\cdot S=S^aS^a~,\qquad
J =\diag \{J^1,J^2, J^3\}~,\qquad(S\wedge~)_{ab}=
2\ep_{abc}S^c,\eq$$ in terms of which the equation reads
$${dS\over dt} =S\wedge(S_{xx}+JS) = S\wedge{\delta H\over \delta S}.\eq$$
The mastersymmetry of degree 1 for this system is [\FUC]
$$\tau_1 = x(S\wedge S_{xx} + S\wedge JS) + S\wedge S_x.\eq$$
Quite remarkably, it is hamiltonian with respect to $P_1\equiv S\wedge$:
$$
\tau_1 = S\wedge{\delta B\over \delta S},\eq$$
with $B$ given by
$$B = \int x(-S_x\cdot S_x+ S\cdot JS)~dx,\eq$$
which is exactly the first moment of the hamiltonian.

The Landau-Lifshitz equation is integrable and it has an infinite number of
conserved
integrals $H_n$ (the density of $H_n$ contains $n~S$-factors).
The corresponding
symmetries are defined by
$$K_n = S\wedge {\delta H_n\over \delta S}.\eq$$
These vector fields can be generated recursively from $\tau_1$ by
$$[\tau_1, K_n]_L = \const~ K_{n+1},\eq$$
which is equivalent to
$$[ S\wedge {\delta B\over \delta S}, S\wedge {\delta H_n\over \delta S}]_L =
\const  ~S\wedge {\delta H_{n+1}\over \delta S}.\eq$$
Using the standard isomorphism between the Lie commutator and the Poisson
bracket
$$[ S\wedge {\delta B\over \delta S}, S\wedge {\delta H_n\over \delta S}]_L
= S\wedge {\delta \over \delta S} (\{B, H_n\}),\eq$$
we end up with [\FUC]:
$$\{B, H_n\} = \const ~H_{n+1}.\eq$$
Therefore, the quantum ladder operator of the XYZ chain model survives in the
continuous classical limit.

\subsec{Higher order mastersymmetries of the classical XX model}

Although higher order ladder operators have not been found for the
quantum XYZ model, they have simple form in the degenerate case of the
XY model, as we have seen in section (6).
One can therefore expect the existence of an
infinite family of hamiltonian mastersymmetries $\{\tau_n\}$ in the classical
continuous limit of the XY model. This is indeed the case, as we now
show.  To simplify discussion we focus on the XX case.

Using the Jordan-Wigner transformation (\JW), the XX conservation laws
\hxypl, \hxym\
can be written, modulo constants, in the form:
$$H_n^{(\pm)}=\sumL a^+_{j+n-1} a_j \mp a^+_j a_{j+n-1},\eq $$
where the fermionic variables satisfy the anticommutation relations
(\anticomf)-(\anticoms) (leaving out unnecessary subscripts).
To obtain the classical continuum limit we represent the
fermionic variables in terms of Grassman (anticommuting) functions
as follows:
$$ a^+_{j+n}=\ksi^+(x+n \delta),\quad a_{j+n}=\ksi(x+n \delta).\eq$$
The Poisson bracket is fixed to be:
$$\{ \ksi(x), \ksi^+(y)\}= \{ \ksi^+(x), \ksi(y)\}=\delta(x-y),\eq$$
$$\{ \ksi(x), \ksi(y)\}= \{ \ksi^+(x), \ksi^+(y)\}=0. \eq$$
The corresponding classical conservation laws become then
(modulo a linear combination of lower order charges):
$$ H_n^{(\pm)}= \int [ (\partial_x^{n-1}\ksi^+ )\ksi \mp
\ksi^+ (\partial_x^{n-1}\ksi) ] dx.\eqlabel\hngrass$$
Integrating by parts, one obtains
$$H_n^{(\pm)}=\int [( 1\pm (-)^n) (\partial_x ^{n-1} \ksi^+)\ksi ]dx,\eq$$
from which one can see that $H^{(+)}_{2n+1}$ and $H_{2n}^{(-)}$ are zero.
In other words, even though the XX model has twice as many charges as
the nondegenerate XYZ model, half of these charges disappear in the
continuous classical limit.

The hamiltonian becomes
$$H_2=\hal H_2^{(+)}=\int \ksi^+ \ksi_x dx, \eq$$
giving the canonical equations of motion:
$$ \ksi_t= \ksi_x, \quad \ksi_t^+=\ksi_x^+\eq $$
The degree zero mastersymmetry is
easily found to be
$$\tau_0=\pmatrix
{x \ksi_x \cr
x \ksi_x^+ + \ksi^+}.\eq$$
Indeed, its Lie commutation with the vector field
$$K=\pmatrix { \ksi_x \cr \ksi_x^+ }\eq$$
gives $[\tau_0, K]_L=0$. Higher order mastersymmetries are easily found  as
$$\tau_n=\pmatrix
{x \partial_x^{n+1} \ksi \cr
\partial_x^{n+1} (x \ksi^+) }.\eqlabel\XXms$$
They satisfy the algebra
$$[\tau_n,\tau_m]=(m-n) \tau_{n+m}.\eq$$
Consider the hamiltonian structure $P_1$:
$$P_1=\pmatrix{ 0&  1\cr 1&  0}, \eq$$
whose components are defined by
$$P_1\delta(x-y)=\pmatrix{
\{\ksi(x), \ksi(y)\}&
\{\ksi(x), \ksi^+(y)\}\cr \{\ksi^+(x), \ksi(y)\}&
\{\ksi^+(x), \ksi^+(y)\} } .\eq$$
All the mastersymmetries in (\XXms) are hamiltonian with respect to
$P_1$, with the quantities
$$B_{n+2}=\int x \ksi^+ \partial_x^{n+1}\ksi dx, \eq$$
playing the r\^ole of the hamiltonians:
$$\tau_n=P_1 \pmatrix{
{\delta\over {\delta\ksi}} \cr
{\delta\over {\delta\ksi^+}} } B_{n+2}.\eq$$

\subsec{Remark on the classical Heisenberg chain}

We conclude this appendix with a remark on the classical lattice Heisenberg
model.
It is interesting to note that the quantum ladder operator does not survive
when only the
classical limit is considered.  Consider for simplicity the isotropic XXX
model, whose
classical version is defined by the hamiltonian [\ref{E. K. Sklyanin,
{\it Funct. Anal. Appl.}
{\bf16} (1982),  263.},
\ref{F. D. M. Haldane, {\it J. Phys. C: Solid State Phys.} {\bf 15} (1982),
L1309.},
\ref{Y. Ishimori, {\it J. Phys. Soc. Japan}, {\bf 51} (1982), 3417.}]
$$H = \sumL \ln(1+S_j^aS_{j+1}^a),\eqlabel\lhm$$
with the constraint
$$ S_j^a S^a_j=1.\eq$$
This expression for the hamiltonian may look somewhat surprising at first
sight.  The naive
choice would have been
$\sumL
S_i^aS_{i+1}^a$, but it turns out to be non-integrable.  In particular, the
candidate $H_3$ having the structure of its quantum relative does
not commute with this naive form of the hamiltonian.  Furthermore,
the canonical equation of motion it induces has no zero-curvature
representation [\FTa].  On
the other hand, recall that the classical limit refers to large values of the
spin, and
that the integrable higher spin version of the quantum Heisenberg model
involves higher
powers of
$S_i^aS_{i+1}^a$, i.e. for spin $s$, it is a particular
polynomial of degree $2s$ in
$S_i^aS_{i+1}^a$ [\ref{L. A. Takhtajan, {\it Phys. Lett.} {\bf 87A} (1982),
 479; H.M. Babujian, {\it Phys. Lett.} {\bf 90A} (1982), 479 .}].  It
thus appears that in the classical
$s\ra
\infty $ limit, the logarithm captures the essential integrability aspect of
this
polynomial.
The lattice Heisenberg hamiltonian (\lhm) with the Poisson bracket
$$\{S^a_i,S^b_j\}=2\ep_{abc}\delta_{ij}S^c_j\eq$$
defines the classical version of (\canqu).  Somewhat surprisingly, there is no
ladder
operator defined with respect to this Poisson structure.  To demonstrate this,
we
introduce
$$B = \sumL  j~f(1+S_j^aS_{j+1}^a),\eq$$
where $f$ is a function to be determined.  By requiring $\{B, H\}$ (with $H$
given by
(\lhm)) to be independent of $j$, $f(X)$ is forced to be $\ln(X)$. The
candidate $H_3$ so produced is of the form
$$H_3 = \sumL  {\ep_{abc}S_j^aS_{j+1}^bS_{j+2}^c\over
(1+S_j^dS_{j+1}^d)(1+S_{j+1}^eS_{j+2}^e)},\eq$$
whose Poisson bracket with $H$ does not vanish.

\appendix{B}{No-go theorem for the existence of a ladder operator for
continuous
integrable systems related to the XYZ model}

\subsec{The nonlinear Schr\"odinger equation}

The first four conservation laws of the quantum nonlinear Schr\"odinger
equation are
$$\eqalign{
H_0 =& \int  \Psi^+ \Psi ~dx,\cr
H_1 =& \int  \Psi^+_x \Psi ~dx,\cr
H_2=&\int (\Psi^+_x \Psi_x +
 \Psi^+\Psi^+\Psi\Psi) dx,\cr
H_3=&\int (\Psi^+_{xx} \Psi_x +
3 \Psi^+_x\Psi^+\Psi\Psi) dx,\cr}\eq$$
and the defining commutation relations are
$$[\Psi^+(x), \Psi(y)]=\delta (x-y)~, \qquad [\Psi(x), \Psi(y)]=0.\eq$$
 For
definiteness, we treat the repulsive case, and the coupling
constant has been rescaled to one.

The problem that we now consider is to try to find an
operator
$B$ such that, modulo lower order charges,
$$[B, H_n] = c_n ~ H_{n+1}\quad {\rm for} \quad n=1,2,
\dots,\eqlabel\searchforb$$
where the coefficients
$c_n$'s are not zero for all $n$ greater then some $n_0$.
Restricting our attention to the four first integrals of motion, which contain
terms with at most two creation and two
annihilation operators,  allows us to reformulate the
problem in the two-particle sector without being restrictive.

Recall
that a complete set of eigenfunctions in the $N$-particle sector is given by
(see e.g.
[\ref{B. Davies, {\it Physica} {\bf A 167} (1990), 433.}])
$$|\la_1,...\la_N\rangle = {1\over \sqrt {N!}}\int
\chi_N(x_1,...,x_N|\la_1,...,\la_N)
\Psi^+(x_1)...\Psi^+(x_N)|0\rangle,\eq$$
where
$$\chi_N = \sum_P (-)^P \prod_{j>k} [\la_{P_j}- \la_{P_k}-i \sgn(x_j-x_k)]
e^{i\sum_{n=1}^N x_n \la_{P_n}} .\eq$$
$P$ is a permutation of the set $(1,2,...,N)$. In the two-particle sector, the
expression of the conservation laws ${\hat H}_{1,2,3}$ is (up to
irrelevant
multiplicative factors)
$$\eqalign{
{\hat H}_1 =&\partial_1+\partial_2,\cr
{\hat H}_2 =&\partial_1^2+\partial_2^2- 2\delta (x_1-x_2) ,\cr
{\hat H}_3 =&\partial_1^3+\partial_2^3- 3\delta
(x_1-x_2)(\partial_1+\partial_2) ,\cr}\eq$$
with $\partial_i \equiv \partial_{x_i}$.

We now look for a differential operator $b$ such that
$$[{\hat H}_1, b] = c_1 {\hat H}_2~, \qquad [{\hat H}_2, b] = c_2 {\hat
H}_3~,\eqlabel\nlsb$$ for some constants $c_2$ and $c_1\ne 0$.
It is simple to see that
the first relation is satisfied by the following choice for $b$:
$$ x_1\partial_1^2+x_2\partial_2^2- (x_1-x_2)\delta (x_1+x_2)\eq$$
(this gives $c_1=1$).  Actually this first commutation relation determines $b$
only up to
a term of the form
$$f(x_1-x_2) M(\partial_1,\partial_2),\eq$$
where $f$ is an arbitrary function of the difference $x_1-x_2$ and $M$ is a
general
polynomial in the derivatives, which does not contain any $x_i$ dependent
terms.
The commutator of
$$b  = x_1\partial_1^2+x_2\partial_2^2- (x_1-x_2)\delta (x_1+x_2)+f(x_1-x_2)
M(\partial_1,\partial_2)\eq$$
with ${\hat H}_2$ yields
$$[{\hat H}_2,b] = 2(\partial_1^3+\partial_2^3)M
-4\delta(\partial_1+\partial_2) +2(\partial_1^2f)M +
2(\partial_1f)M(\partial_1-\partial_2) + 2f(M\delta -
\delta M),\eq$$
where $\delta$ is a shorthand for $\delta(x_1+x_2)$.
We want to write the rhs of
the last expression in the form
$$c_2{\hat H}_3 = c_2 [\partial_1^3+\partial_2^3- 3\delta
(x_1-x_2)(\partial_1+\partial_2)].\eq$$
This leads to the following equation for the functions $f$ and $M$:
$$\eqalign{ 2(\partial_1^2f)M + 2(\partial_1f)M(\partial_1-\partial_2)& +
2f(M\delta -
\delta M) = (c_2-2) (\partial_1^3+\partial_2^3)\cr
&~~ +(4-3c_2) \delta
(x_1-x_2)(\partial_1+\partial_2).\cr}\eqlabel\fmeqa$$
  First suppose that
$c_2\not=2$. Given that the higher order operators on each side of this
equation must be
the same, under the assumption that $(\partial_1f)\not=0$, it forces
$$2(\partial_1f)M(\partial_1-\partial_2)=  (c_2-2)
(\partial_1^3+\partial_2^3).\eq$$
Since there is no $x_i$ dependence on the rhs, $(\partial_1f)$ has to be a
constant. With
$$f =
a_1(x_1-x_2)+a_2,\eq$$
where by hypothesis $a_1\not=0$, $M$ is found to be
$$M={c_2-2\over a_1-1}
(\partial_1^3+\partial_2^3)(\partial_1-\partial_2)^{-1}.\eq$$
Equation (\fmeqa) reduces to
$$2f(M\delta -
\delta M) =(4-3c_2) \delta (\partial_1+\partial_2).\eq$$
With the expressions just obtained for $f$ and $M$, the last equation is
not satisfied.
There is thus no solution when $c_2\not=2$ and $(\partial_1f)\not=0$.  A
similar argument shows
with the assumption
$(\partial_1f)=0$ there is no solution either.  With $c_2=2$, (\fmeqa) becomes
$$ 2(\partial_1^2f)M + 2(\partial_1f)M(\partial_1-\partial_2) +
2f(M\delta -
\delta M) =  -2 \delta (x_1-x_2)(\partial_1+\partial_2),\eq$$
for which it is easily seen that again there is no solution.
This demonstrates the
non-existence of an operator $b$ that could satisfy (\nlsb). In other
words, (\searchforb) is impossible for $c_1\ne 0$. A similar argument,
making use of the Jacobi identity, excludes the possibility that
$c_1=\dots=c_{n_0-1}=0$ with $c_{n_0}\ne 0$ in (\searchforb).

\subsec{The quantum KdV equation}

In terms of the quantum extension of the classical KdV field, denoted by $T$,
the quantum
KdV equation reads [\KM]:
$$ \partial_t T = [T,H] \quad,\quad H = {1\over 2\pi i}\oint dw~(TT)(w)
,\eqlabel\qkdv$$
The parentheses stand for the usual normal
ordering [\ref{F. A. Bais, P.
Bouwknegt, K. Schoutens and M. Surridge, {\it Nucl. Phys.} {\bf B304} (1988),
348.}]
 i.e.  $$(AB)(w) = {1\over 2\pi i}\oint_w {dx\over x-w} A(x)B(w),\eq$$
and the subscript $w$ indicates that the contour integration circulates once
around the
singular point $w$.  The second Poisson structure is now replaced by its
quantized form,
which we take as the radial operator product expansion [\ref{A. A. Belavin, A.
M. Polyakov
and A. B. Zamolodchikov, {\it Nucl. Phys.} {\bf B 241} (1984), 333.}] (the
global form of the
Virasoro algebra [\ref{J. L. Gervais, {\it Phys. Lett.} {\bf B160} (1985),
277.}])
$$T(z)T(w) = {c/2\over (z-w)^4} + {2T(w)\over (z-w)^2} + {2\partial_wT(w)\over
z-w} +...,\eq$$
where the dots indicate irrelevant regular terms and $c$ is the central charge,
a free
parameter of the quantum theory that cannot be scaled out.
The commutator in (\qkdv) translates into an
operator product expansion
$$[T(z),H] = {1\over 2\pi i}\oint_z dw~T(z)(TT)(w).\eq$$
This equation can be written as
$$\partial_t T = {(1-c)\over 6} \partial_z^3T -3\partial_z(TT).\eq$$
The first three conservation laws are [\KM]
$$\eqalign{
H_1 =& \oint T~dz,\cr
H_2 =& \oint (TT)~dz,\cr
H_3 =& \oint [(T(TT))-{(2+c)\over 12}(\partial_zT\partial_zT)]~dz.\cr}\eq$$

Following the same strategy as in the nonlinear Schr\"odinger case, we look for
an
integral $B$ satisfying (\searchforb).  The first commutation relation is
satisfied with
the choice
$$B = \oint z (TT)~dz + B_0,\eq$$
where $B_0$ is in the commutant of $H_1$.  With $B_0=0$, it is simple to check
that
$[B, H_2]$ is not proportional to $H_3$.

Can we find a non-zero $B_0$ that commutes with $H_1$?
Clearly $B_0 = \oint z^n A~dz$ is not in the commutant of $H_1$ for any $n>0$.
If $n=0$,
the only dimensionally possible choice for a local expression for $A$ is
proportional to $
\partial_zT$, which makes
$B_0$ vanish. The remaining possibility is to have a non-local expression for
$A$.
Because the pole term in the operator product $T(z)A(w)$ has residue
$\partial_wA(w)$,
we see that this will indeed  commute with $H_1$.  However, in the commutation
with
$H_2$, it will produce non-local terms; since there are no such terms in $H_3$,
this
last possibility is ruled out.

These arguments can be made even more explicit at $c=-2$ [\ref{P. Di Francesco,
P. Mathieu
and D. Senechal, {\it Mod. Phys. Lett.} {\bf A 7} (1992), 701.}\refname\FMS].
For this value of
the central charge,
$T$ can be represented by the bilinear
$$T = (\phi\psi),\eq$$
where $\phi$ and $\psi$ are both fermions of spin 1 with operator product
expansion
$$\phi(z)\psi(w) = {-1\over (z-w)^2}\quad,\quad
\psi(z)\phi(w) = {1\over (z-w)^2}.\eq$$
In terms of these fields, the infinite set of conserved quantities is
$$ H_{k/2+1} = \oint dz(\phi^{(k)}\psi)(z),\eq$$
where $\phi^{(k)}=\partial_z^k\phi$.
For $k$ even these integrals can be reexpressed in terms of $T$ as follows
[\FMS, \ref{R. Sasaki and I. Yamanaka, {\it Adv. Stud. in Pure Math.} {\bf 16}
(1988), 271.}]:
$$H_{n} = {2^{n-1}\over n}\oint dz \Tn{n}(z),\eq$$
where
$$\Tn{n} = (\dots(((TT)T)T)\dots T)\qquad (n~\hbox{factors}).\eq$$
{}From
$$ (TT)(z) = \hal~(\phi''\psi + \phi\psi'')(z),\eq$$
where a prime stands for a derivative w.r.t. $z$, we have
$$B= B_1+B_0\qquad {\rm with} \quad B_1=\oint z \phi''\psi dz .\eq$$
A simple calculation gives
$$[H_2,B_1] = (2\pi i)\oint w[\phi''\psi'''+\phi^{(5)}\psi](w)~dw
,\eq$$
which is clearly not proportional to $H_3 = \oint \phi^{(4)}\psi~dw$ because
$$[\phi''\psi'''+\phi^{(5)}\phi]\not=\partial_z (...).\eq$$
At $c=-2$, all conservation laws are bilinear in the two fermionic fields so
$B_0$ is also
necessarily bilinear (otherwise non-bilinear terms would be generated from
single
contractions).  The most general non-local and bilinear form for
$B_0$ is
$$B_0 = a\oint z (\partial_z^{-m}\phi)( \partial_z^{m+2}\phi)~,\qquad a=\const
.\eq$$
This is indeed in the commutant of $H_1$.
By integration by parts, this can be rewritten in the form
$$B_0 = a(-)^m\oint z\phi''\psi~dz + a(-)^m (m+2) \oint \phi'\psi~dz.\eq$$
The first term is proportional to $B_1$, while the second one is simply
$H_{3/2}$, which
commutes with all $H_n$.
Thus, with the most general form for $B$, we must still conclude that
$[H_2,B]$ cannot be proportional to $H_3$.

\vskip2cm
\centerline{\bf Acknowledgment}
We thank W. Oevel for very useful discussions concerning the material of
Appendix A.
\bigskip \hrule \bigskip \centerline{\sc{references}}

\immediate\closeout\refs \vskip 0.5cm
  \message{References}\input references

\end